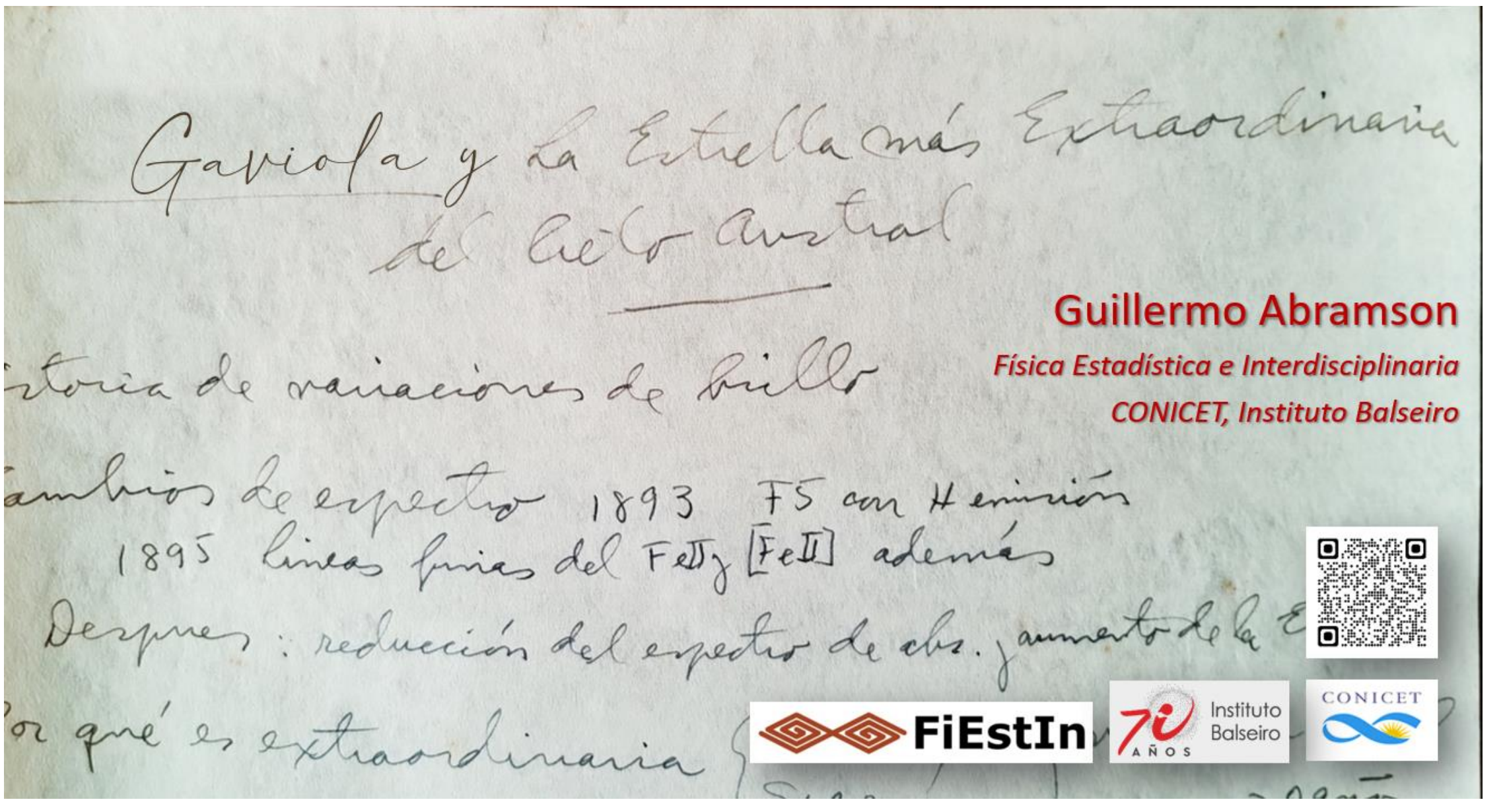

# Gaviola y la estrella más extraordinaria del cielo austral

Guillermo Abramson

*División Física Estadística e Interdisciplinaria, Centro Atómico Bariloche,*
*CONICET, Instituto Balseiro*

El presente texto se basa en la conferencia de homenaje a Enrique Gaviola, dictada el 14 de agosto de 2025 en el Observatorio Astronómico de Córdoba, con motivo del 125º aniversario de su nacimiento. La actividad fue organizada por el OAC, el IATE, el Instituto Gaviola y la FAMAF de la Universidad Nacional de Córdoba. El trabajo ofrece una reflexión personal sobre su vida, su obra y su legado, a partir de los documentos del Archivo Gaviola del Instituto Balseiro y de los testimonios de colegas que lo conocieron.

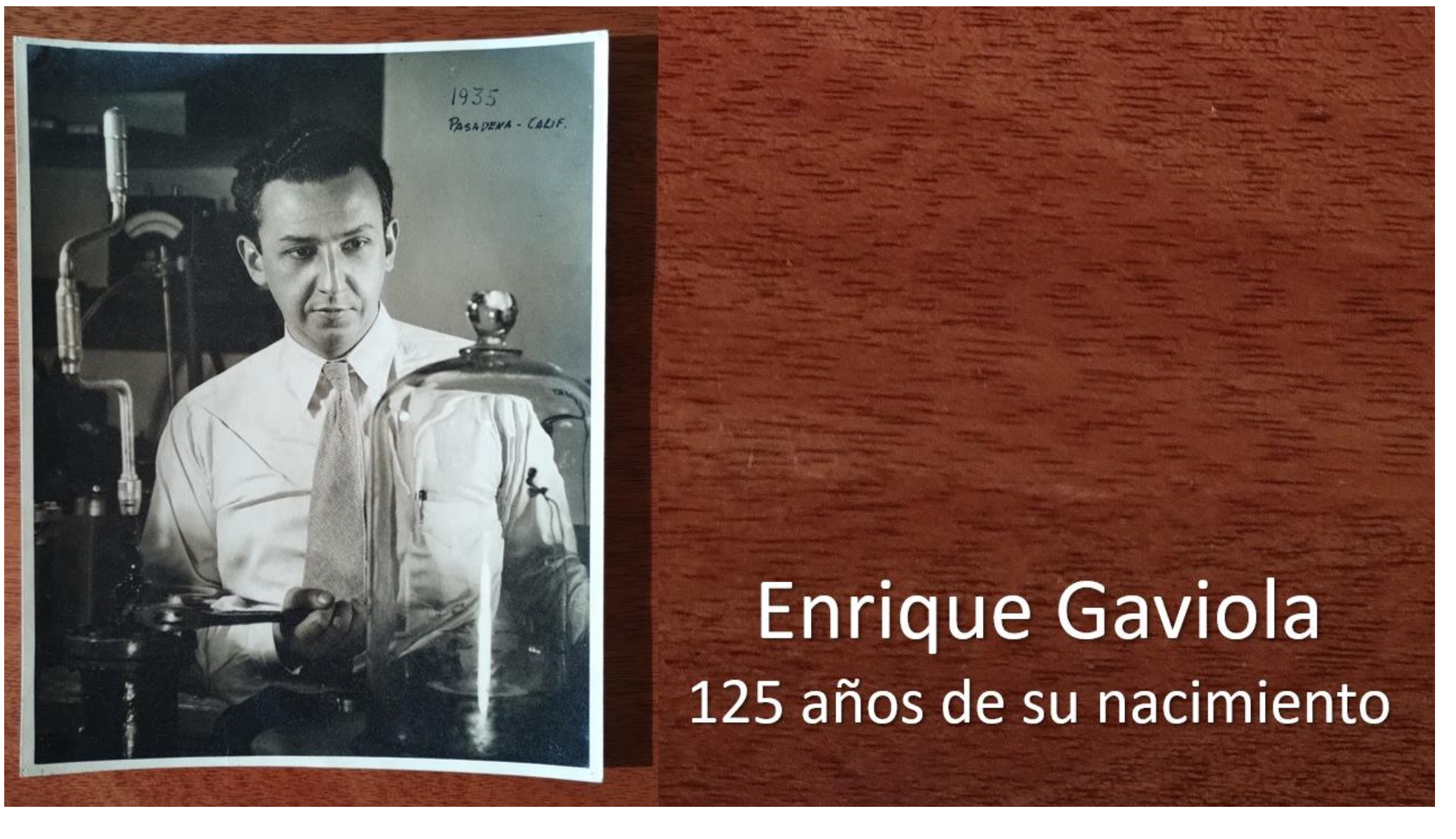

Enrique Gaviola tiene un vínculo muy cercano con el Instituto Balseiro, y es además uno de mis ídolos personales, así que estoy encantado de que hayan invitado a este homenaje. Pero no quiero hacer una biografía exhaustiva de Gaviola, ni un elogio de su carrera. Para eso está el libro de Bernaola (Enrique Gaviola, y el Observatorio Astronómico de Córdoba). Quiero más bien darles la perspectiva que tengo desde Bariloche, como físico y aficionado a la astronomía, y en particular la sensación especial que me ha dado, cada vez que lo consulté, el extenso archivo de Gaviola en nuestra Biblioteca, y las cosas que me contaron quienes lo conocieron. El archivo sólo en parte está catalogado y organizado, en las carpetas que se ven en la figura. Una buena parte está todavía en cajas, que contienen verdaderos tesoros desconocidos.

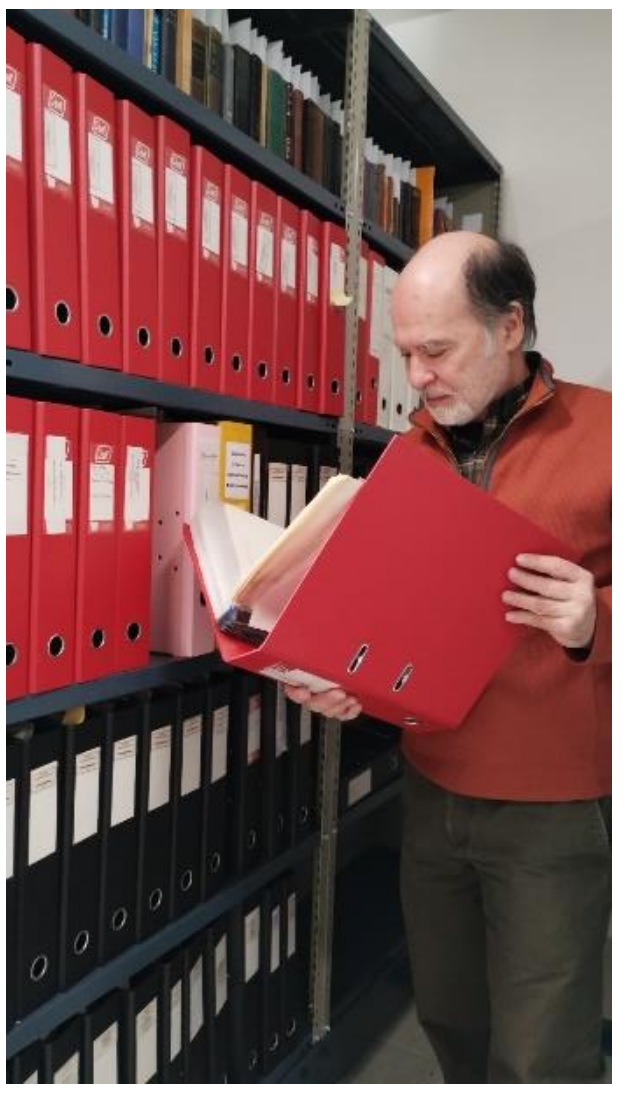

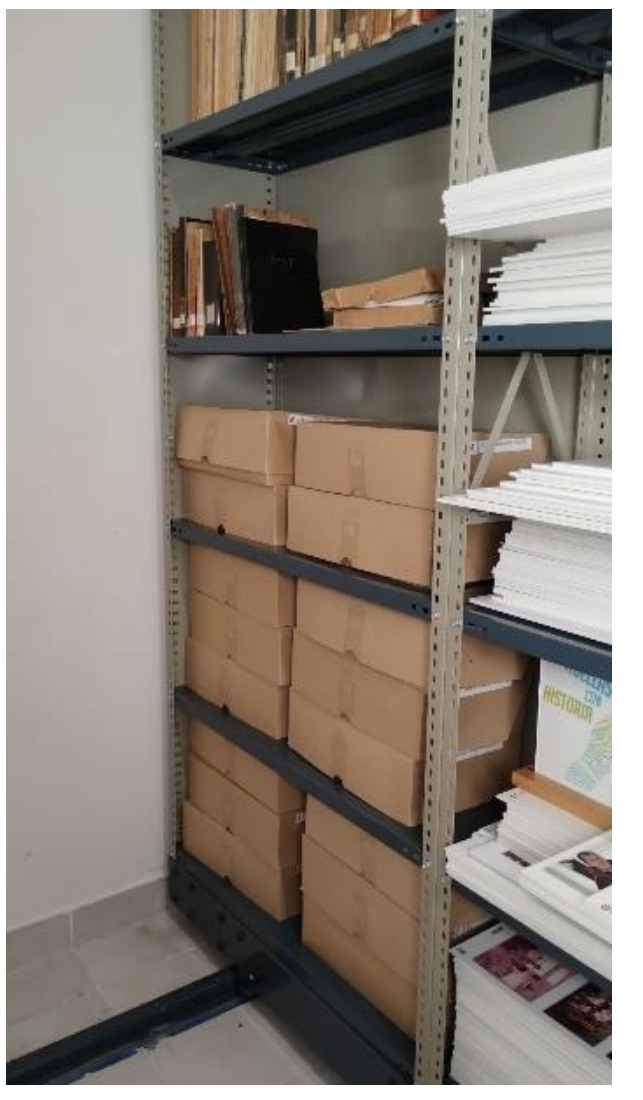

Hace poco se pusieron en exhibición algunos objetos personales, sus patines, y la raqueta de tenis. Carlos Balseiro, que lo conoció de niño y después fue su alumno, me contó que Elena Gaviola era una mujer encantadora (a diferencia de Enrique, que era "intimidante") y que jugaba al tenis con ella cuando era chico, así que tal vez sea la raqueta de ella.

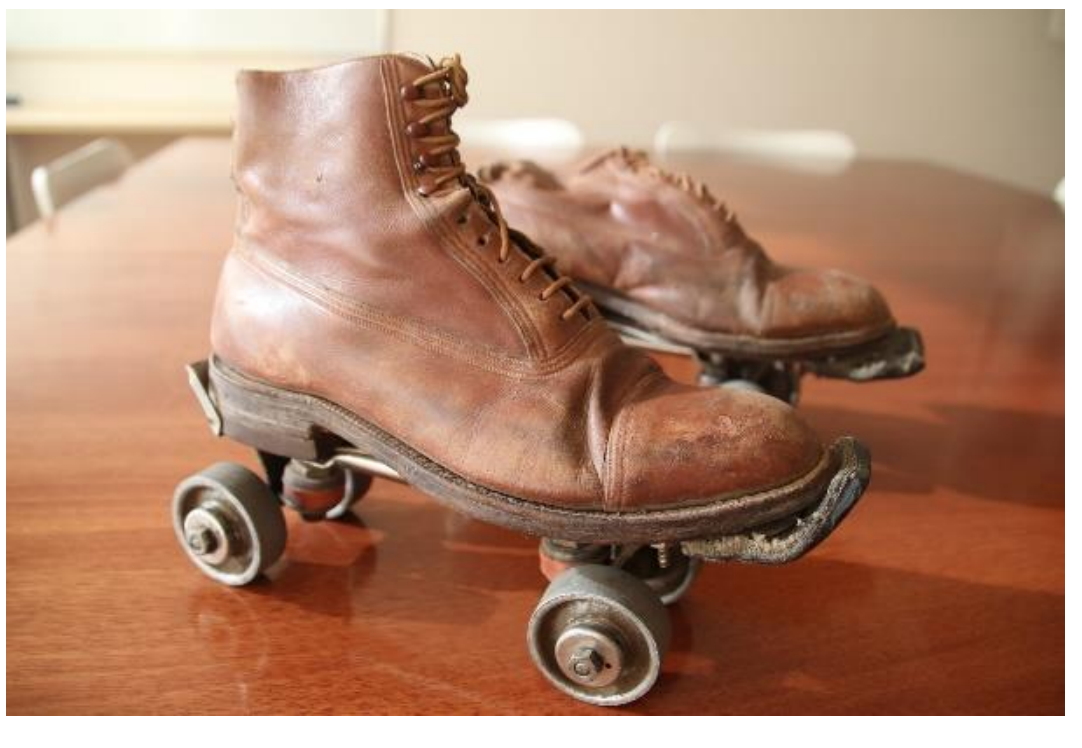

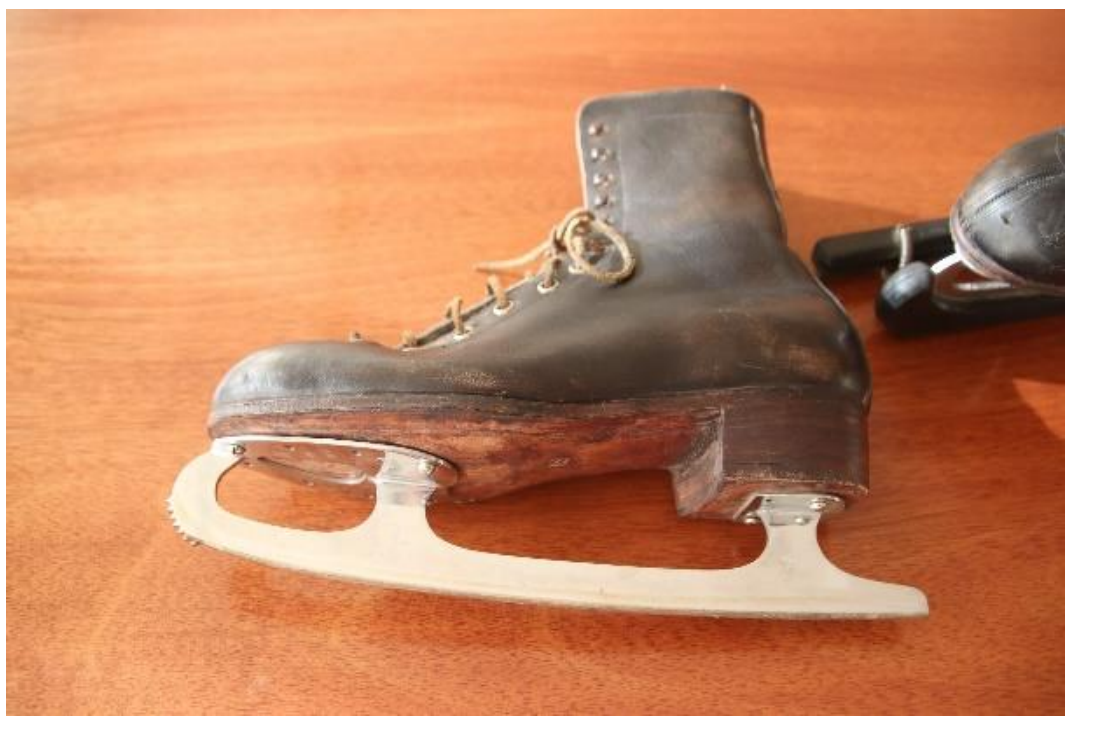

En las cajas hay también muchos libros personales, casi todo ensayo, casi nada de literatura. Su gran colección de revistas científicas antiguas sí estuvo en uso en la hemeroteca durante décadas, yo la he consultado cuando era estudiante.

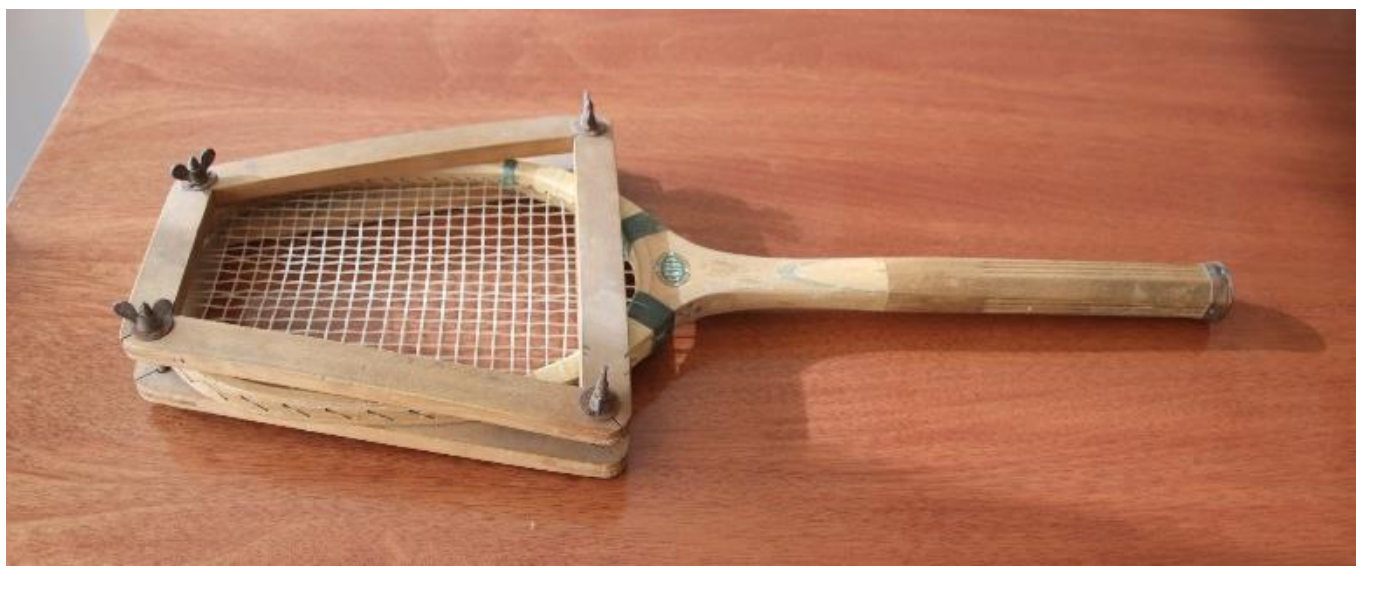

En 1917, junto con su hermano, Gaviola se fue a La Plata a estudiar ingeniería. La Universidad Nacional de La Plata, fundada por Joaquín V. González algunos años antes, era la primera universidad argentina dedicada principalmente a la ciencia. Tenía un Departamento de Física buenísimo, con un presupuesto del primer mundo, dirigido por un excelente físico alemán, Richard Gans. En los cursos de Gans, Gaviola descubrió que además de la ingeniería existía la física. El curso lo impresionó tanto que, al año siguiente, junto con un par de compañeros, decidieron tomar apuntes detallados del curso. Son estos tres tomos.

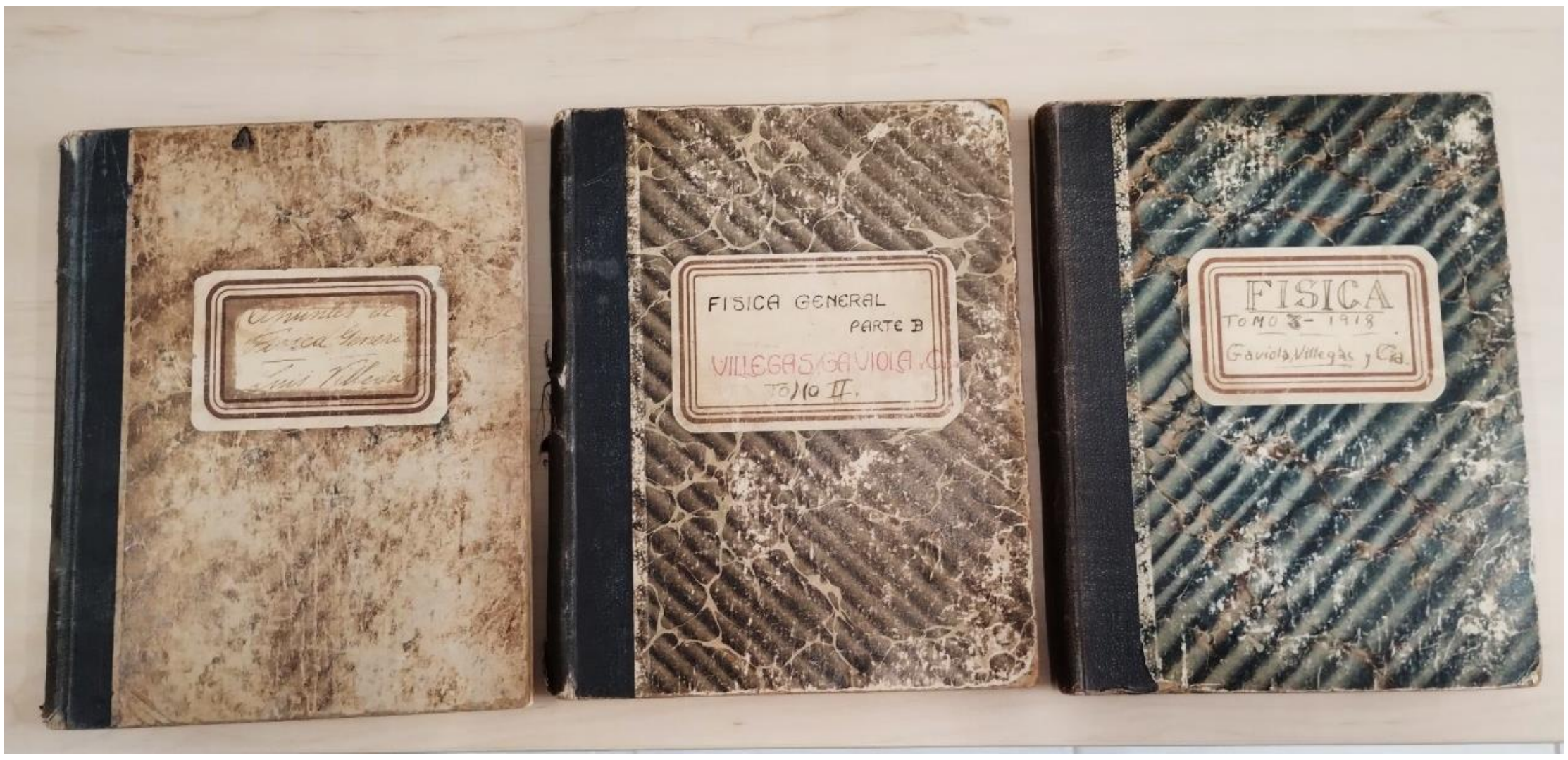

Los cuadernos son encantadores. Gaviola se lamentaba de no haber hecho lo mismo en 1917, dice que todavía no había descubierto el valor de tomar buenos apuntes de clase. Hay que imaginar que los chicos tomaban estos apuntes con pluma y tintero, y en la traza se aprecia cómo se iba acabando la tinta a medida que escribían.

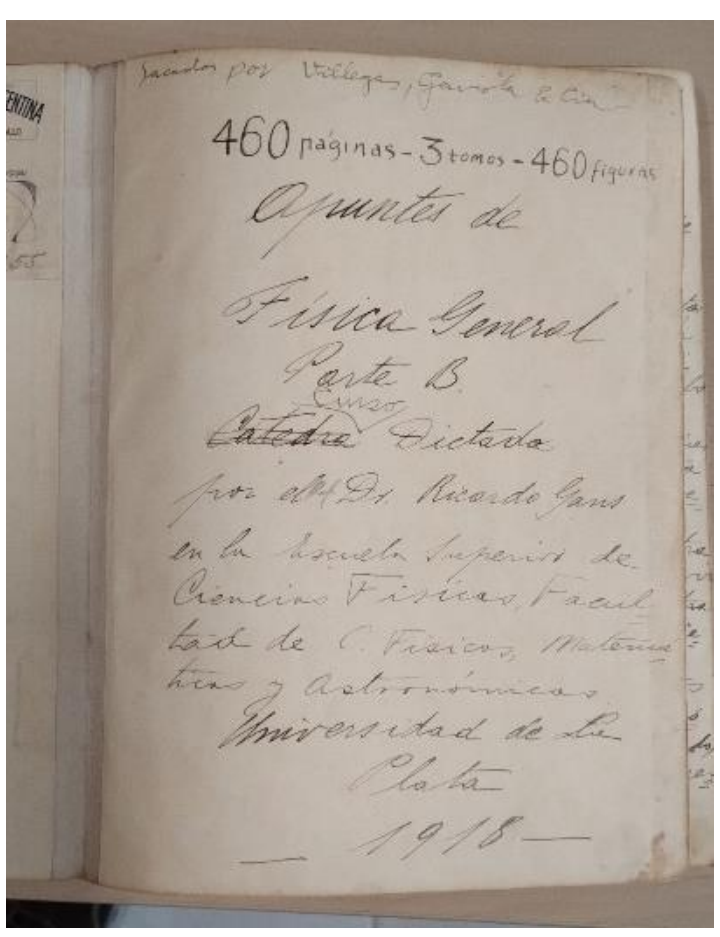

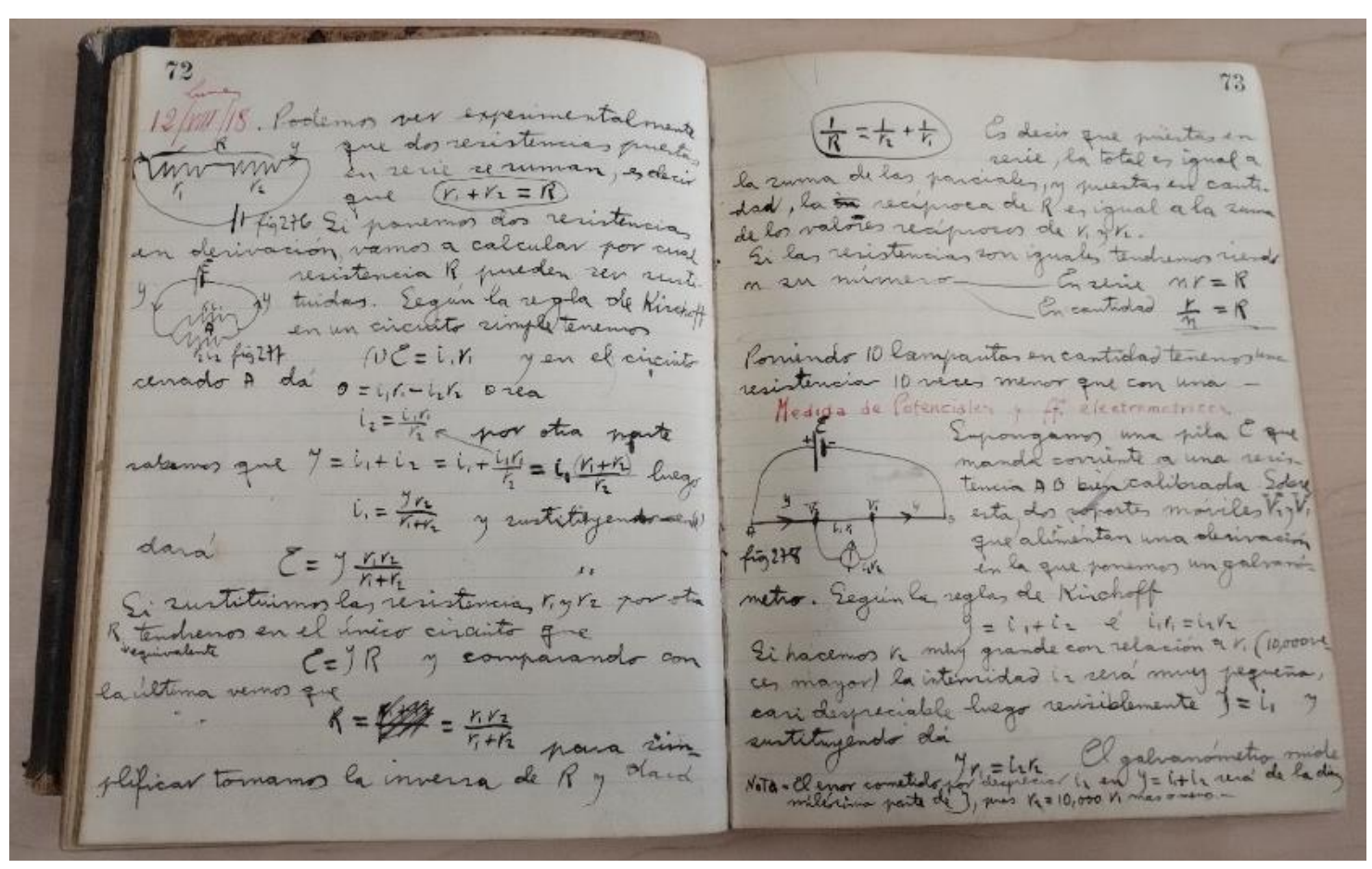

A fin de 1918 se los mostró a Gans, que se quedó muy impresionado, y le dijo:

*«Si Ud. quiere estudiar física, tiene que irse a Alemania».*

Gaviola lo consideró. Tenía 18 años, la familia no tenía fortuna, no era fácil conseguir una beca, así que volvió a La Plata en 1919 y se recibió de Agrimensor. Dice que se le atrasó la cursada por la Reforma Universitaria.  Éste es el título, con fecha abril de 1921.

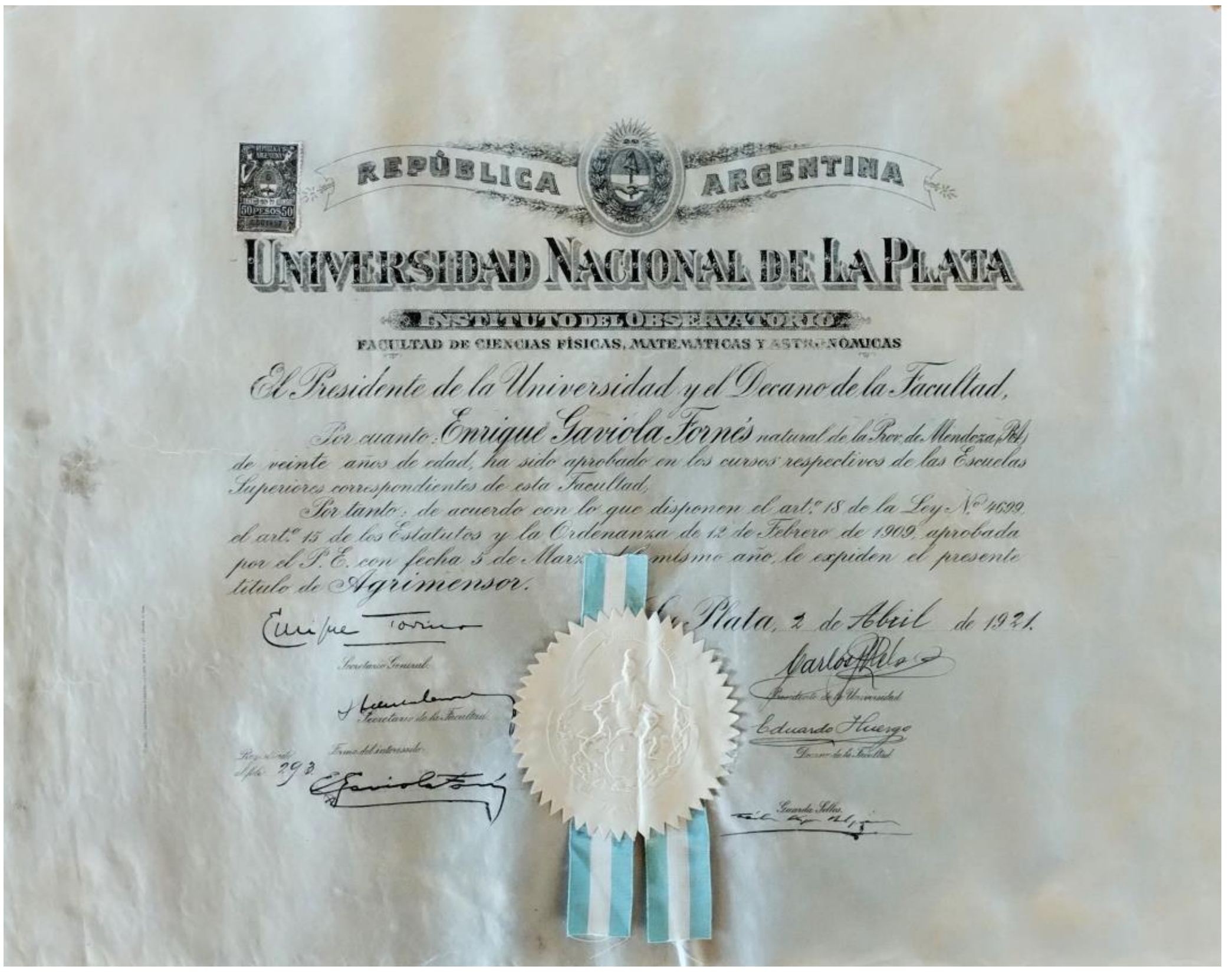

REPÚBLICA ARGENTINA

UNIVERSIDAD NACIONAL DE LA PLATA

INSTITUTO DEL OBSERVATORIO

FACULTAD DE CIENCIAS FÍSICAS, MATEMÁTICAS Y ASTRONÓMICAS

El Presidente de la Universidad y el Decano de la Facultad,

Por cuanto Enrique Gaviola Fornés natural de la Prov. de Mendoza, R.A. de veinte años de edad, ha sido aprobado en los cursos respectivos de las Escuelas Superiores correspondientes de esta Facultad,

Por tanto, de acuerdo con lo que disponen el art.º 18 de la Ley Nº 4699, el art.º 15 de los Estatutos y la Ordenanza de 12 de Febrero de 1909, aprobada por el P.E. con fecha 5 de Marzo mismo año, le expiden el presente título de Agrimensor.

La Plata, 2 de Abril de 1921.

Secretario General

Secretario de la Facultad

Firma del interesado

Presidente de la Universidad

Eduardo Huergo

Decano de la Facultad

Guarda Sellos

Volvió a Mendoza, trabajó de agrimensor un año, y con lo que ahorró, en 1922 se fue a Alemania. Se embarcó en el Cap Polonio, un vapor transatlántico bastante lujoso, que hizo el trayecto de Hamburgo a Buenos Aires durante muchos años. Había sido botado en 1914, pero por la guerra recién en 1922 empezó a cubrir la ruta para la cual fue construido. El de Gaviola debe haber sido uno de los primeros viajes, estaría flamante. Es el mismo barco que trajo a Einstein a la Argentina en 1925, hace 100 años, pero Einstein viajó en 1ª, obviamente, mientras que Gaviola viajó en 3ª.

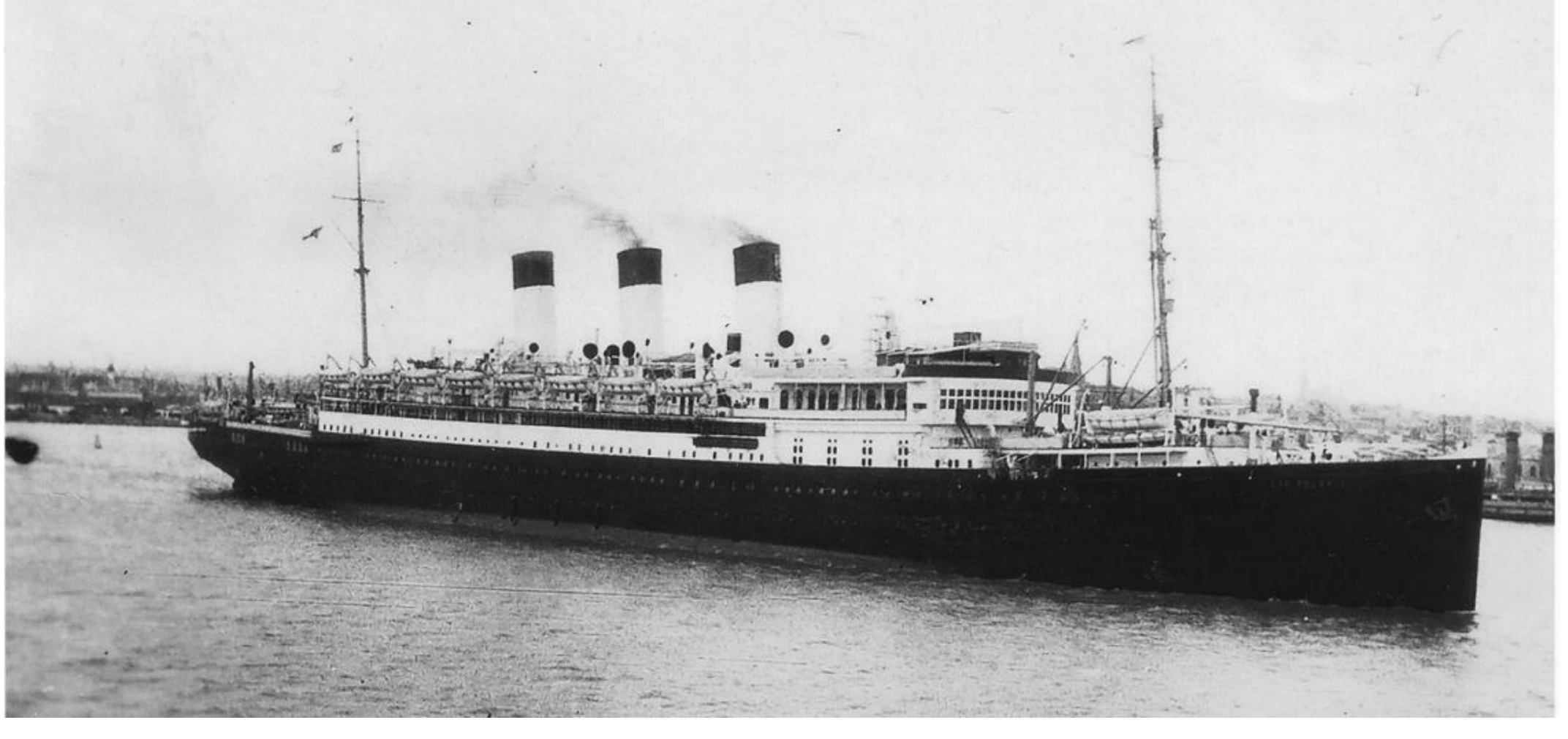

En abril del 22 llegó a Göttingen, con una carta de Gans dirigida a Robert Wichard Pohl, que era el director del Instituto de Física. En esta foto aparecen los cuatro "peces gordos" (*die*

*Bonzen*) del Instituto: Pohl (a la derecha) y sus amigos James Frank (tomándole el brazo), Max Born junto a él, y Max Reich a la izquierda. Nótese qué bien vestidos están.

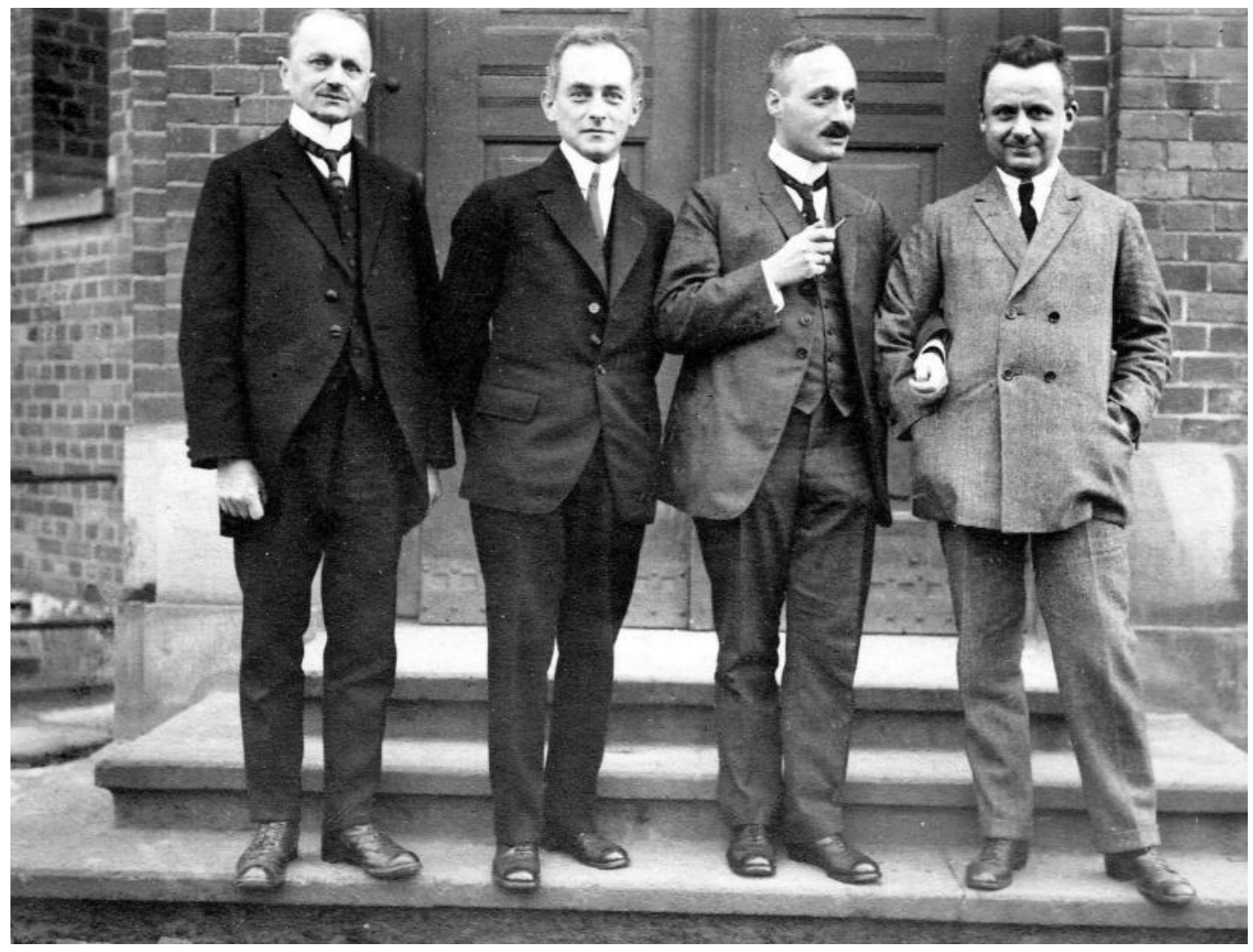

Göttingen en ese momento era el centro de la física mundial. Era donde se estaba haciendo la revolución de la mecánica cuántica. Pohl y sus amigos eran los peces gordos, pero estaban también los "pibitos": Pauli, Heisenberg, Jordan, Oppenheimer, todos los que en esos años revolucionaron la física. Y además los que venían de visita, especialmente en verano: venían Dirac de Inglaterra, Bohr de Dinamarca, Schroedinger de Zurich, Fermi de Pisa, Gamow de Leningrado. Todos los que querían hacer algo en física venían a Göttingen.

Pohl leyó la carta de Gans y se lo presentó al Secretario. El Secretario le dio una libreta con su nombre y le dijo que se anotara en las materias que quisiera. Así nomás. Nadie miró siquiera sus diplomas de Bachiller o de Agrimensor.

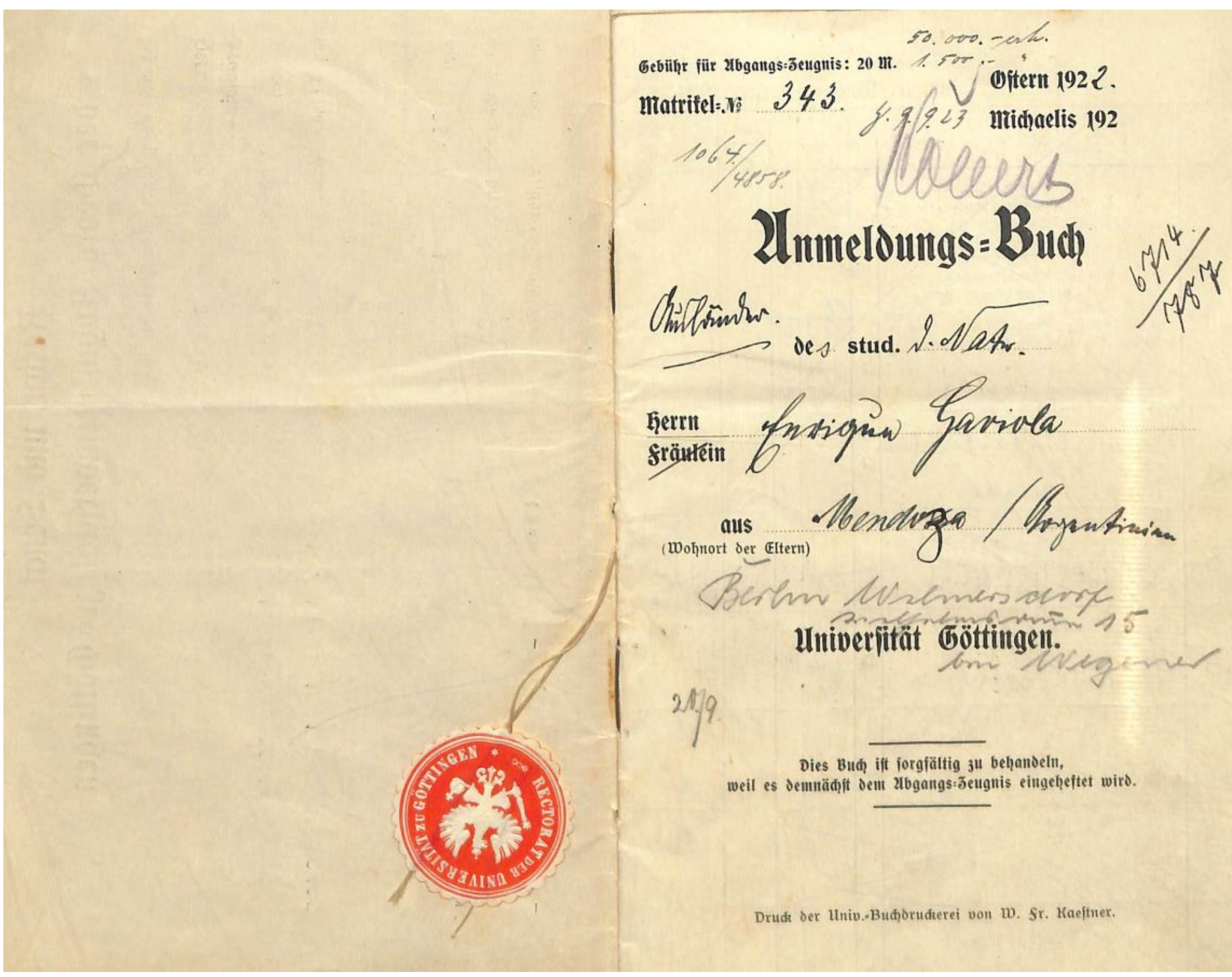

Gebühr für Abgangs-Zeugnis: 20 M. 1.500.– 50.000.–

Ostern 1922.

Matrikel-№ 343. 8.9.923

Michaelis 192

1064/4858.

# Anmeldungs-Buch

Ausländer. des stud. d. Natw.

Herrn Enrique Gaviola

aus Mendoza / Argentinien
(Wohnort der Eltern)

Universität Göttingen.

Dies Buch ist sorgfältig zu behandeln,
weil es demnächst dem Abgangs-Zeugnis eingeheftet wird.

Druck der Univ.-Buchdruckerei von W. Fr. Kaestner.

Durante sus años de estudiante Gaviola siguió escribiéndose con Gans. Se conservan muchas cartas y tarjetas postales, pero sólo la primera está en castellano. Es la que aparece en la imagen de aquí al lado. Allí Gans le recomienda que se quede por lo menos dos años y que aprenda la lengua. Y que si tiene problemas económicos le avise, que el le va a escribir a sus profesores, que los conoce a todos.

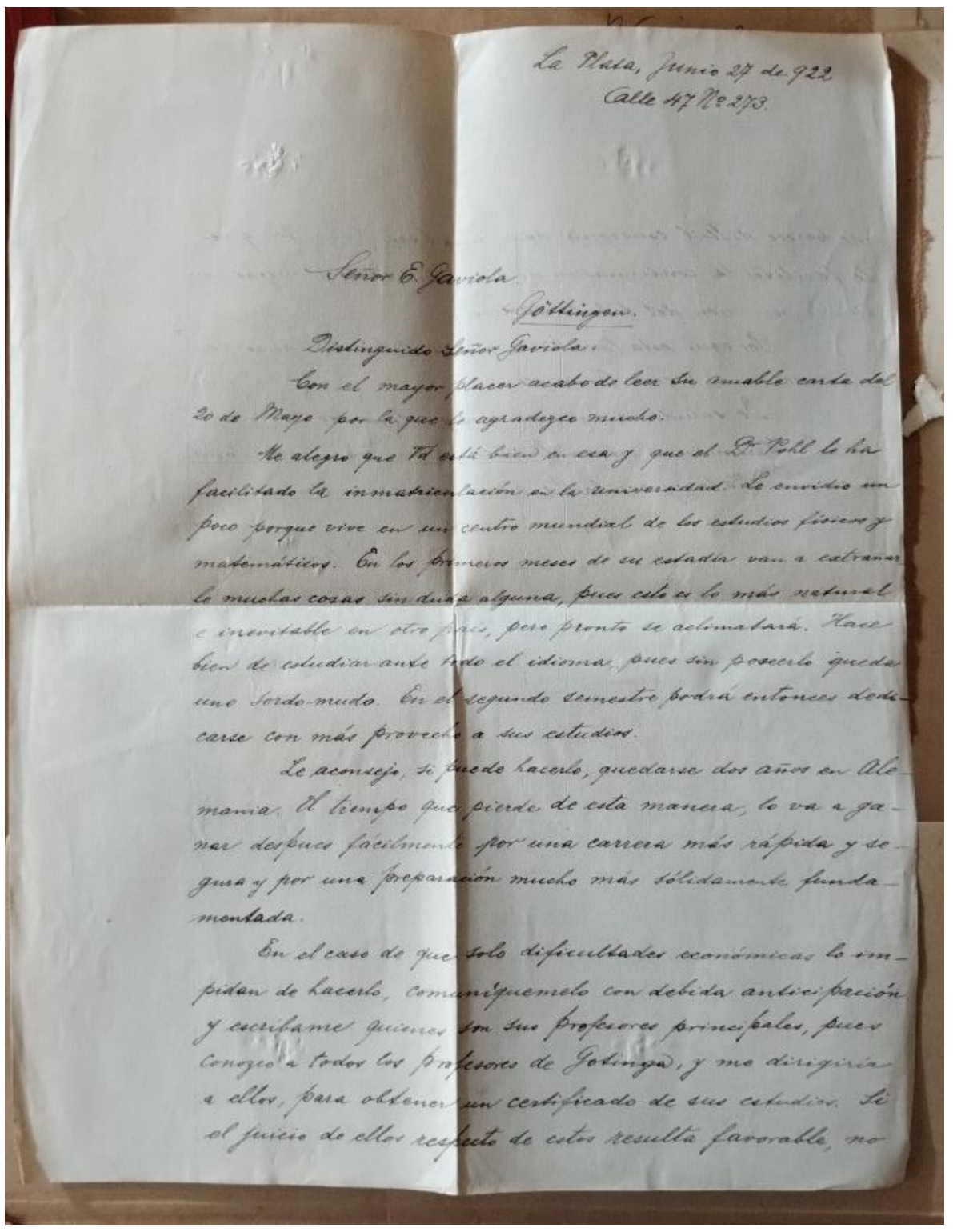

La Plata, Junio 27 de 922
Calle 47 Nº 273.

Señor E. Gaviola
Göttingen.

Distinguido Señor Gaviola:

Con el mayor placer acabo de leer su amable carta del 20 de Mayo por la que le agradezco mucho.

Me alegro que Vd esté bien en esa y que el Dr. Pohl le ha facilitado la inmatriculación en la universidad. Le envidio un poco porque vive en un centro mundial de los estudios físicos y matemáticos. En los primeros meses de su estadía van a extrañar le muchas cosas sin duda alguna, pues esto es lo más natural e inevitable en otro país, pero pronto se aclimatará. Hace bien de estudiar ante todo el idioma, pues sin poseerlo queda uno sordo-mudo. En el segundo semestre podrá entonces dedicarse con más provecho a sus estudios.

Le aconsejo, si puede hacerlo, quedarse dos años en Alemania. El tiempo que pierde de esta manera, lo va a ganar después fácilmente por una carrera más rápida y segura y por una preparación mucho más sólidamente fundamentada.

En el caso de que solo dificultades económicas lo impidan de hacerlo, comuníquemelo con debida anticipación y escríbame quienes son sus profesores principales, pues conozco a todos los profesores de Gotinga, y me dirigiría a ellos, para obtener un certificado de sus estudios. Si el juicio de ellos respecto de estos resulta favorable, no

En el 25 Gans regresó a Europa, y años después fue perseguido por el nazismo. Le pidió ayuda a Gaviola, que al principio no pudo hacerlo. Pero finalmente, después de la Guerra, cuando estaba en una mejor posición, Gaviola logró sacarlo de Alemania y traerlo de nuevo a Argentina.

Estos son los primeros cursos, verano de 1922 (abril a septiembre). Probablemente son sólo 3 porque, además, ¡estaba aprendiendo alemán! (El curso tachado aparece al año siguiente.)

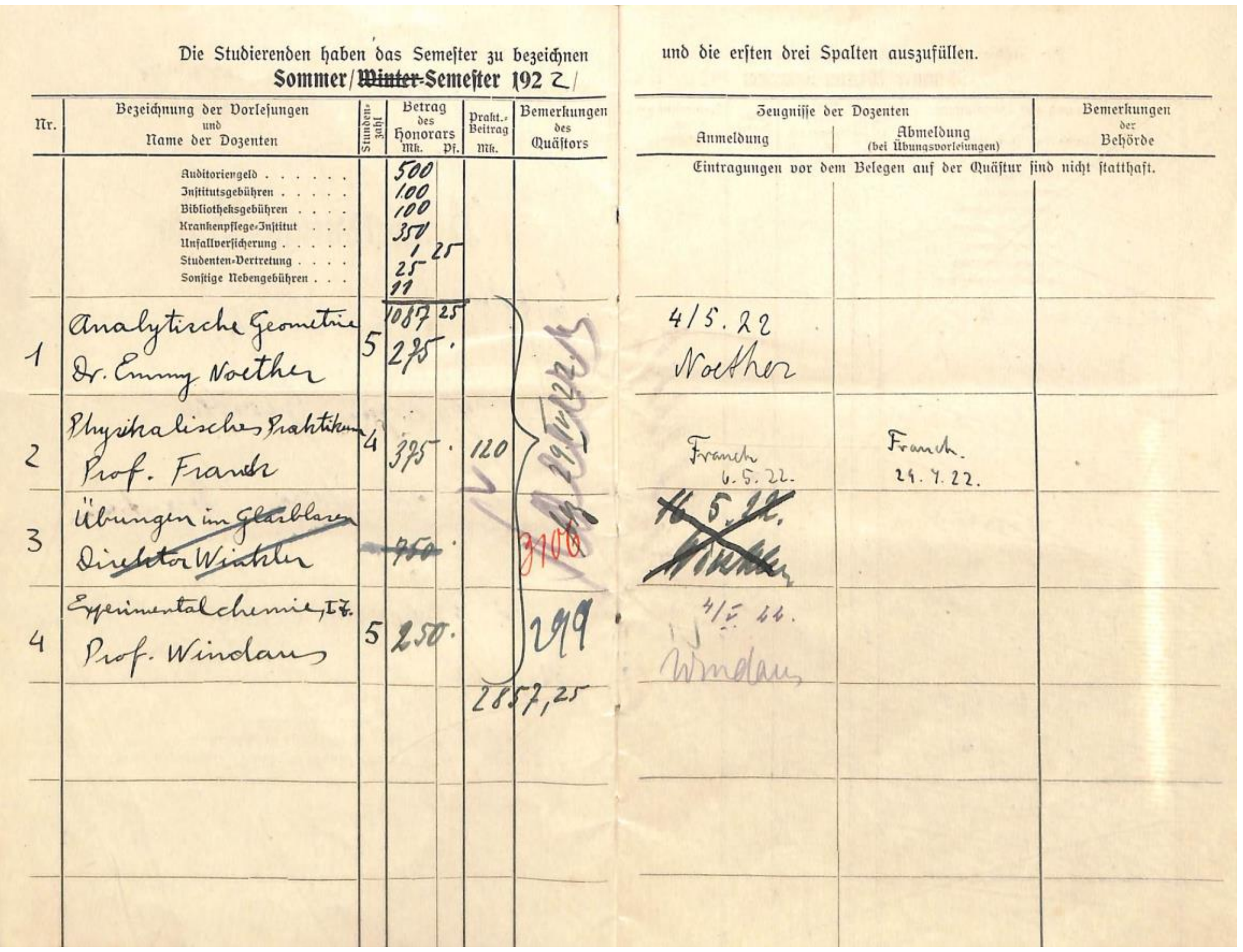

Die Studierenden haben das Semester zu bezeichnen und die ersten drei Spalten auszufüllen.

Sommer/~~Winter~~-Semester 1922

| Nr. | Bezeichnung der Vorlesungen und Name der Dozenten | Stundenzahl | Betrag des Honorars Mk. | Pf. | Prakt.-Beitrag Mk. | Bemerkungen des Quästors | Zeugnisse der Dozenten: Anmeldung | Abmeldung (bei Übungsvorlesungen) | Bemerkungen der Behörde |
|---|---|---|---|---|---|---|---|---|---|
| | Eintragungen vor dem Belegen auf der Quästur sind nicht statthaft. | | | | | | | | |
| | Auditoriengeld | | 500 | | | | | | |
| | Institutsgebühren | | 100 | | | | | | |
| | Bibliotheksgebühren | | 100 | | | | | | |
| | Krankenpflege-Institut | | 350 | | | | | | |
| | Unfallversicherung | | 1 | 25 | | | | | |
| | Studenten-Vertretung | | 25 | | | | | | |
| | Sonstige Nebengebühren | | 11 | | | | | | |
| | | | 1087 | 25 | | | | | |
| 1 | Analytische Geometrie Dr. Emmy Noether | 5 | 275 | | | | 4/5.22 Noether | | |
| 2 | Physikalisches Praktikum Prof. Franck | 4 | 375 | | 120 | | Franck 6.5.22. | Franck 21.7.22. | |
| 3 | ~~Übungen im Glasblasen Direktor Wichler~~ | | ~~150~~ | | | 3106 | ~~16.5.22. Wichler~~ | | |
| 4 | Experimentalchemie, I.T. Prof. Windaus | 5 | 250 | | | 299 | 4/5 22. Windaus | | |
| | | | | | 2857,25 | | | | |

Vemos que cursó *Geometría Analítica* con la gran matemática Emmy Noether, autora de uno de los resultados más profundos de la física teórica. Hizo también *Laboratorio* con James Franck (Nobel 1926), y *Química* con Adolf Windaus (Nobel de Química 1928). James Franck ganó el premio Nobel de Física en por su trabajo con Gustav Hertz sobre la descarga electrónica en un vapor de mercurio, que demostraba la validez del modelo atómico de Bohr. En Bariloche tenemos este aparato, que posiblemente haya sido diseñado y construido por Gaviola en los años que estuvo a cargo de Física Experimental, en la década del 60. La bomba de vacío, de vidrio, está hecha a mano, y el diseño seguro que es de los años 20, de hace 100 años. Yo llegué a usarlo en 1987. Era dificilísimo, me produjo tal frustración que decidí dedicarme a la física teórica. Tal vez porque no conté con la ayuda de Gaviola.

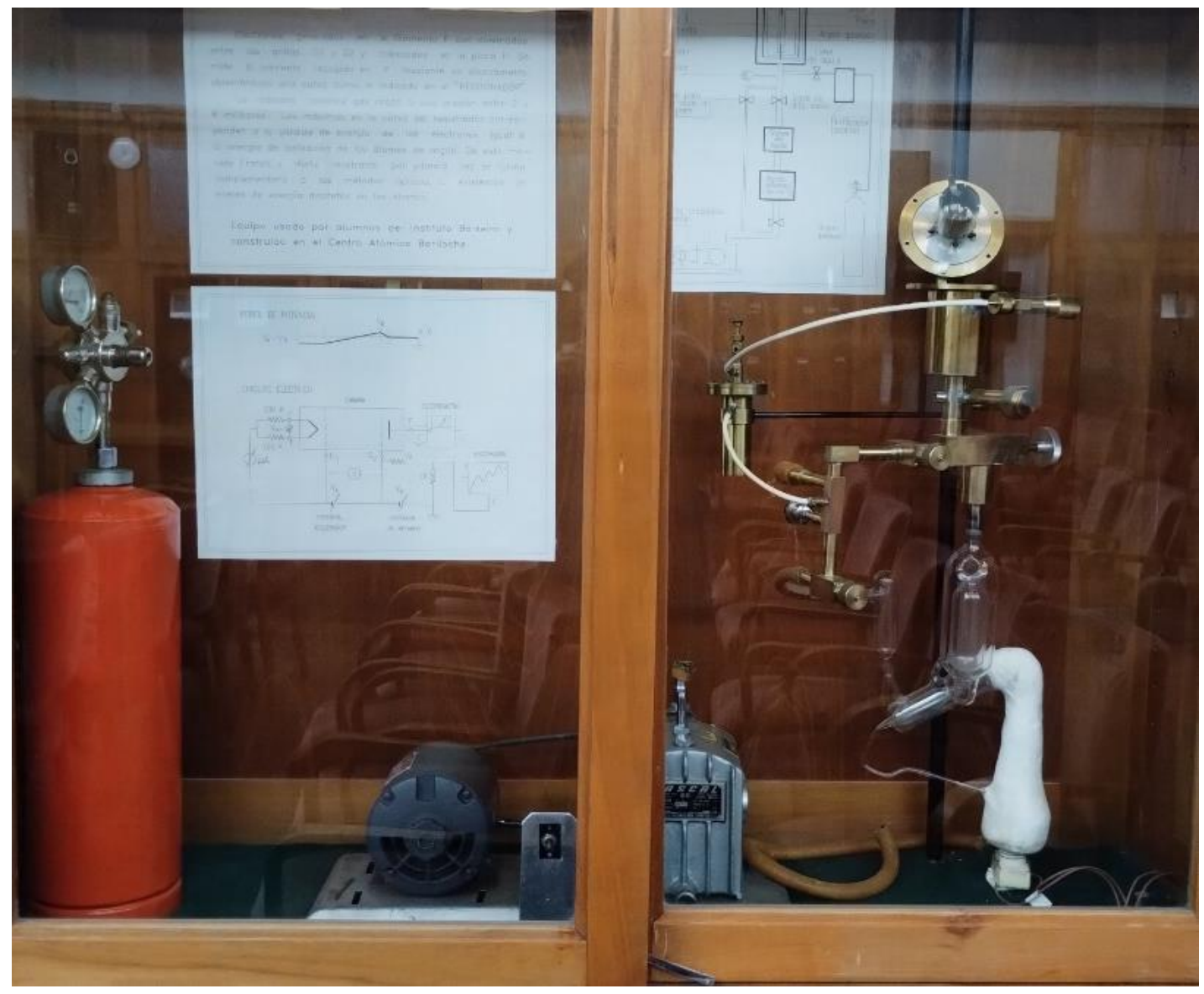

En el invierno del 22/23 (octubre a marzo) hay 11 materias. Estas son las 6 que aparecen primero:

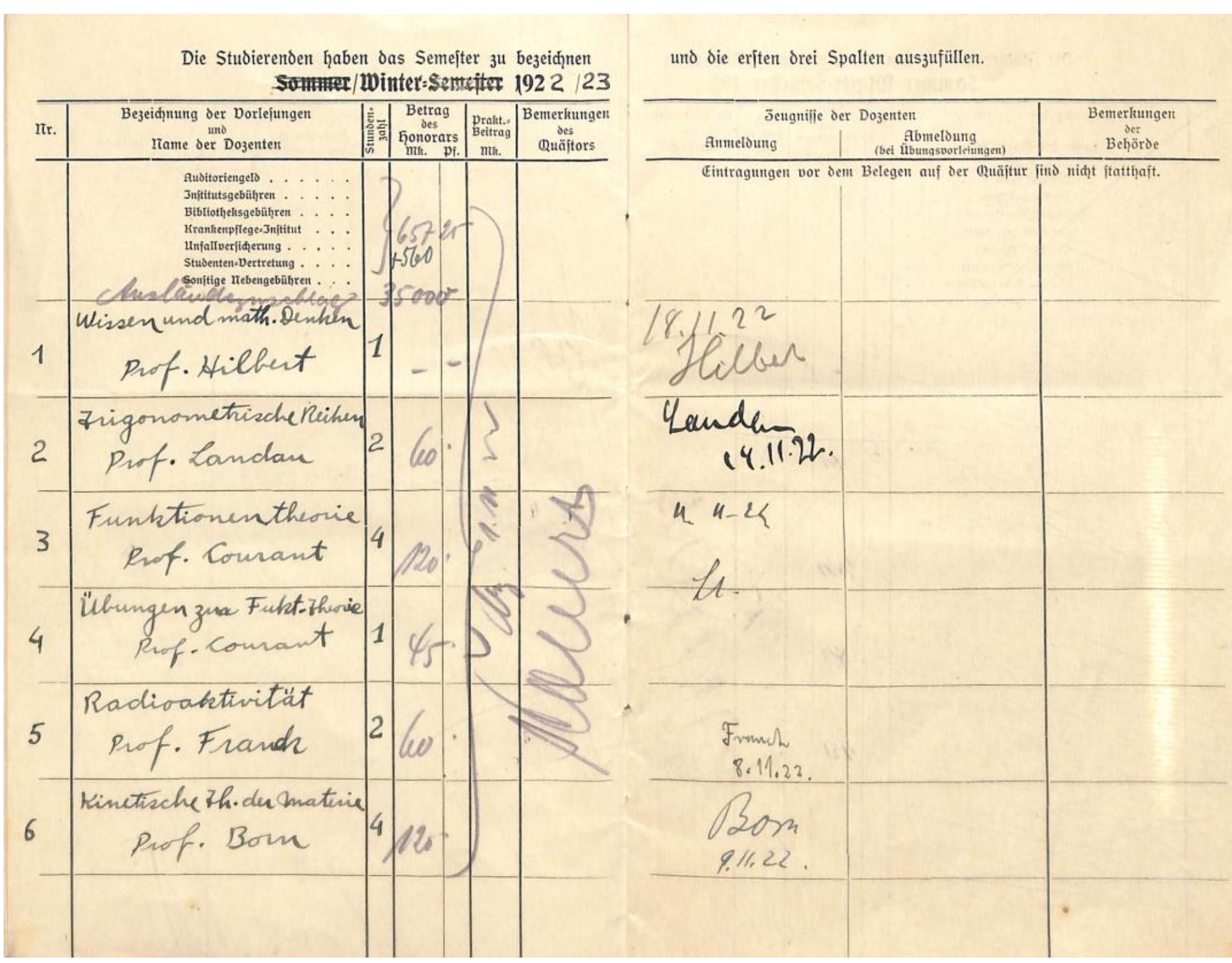

Die Studierenden haben das Semester zu bezeichnen und die ersten drei Spalten auszufüllen.

~~Sommer~~/Winter-Semester 1922/23

Eintragungen vor dem Belegen auf der Quästur sind nicht statthaft.

| Nr. | Bezeichnung der Vorlesungen und Name der Dozenten | Stundenzahl | Betrag des Honorars Mk. Pf. | Prakt. Beitrag Mk. | Bemerkungen des Quästors | Zeugnisse der Dozenten: Anmeldung | Abmeldung (bei Übungsvorlesungen) | Bemerkungen der Behörde |
|---|---|---|---|---|---|---|---|---|
| | Auditoriengeld, Institutsgebühren, Bibliotheksgebühren, Krankenpflege-Institut, Unfallversicherung, Studenten-Vertretung, Sonstige Nebengebühren | | 657 25 +560 | | | | | |
| | Ausländerzuschlag | | 35000 | | | | | |
| 1 | Wissen und math. Denken Prof. Hilbert | 1 | – – | | | 18.11.22 Hilbert | | |
| 2 | Trigonometrische Reihen Prof. Landau | 2 | 60 | | | Landau 14.11.22. | | |
| 3 | Funktionentheorie Prof. Courant | 4 | 120 | | | [illegible] | | |
| 4 | Übungen zur Funkt.-theorie Prof. Courant | 1 | 45 | | | [illegible] | | |
| 5 | Radioaktivität Prof. Franck | 2 | 60 | | | Franck 8.11.22. | | |
| 6 | Kinetische Th. der Materie Prof. Born | 4 | 120 | | | Born 9.11.22. | | |

Cursó *Conocimiento y Pensamiento Matemático* con David Hilbert, que era el más destacado e influyente matemático del momento. Göttingen había sido un importante centro de matemática durante siglos (una tradición que establecieron Gauss, Dirichlet, Riemann, Minkowski y otros), y recién con la llegada de Pohl y sus amigos se transformó en un importante centro de física.

Edmund Landau le dio *Trigonometría*.  Era un matemático destacado en análisis complejo y teoría de números, uno de los fundadores de la Universidad Hebrea de Jerusalén, yerno del premio nobel Paul Ehrlich.

Por si fuera poco con Hilbert y Noether, lo tuvo a Richard Courant en *Teoría de Funciones*. Era un destacadísimo colaborador de Hilbert, inventor del método de elementos finitos. El libro de texto de Métodos de la Física Matemática de Courant y Hilbert, publicado en 1924, se sigue editando y usando hasta hoy en día.

Max Born (Nobel 1954), quien era uno de los físicos más destacados del desarrollo inicial de la Mecánica Cuántica, fue su profesor de *Teoría Cinética de la Materia*. Y aparece nuevamente James Franck, esta vez en *Radioactividad*. Los apuntes de Radioactividad están en la parte que le había quedado libre del Tomo 3 de sus cuadernos de La Plata. Ahí están los niveles del átomo de Bohr que le valieron a Franck su premio Nobel.

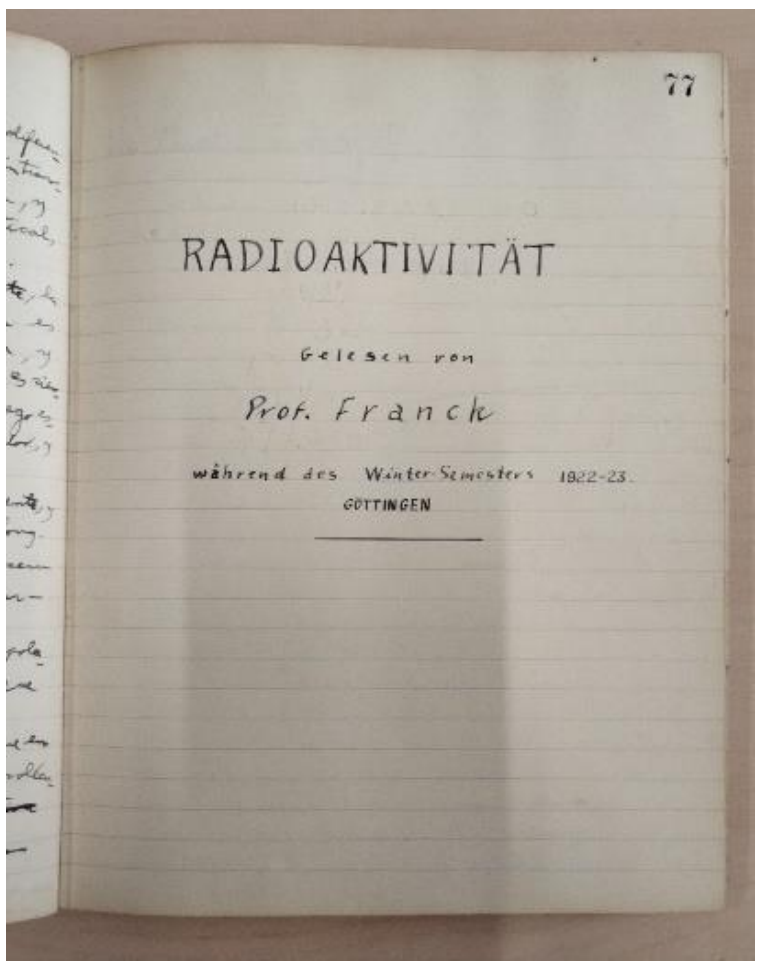

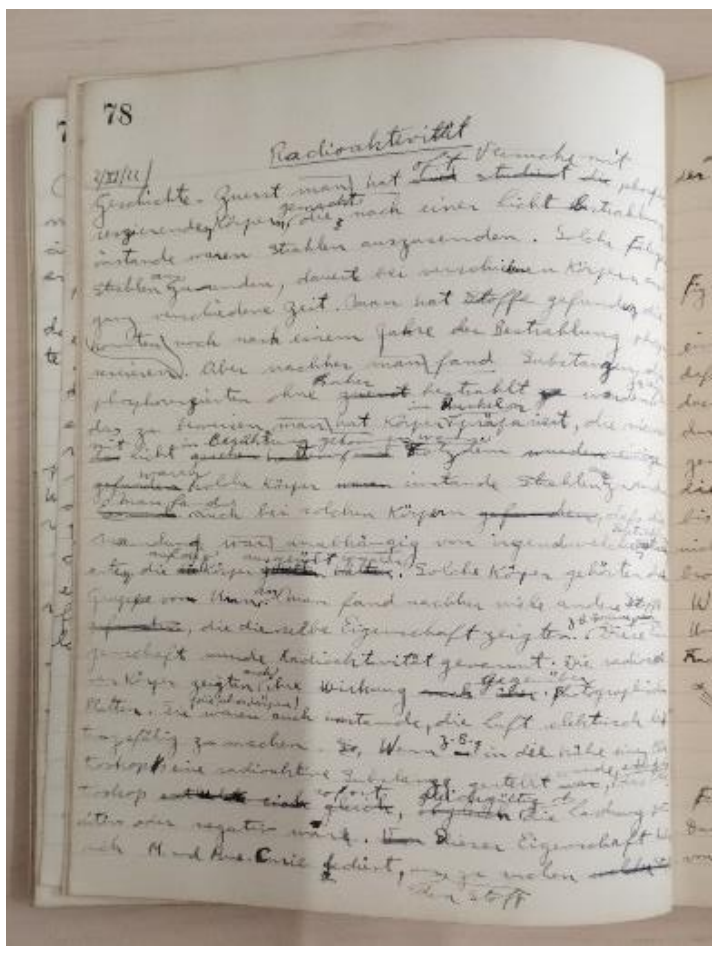

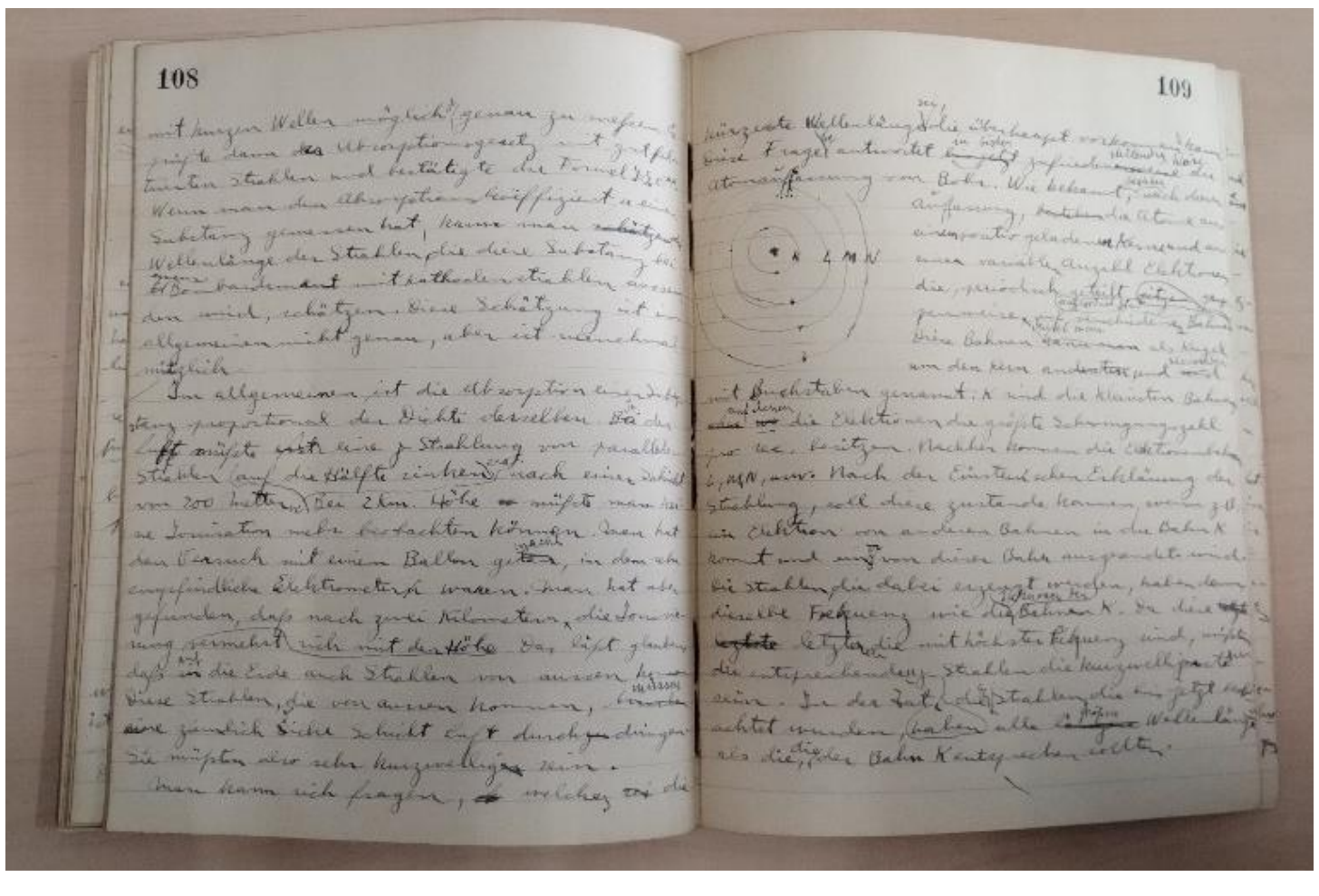

Sigue el semestre de invierno 1922/23:

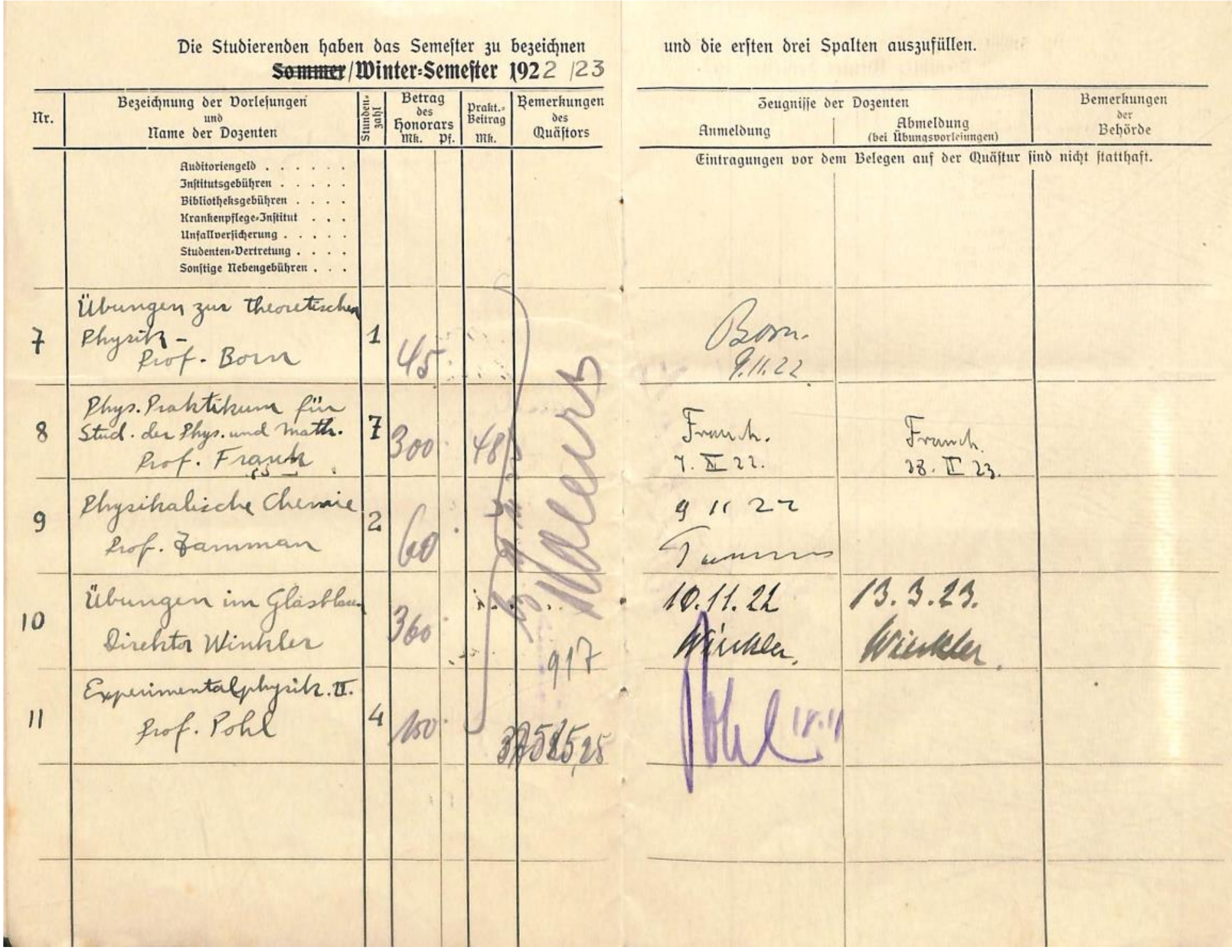
Die Studierenden haben das Semester zu bezeichnen und die ersten drei Spalten auszufüllen.

~~Sommer~~/Winter-Semester 1922/23

| Nr. | Bezeichnung der Vorlesungen und Name der Dozenten | Stundenzahl | Betrag des Honorars Mk. | Pf. | Prakt.-Beitrag Mk. | Bemerkungen des Quästors | Zeugnisse der Dozenten: Anmeldung | Abmeldung (bei Übungsvorlesungen) | Bemerkungen der Behörde |
|---|---|---|---|---|---|---|---|---|---|
| | Auditoriengeld, Institutsgebühren, Bibliotheksgebühren, Krankenpflege-Institut, Unfallversicherung, Studenten-Vertretung, Sonstige Nebengebühren | | | | | | Eintragungen vor dem Belegen auf der Quästur sind nicht statthaft. | | |
| 7 | Übungen zur theoretischen Physik – Prof. Born | 1 | 45 | | | | Born. 9.11.22 | | |
| 8 | Phys. Praktikum für Stud. der Phys. und Math. Prof. Franck | 7 | 300 | | 48 | | Franck. 7.XI.22. | Franck 28.II.23. | |
| 9 | Physikalische Chemie Prof. Tammann | 2 | 60 | | | | 9 11 22 Tammann | | |
| 10 | Übungen im Glasblasen Direktor Winkler | | 360 | | | 917 | 10.11.22 Winkler | 13.3.23. Winkler | |
| 11 | Experimentalphysik II. Prof. Pohl | 4 | 150 | | | 3375,25 | Pohl 17.11 | | |

Vemos nuevamente a Born, con *Prácticas de Física Teórica*, y nuevamente a Franck, con *Laboratorio de Física para Estudiantes de Física y Matemática*.

Gustav Tammann, destacadísimo físico-químico en el área de las transiciones orden-desorden en aleaciones, le dio *Físicoquímica*.

Con alguien llamado Winkler (nombre no identificado) hizo *Prácticas de Soplado de Vidrio*. Muchos años después, en Bariloche, Gaviola hacía que los estudiantes de Física Experimental siguieran talleres de carpintería, soplado de vidrio, soldadura, tornería, todas las artesanías útiles en el laboratorio. Su última actividad profesional, de hecho, fue estar a cargo de esos talleres, cuando ya estaba jubilado.

Y el director del Departamento, Robert Pohl, que es el padre de la física del estado sólido, le dio *Física Experimental II*.

En el verano del 23 vemos un curso de *Ecuaciones Diferenciales*, a cargo de Hellmuth Kneser, y uno de *Magneto-óptica*, dictado por el Prof. Aldenberg (cuyo nombre tampoco pude identificar.

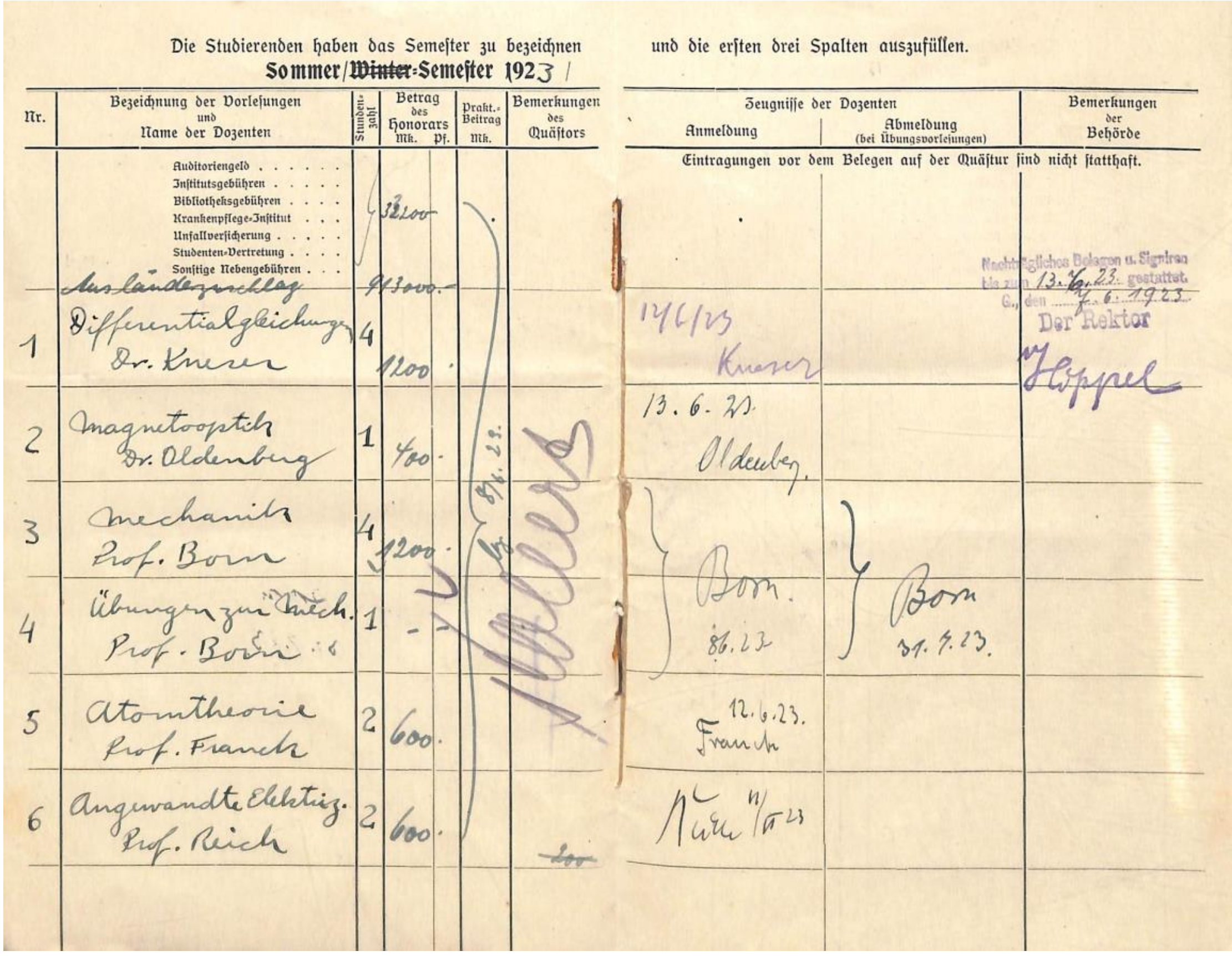
Die Studierenden haben das Semester zu bezeichnen und die ersten drei Spalten auszufüllen.

Sommer/~~Winter~~-Semester 1923

| Nr. | Bezeichnung der Vorlesungen und Name der Dozenten | Stundenzahl | Betrag des Honorars Mk. Pf. | Prakt.-Beitrag Mk. | Bemerkungen des Quästors | Zeugnisse der Dozenten: Anmeldung | Abmeldung (bei Übungsvorlesungen) | Bemerkungen der Behörde |
|---|---|---|---|---|---|---|---|---|
| | Auditoriengeld, Institutsgebühren, Bibliotheksgebühren, Krankenpflege-Institut, Unfallversicherung, Studenten-Vertretung, Sonstige Nebengebühren | | 32200 | | | | | Nachträgliches Belegen u. Signiren bis zum 13.6.23. gestattet. G., den 7.6.1923. Der Rektor Hoppel |
| | Ausländerzuschlag | | 913000.– | | | | | |
| 1 | Differentialgleichungen Dr. Kneser | 4 | 1200. | | | 12/6/23 Kneser | | |
| 2 | Magnetooptik Dr. Oldenburg | 1 | 400. | | | 13.6.23. Oldenberg | | |
| 3 | Mechanik Prof. Born | 4 | 1200. | | | Born 8.6.23 | Born 31.7.23. | |
| 4 | Übungen zur Mech. Prof. Born | 1 | – | | | | | |
| 5 | Atomtheorie Prof. Franck | 2 | 600. | | | 12.6.23. Franck | | |
| 6 | Angewandte Elektriz. Prof. Reich | 2 | 600. | | | | | |

Lo tuvo a Max Born de nuevo, esta vez en *Mecánica*, que es la materia que estoy dando yo ahora. No tenían el Goldstein, el libro de texto que hemos usado tantas generaciones de físicos en los siglos XX y XXI; seguro que usaban el libro de Sommerfeld. Y también de nuevo a Franck, esta vez con *Teoría Atómica*.

Max Reich, que era otro de los “peces gordos”, le dio un curso de *Electricidad Aplicada*. Era un especialista en el tema, y también en radiocomunicaciones, y fue director en Göttingen del Instituto de Electricidad Aplicada que se creó en esos años (sospecho que es lo que luego llamaríamos “electrónica”).

En este semestre Franck le dio también otro Laboratorio de Física. Y está anotado en dos cursos avanzados de Born, el *Proseminar de Física* y el *Coloquio*.

En las páginas sucesivas de la libreta puede observarse cómo fue aumentando la matrícula de los cursos. Me contó Bressan que Gaviola cambiaba 1 peso por semana, y que cada semana le daban más marcos. Así hizo rendir sus ahorros de agrimensor, gracias a la inflación de la república de Weimar.

Finalmente llegamos al verano de 1923:

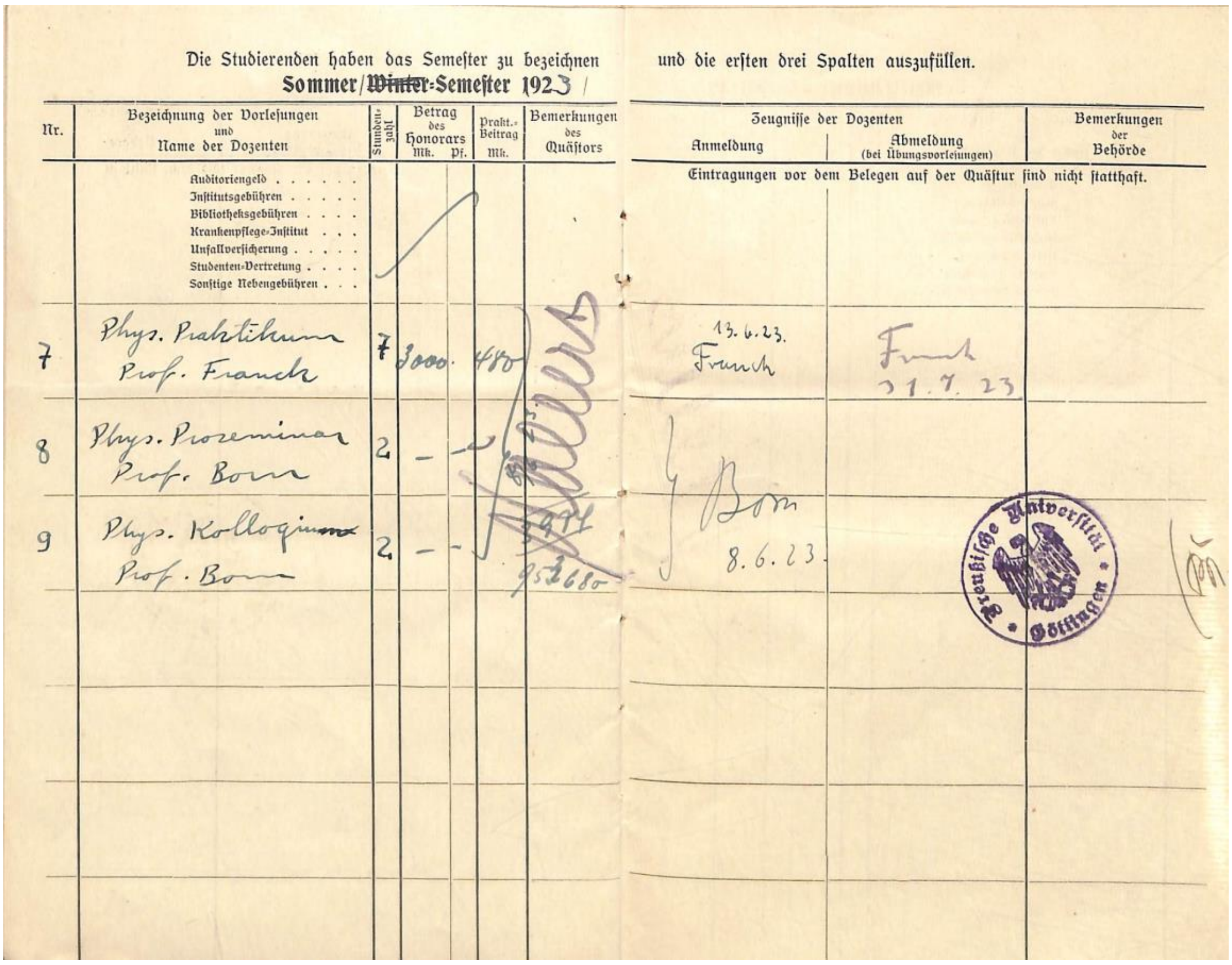

Die Studierenden haben das Semester zu bezeichnen und die ersten drei Spalten auszufüllen.

Sommer/~~Winter~~-Semester 1923

| Nr. | Bezeichnung der Vorlesungen und Name der Dozenten | Stundenzahl | Betrag des Honorars Mk. Pf. | Prakt.-Beitrag Mk. | Bemerkungen des Quästors | Zeugnisse der Dozenten: Anmeldung | Zeugnisse der Dozenten: Abmeldung (bei Übungsvorlesungen) | Bemerkungen der Behörde |
|---|---|---|---|---|---|---|---|---|
| | Auditoriengeld; Institutsgebühren; Bibliotheksgebühren; Krankenpflege-Institut; Unfallversicherung; Studenten-Vertretung; Sonstige Nebengebühren | | | | | Eintragungen vor dem Belegen auf der Quästur sind nicht statthaft. | | |
| 7 | Phys. Praktikum Prof. Franck | 7 | 3000 | 480 | | 13.6.23. Franck | Franck 31.7.23 | |
| 8 | Phys. Proseminar Prof. Born | 2 | — | — | | | | |
| 9 | Phys. Kolloquium Prof. Born | 2 | — | — | | Born 8.6.23 | | |

Preußische Universität Göttingen

Este fue su último semestre en Göttingen. Años más tarde, en un discurso en el que relataba su experiencia en Alemania, dijo:

*«Después de tres semestres cometí el error de pedir el pase a Berlín»*.

Dice que se cansó de la vida en un pueblo chico, y que lo atraía la gran ciudad. Yo le pregunté a su discípulo y amigo Oscar Bressan por qué se había ido a Berlín, y me dijo “porque había más chicas”. Göttingen era una ciudad pequeña, de 100 mil habitantes, más o menos como La Plata, y además era eminentemente universitaria, y en esa época eso significa sobrepoblación de varones. Berlín, en cambio, era una ciudad de 4 millones de habitantes. En cuanto a por qué lo consideraba un error, seguramente es porque justo al año siguiente explotó la mecánica cuántica en Göttingen, con la publicación de los trabajos de Heisenberg, Born y Jordan.

Por cierto, se fue a Berlín a una universidad buenísima, la Universidad Friedrich Wilhelm (hoy Universidad Humboldt, establecida por el Rey Federico Guillermo III de Prusia a iniciativa de Alexander von Humboldt). Muchos grandes físicos trabajaban allí y fueron sus profesores, pero probablemente en ese momento Göttingen era “el centro del mundo mundial” en física cuántica. Así que se tomó vacaciones en la gran ciudad en el verano de 1923, y en invierno empezó a cursar. También tenemos la libreta de Berlín, por supuesto.

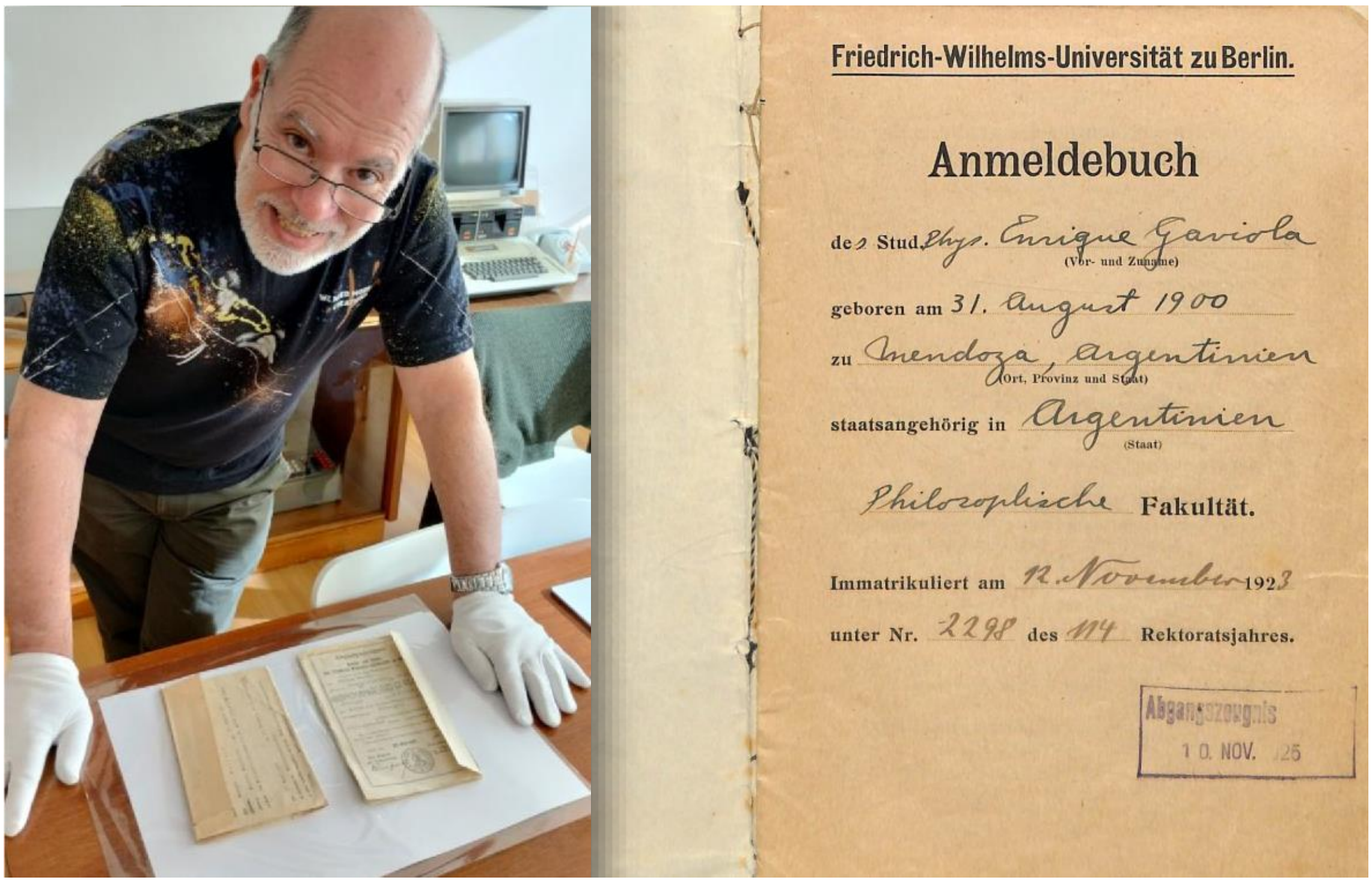

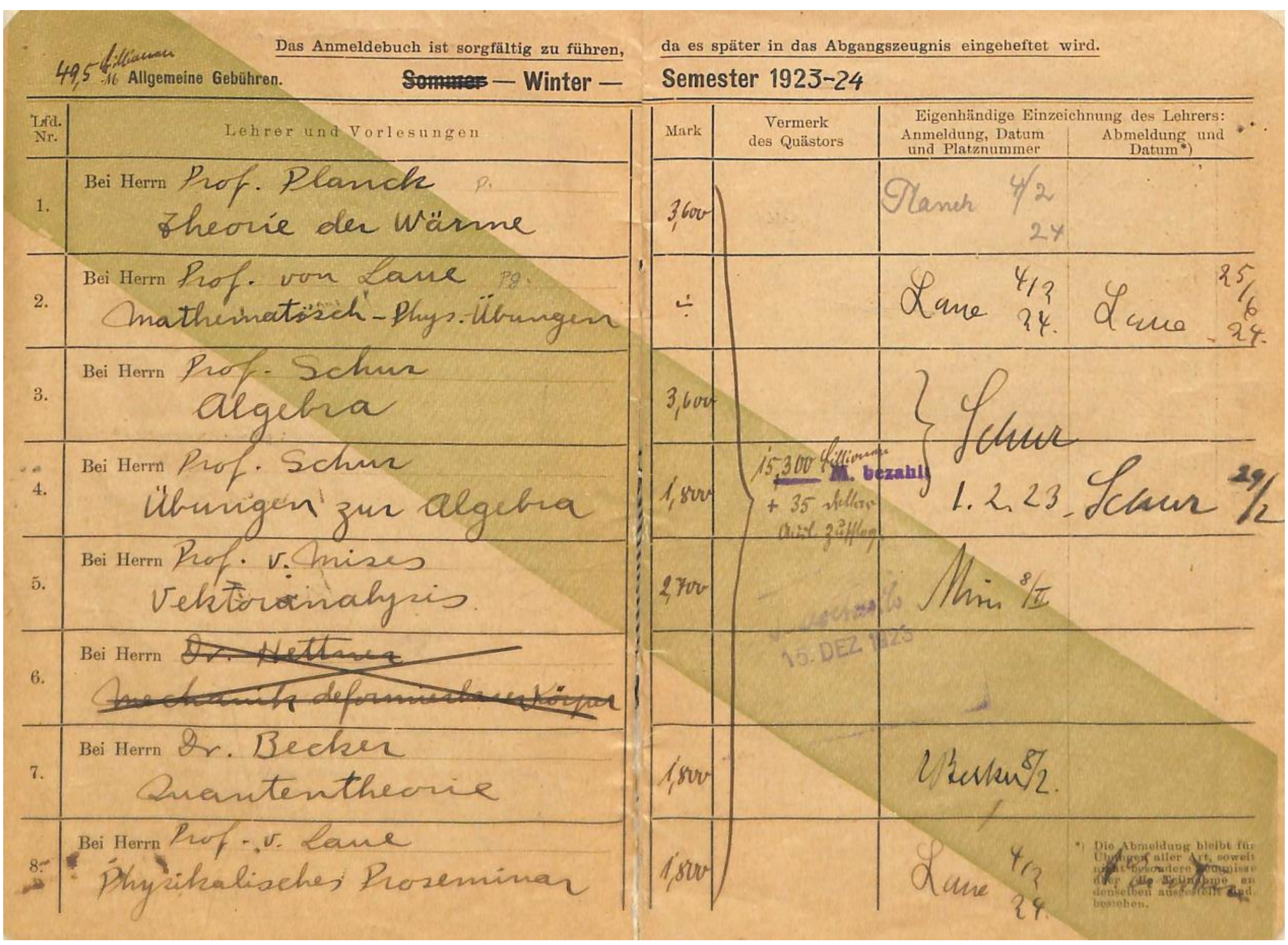

En su primer semestre en Berlín se inscribió en 12 materias. Max Planck, premio Nobel 1918 y físico extraordinario de la generación anterior a los pibitos que estaban haciendo la revolución cuántica, pero que había dado el puntapié inicial con su postulado de los cuantos de energía, le dio *Teoría del Calor*. Imagino que dio en clase su teoría del cuerpo negro.

Max Von Laue, Nobel 1914 y descubridor de la difracción de los rayos X en cristales, le dio *Prácticas de Física Matemática*. Von Laue fue un gran objetor del nazismo, y fue uno de los personajes cruciales en la reorganización de la ciencia alemana después de la 2a Guerra. Con Laue también empezó un nuevo *Proseminar de Física* (última línea), que era un trabajo de investigación de varios semestres, donde tenía que estudiar sobre un tema, presentar un informe y exponerlo de manera crítica. Era un requisito para el doctorado, y estaba dirigido por von Laue, e integrado por Lise Meitner, Albert Einstein y Peter Pringsheim, quien luego dirigiría su trabajo de doctorado. La firma de Einstein se puede ver en el último casillero, superpuesta a un texto impreso.

Con Issai Schur cursó *Álgebra*. Schur trabajaba en teoría de grupos; es el de la descomposición de Schur, y el lema de Schur, que hemos estudiado todos.

Richard von Mises fue su profesor de *Análisis Vectorial*. Era hermano del famoso economista austríaco, y un destacado matemático muy influyente en la ingeniería aeronáutica.

Richard Becker, quien le dio *Teoría Cuántica*, en 1922 había completado su *Habilitation* con Planck.

Sigue el invierno del 23-24:

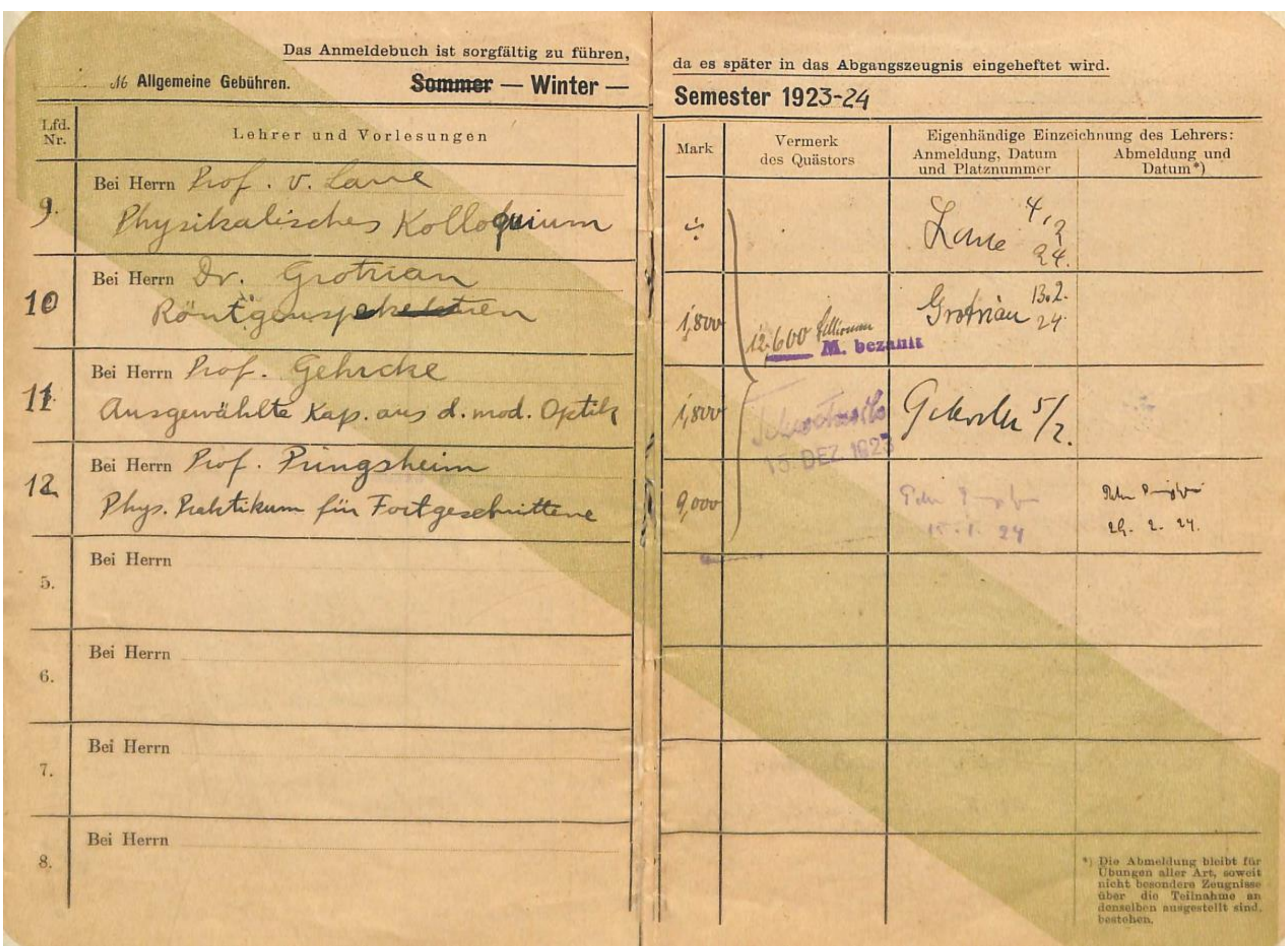
Das Anmeldebuch ist sorgfältig zu führen, da es später in das Abgangszeugnis eingeheftet wird.

ℳ Allgemeine Gebühren. ~~Sommer~~ — Winter — Semester 1923-24

| Lfd. Nr. | Lehrer und Vorlesungen | Mark | Vermerk des Quästors | Eigenhändige Einzeichnung des Lehrers: Anmeldung, Datum und Platznummer | Abmeldung und Datum*) |
|---|---|---|---|---|---|
| 9. | Bei Herrn Prof. v. Laue Physikalisches Kolloquium | ÷ | | Laue 4.1.24 | |
| 10 | Bei Herrn Dr. Grotrian Röntgenspektren | 1,800 | 12.600 Millionen M. bezahlt | Grotrian 13.2.24 | |
| 11 | Bei Herrn Prof. Gehrcke Ausgewählte Kap. aus d. mod. Optik | 1,800 | 15. DEZ. 1923 | Gehrcke 5/2 | |
| 12 | Bei Herrn Prof. Pringsheim Phys. Praktikum für Fortgeschrittene | 9,000 | | Peter Pringsheim 15.1.24 | Peter Pringsheim 29.2.24 |
| 5. | Bei Herrn | | | | |
| 6. | Bei Herrn | | | | |
| 7. | Bei Herrn | | | | |
| 8. | Bei Herrn | | | | |

*) Die Abmeldung bleibt für Übungen aller Art, soweit nicht besondere Zeugnisse über die Teilnahme an denselben ausgestellt sind, bestehen.

Vemos el *Coloquio de Física* dirigido por von Laue, *Espectroscopía de Rayos X* con Walter Grotrian (un destacado astrofísico y espectroscopista), y con Ernst Gehrcke, *Capítulos Seleccionados de Óptica Moderna*. Gehrcke era un antirrelativista, como muchos físicos experimentales de la época (inclusive Lorentz), que no querían abandonar el éter luminífero.

En este semestre se inscribió en un curso de *Laboratorio de Física para Estudiantes Avanzados* con Peter Pringsheim, quien se había doctorado con Röntgen en Munich, era amigo de Franck y de Pohl, y especialista en fluorescencia y fosforescencia cuánticas. Cuando empezó el curso, Pringsheim le dijo:

> *«He recibido una carta de Franck. Ud. no necesita hacer Laboratorio para Avanzados. Ud. empezará a hacer investigación conmigo.»*

Se ve en las fechas que no llegó a cursar un mes. No sé si le devolvieron los 9000 marcos, parece un curso carito.

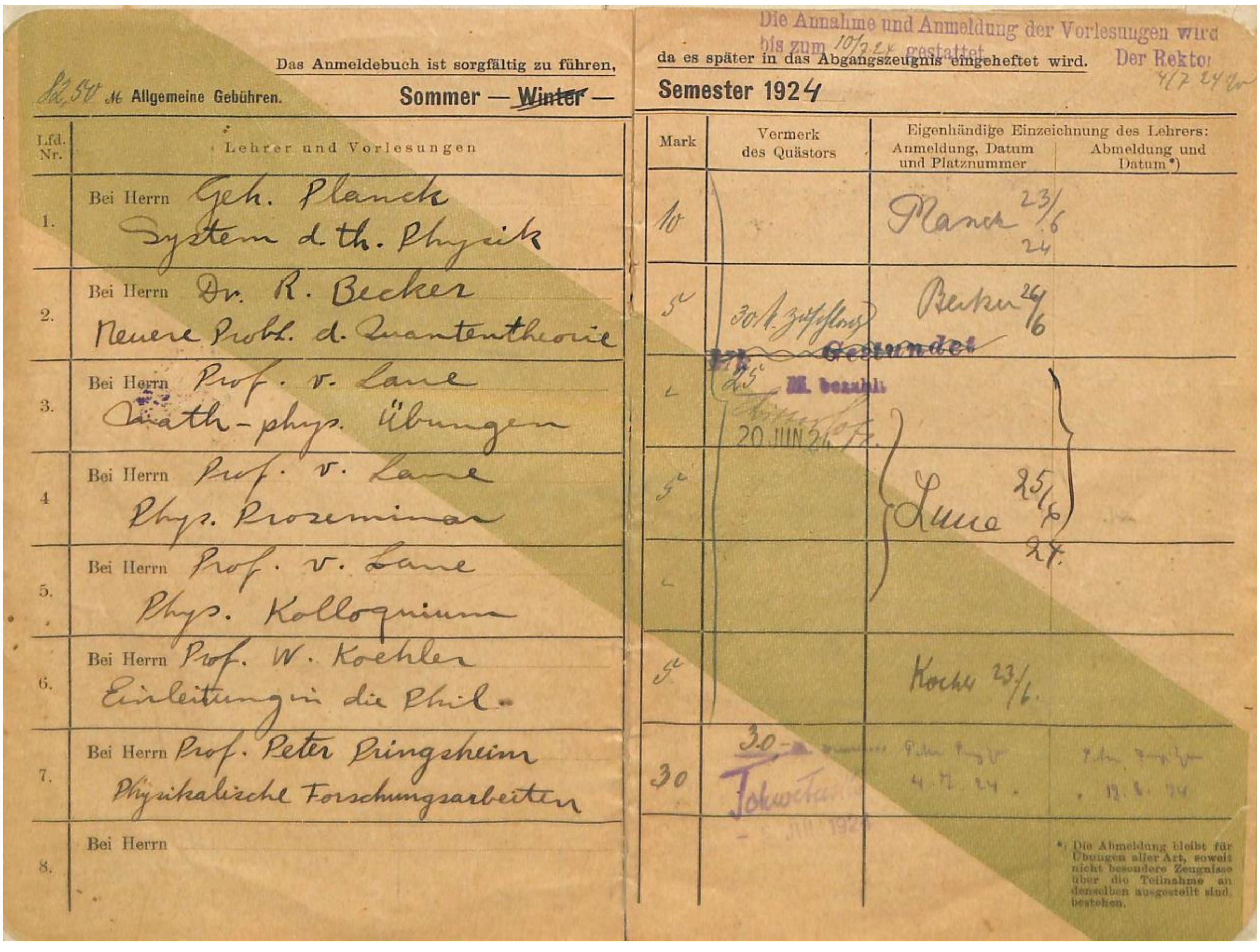

Das Anmeldebuch ist sorgfältig zu führen, da es später in das Abgangszeugnis eingeheftet wird.

Die Annahme und Anmeldung der Vorlesungen wird bis zum 10/7.24 gestattet. Der Rektor

82,50 M. Allgemeine Gebühren. Sommer — ~~Winter~~ — Semester 1924

| Lfd. Nr. | Lehrer und Vorlesungen | Mark | Vermerk des Quästors | Eigenhändige Einzeichnung des Lehrers: Anmeldung, Datum und Platznummer | Abmeldung und Datum*) |
|---|---|---|---|---|---|
| 1. | Bei Herrn Geh. Planck System d. th. Physik | 10 | | Planck 23/6 24 | |
| 2. | Bei Herrn Dr. R. Becker Neuere Probl. d. Quantentheorie | 5 | 30 M. Gestundet | Becker 26/6 | |
| 3. | Bei Herrn Prof. v. Laue Math.-phys. Übungen | – | 25 M. bezahlt 20 JUN 24 | | |
| 4. | Bei Herrn Prof. v. Laue Phys. Proseminar | 5 | | Laue 25/6 27. | |
| 5. | Bei Herrn Prof. v. Laue Phys. Kolloquium | – | | | |
| 6. | Bei Herrn Prof. W. Koehler Einleitung in die Phil. | 5 | | Koehler 23/6. | |
| 7. | Bei Herrn Prof. Peter Pringsheim Physikalische Forschungsarbeiten | 30 | 30.- | Peter Pringsheim 4.7.24. | Peter Pringsheim 19.8.24 |
| 8. | Bei Herrn | | | | |

*) Die Abmeldung bleibt für Übungen aller Art, soweit nicht besondere Zeugnisse über die Teilnahme an denselben ausgestellt sind, bestehen.

En el verano del 24 vemos de nuevo Planck (*Sistemas de Física Teórica*); de nuevo a Becker, con lo último de lo último: *Nuevos Problemas de Teoría Cuántica*; sigue von Laue, con *Prácticas de Física Matemática* y el final del *Proseminar*. Cuenta Gaviola:

> *«Me dieron como tema para exponer en el semestre siguiente los espectros de rayos beta y una lista de publicaciones recientes que debía estudiar y presentar en forma crítica. En esa época reinaba una aguda controversia entre los autores norteamericanos* *capitaneados por Ellis, y los alemanes capitaneados por Lise Meitner. Cuando llegó mi turno, expuse el tema y sostuve que los resultados experimentales norteamericanos eran mejores que los alemanes, pero que la teoría de los autores alemanes era superior a la de los norteamericanos. Hablé como una hora. Al final Lise Meitner me felicitó. »*

(Creo haber identificado correctamente a Charles Ellis, por el tema, pero era de Cavendish, no norteamericano.)

En este semestre hay una rareza: un curso de *Introducción a la Filosofía* con Wolfgang Koehler, uno de los fundadores de la psicología Gestalt. Se interesaba por la relación entre la física y la psicología, y había estudiado con Planck. Posiblemente en este curso conoció a Katri Nieminenn Vaukkari, estudiante de filosofía, con quien se casaría.

Por último, vemos el *Trabajo de Investigación en Física* que empezó a hacer con Pringsheim.

El semestre de invierno 24/25, tuvo una interrupción porque a fines de 1924 tuvo que volver por unos meses a Argentina para cumplir con el servicio militar. Volvió a Berlín en abril del 25.

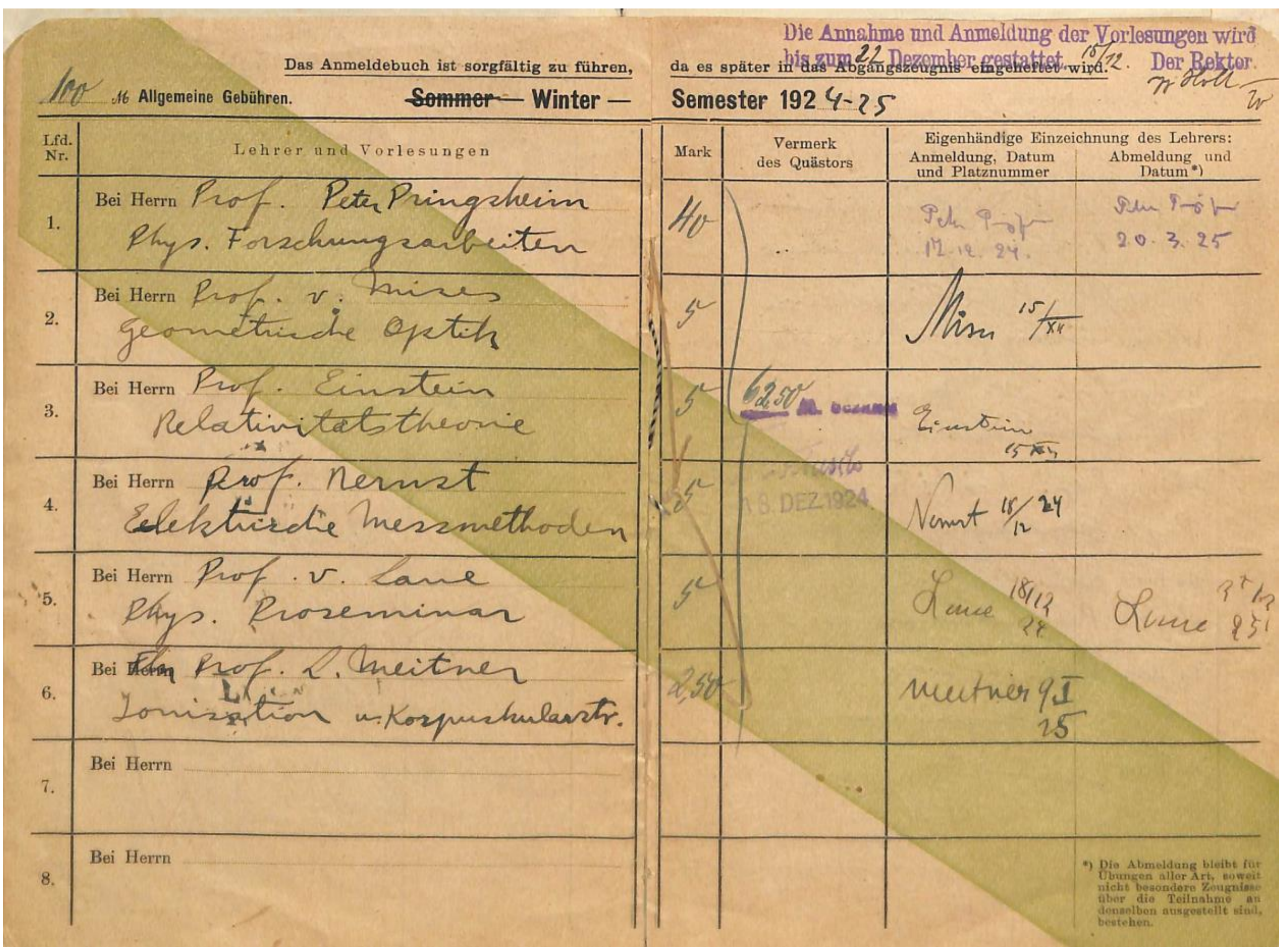

Das Anmeldebuch ist sorgfältig zu führen, da es später in das Abgangszeugnis eingeheftet wird.

Die Annahme und Anmeldung der Vorlesungen wird bis zum 22 Dezember gestattet. 15/12. Der Rektor.

100 M Allgemeine Gebühren. ~~Sommer~~ — Winter — Semester 1924-25

| Lfd. Nr. | Lehrer und Vorlesungen | Mark | Vermerk des Quästors | Eigenhändige Einzeichnung des Lehrers: Anmeldung, Datum und Platznummer | Abmeldung und Datum *) |
|---|---|---|---|---|---|
| 1. | Bei Herrn Prof. Peter Pringsheim Phys. Forschungsarbeiten | 40 | | Peter Pringsheim 17.12.24. | 20.3.25 |
| 2. | Bei Herrn Prof. v. Mises Geometrische Optik | 5 | | Mises 15/XII | |
| 3. | Bei Herrn Prof. Einstein Relativitätstheorie | 5 | 62,50 | Einstein 15 XII | |
| 4. | Bei Herrn Prof. Nernst Elektrische Messmethoden | 5 | 18. DEZ. 1924 | Nernst 18/12 24 | |
| 5. | Bei Herrn Prof. v. Laue Phys. Proseminar | 5 | | Laue 18/12 24 | Laue |
| 6. | Bei Herrn Prof. L. Meitner Ionisation u. Korpuskularstr. | 2,50 | | Meitner 9.I 25 | |
| 7. | Bei Herrn | | | | |
| 8. | Bei Herrn | | | | |

*) Die Abmeldung bleibt für Übungen aller Art, soweit nicht besondere Zeugnisse über die Teilnahme an denselben ausgestellt sind, bestehen.

Sigue el trabajo con Pringsheim y, habiendo terminado el Proseminar, empezó a preparar su tesis. Dice sobre el tema:

*«Von Laue quería que me dedicara a la Física Teórica y me dio como tema el estudio del comportamiento de electrones emitidos por un filamento a potencial cero, rechazados por una placa circundante a potencial negativo. Pringsheim me propuso que construyera un aparato para medir los tiempos de extinción de la fluorescencia.»*

Avanzó un poco con Von Laue, pero se le complicó matemáticamente y prefirió el trabajo experimental con Pringsheim. El trabajo teórico era sobre el sistema que algunos años después se convertiría en el klystron, el primer amplificador de radio, que permitiría el desarrollo del radar en los años 30.

Encontramos nuevamente a von Mises, ahora con *Óptica Geométrica*. Y quién le iba a dar *Relatividad* sino Albert Einstein (Nobel 1921).

*Mediciones Eléctricas* le dio Walther Nernst, que era otro de los peces gordos, gran químico que acababa de ganar el Nobel de Química en 1920. A propósito de Nernst, en otro discurso de Gaviola leemos: *«Años después he asistido a las clases de Richard Pohl en Göttingen, de Walter Nernst en Berlín, de Robert William Wood en Baltimore, y me he dado cuenta que curso de Gans en La Plata era mejor que el de cualquiera de ellos. La Plata tuvo el privilegio de tener el mejor curso de física experimental del mundo desde 1913 hasta 1925.»*

Lise Meitner, su examinadora del Proseminar, que no ganó el Nobel por el descubrimiento de la fisión del uranio porque la desplazaron, pero era una gran física experimental, le dio *Radiaciones Ionizantes*. Meitner fue la primera mujer en obtener un cargo de profesora titular en Alemania. Después se tuvo que escapar, perseguida por el nazismo. Fue nominada al Nobel de Química 19 veces, y 30 veces al de Física. En 1944 se lo dieron a Otto Hahn, su colaborador.

Llegamos al verano de 1925, último semestre, hace exactamente 100 años.

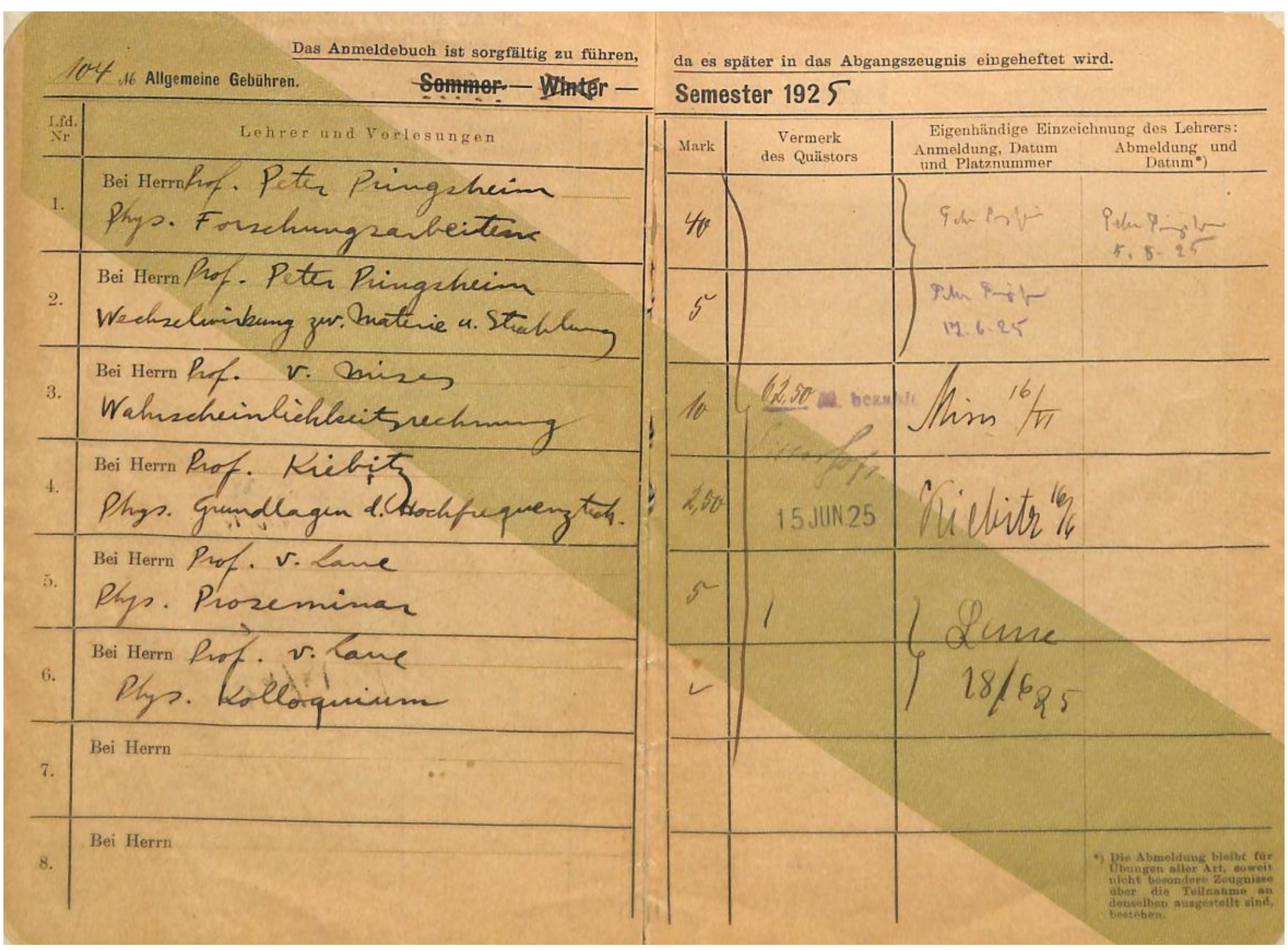

Das Anmeldebuch ist sorgfältig zu führen, da es später in das Abgangszeugnis eingeheftet wird.

104 M Allgemeine Gebühren. ~~Sommer~~ — ~~Winter~~ — Semester 1925

| Lfd. Nr | Lehrer und Vorlesungen | Mark | Vermerk des Quästors | Eigenhändige Einzeichnung des Lehrers: Anmeldung, Datum und Platznummer | Abmeldung und Datum*) |
|---|---|---|---|---|---|
| 1. | Bei Herrn Prof. Peter Pringsheim<br>Phys. Forschungsarbeiten | 40 | | Peter Pringsheim | Peter Pringsheim 5.8.25 |
| 2. | Bei Herrn Prof. Peter Pringsheim<br>Wechselwirkung zw. Materie u. Strahlung | 5 | | Peter Pringsheim 17.6.25 | |
| 3. | Bei Herrn Prof. v. Mises<br>Wahrscheinlichkeitsrechnung | 10 | 62,50 M. bezahlt | Mises 16/VI | |
| 4. | Bei Herrn Prof. Kiebitz<br>Phys. Grundlagen d. Hochfrequenztech. | 2,50 | 15 JUN 25 | Kiebitz 16/6 | |
| 5. | Bei Herrn Prof. v. Laue<br>Phys. Proseminar | 5 | / | Laue 28/6 25 | |
| 6. | Bei Herrn Prof. v. Laue<br>Phys. Kolloquium | – | | | |
| 7. | Bei Herrn | | | | |
| 8. | Bei Herrn | | | | |

*) Die Abmeldung bleibt für Übungen aller Art, soweit nicht besondere Zeugnisse über die Teilnahme an denselben ausgestellt sind, bestehen.

Sigue el trabajo con Pringsheim, y hay también un curso con él sobre *Interacción de la Radiación con la Materia*. De nuevo está von Mises, con una materia cuyo título es una procesión alfabética: *Wahrscheinlichkeitsrechnung* (*Cálculo de Probabilidades*).

Franz Kiebitz, un ingeniero de radio que Planck elogiaba por esos años, le dio un curso de *Bases Físicas de la Técnica de Altas Frecuencias* (radio, probablemente).

Cierran la libreta dos cursos más correspondientes al Proseminar de von Laue.

En este semestre terminó su tesis, e hizo una presentación en la Sociedad Alemana de Física el 31 de octubre. Este es el manuscrito de su conferencia.

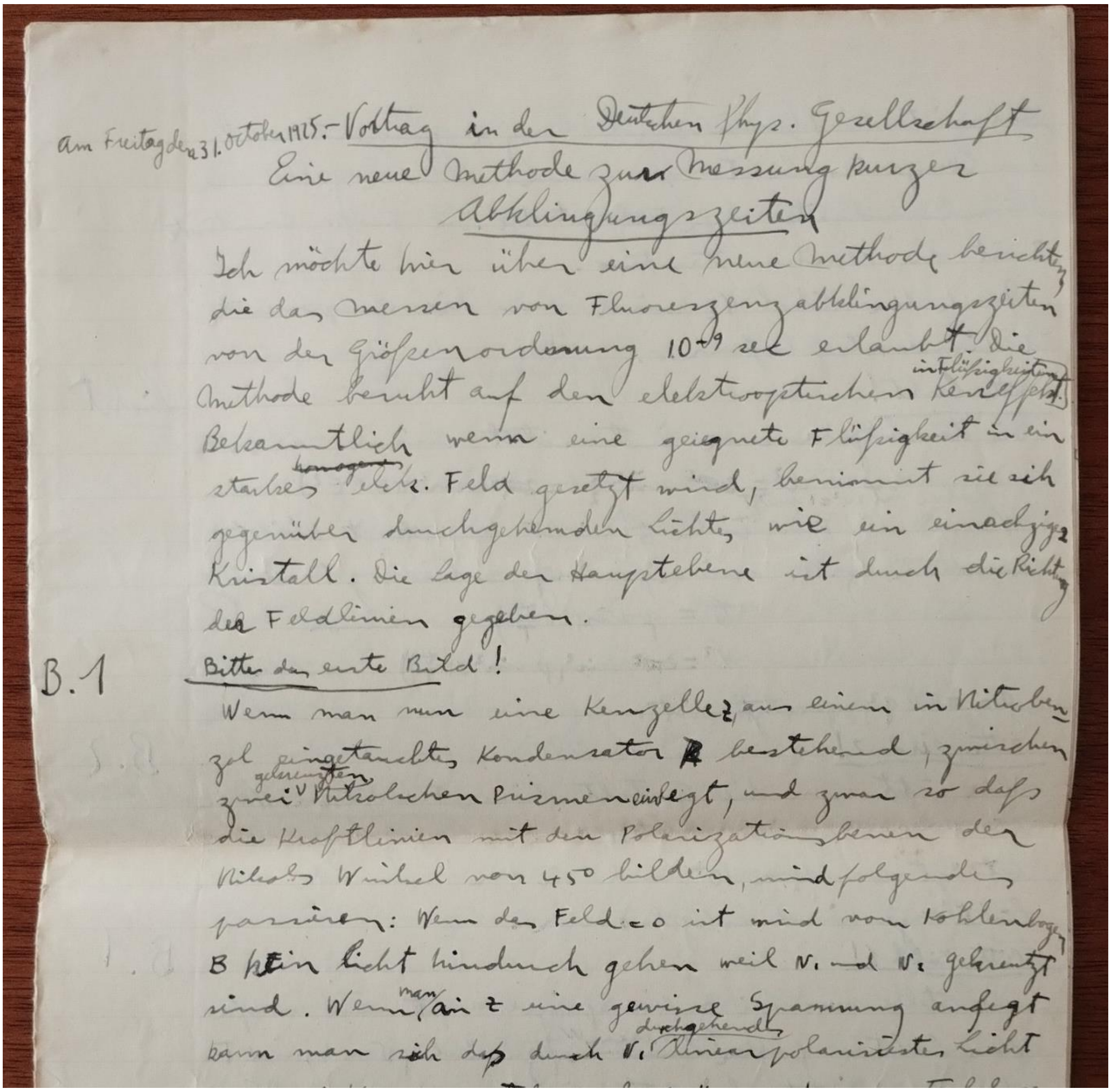
Am Freitag den 31. October 1925. – Vortrag in der Deutschen Phys. Gesellschaft

Eine neue Methode zur Messung kurzer Abklingungszeiten

Ich möchte hier über eine neue Methode berichten, die das Messen von Fluoreszenzabklingungszeiten von der Größenordnung $10^{-9}$ sek erlaubt. Die Methode beruht auf dem elektrooptischen Kerreffekt in Flüssigkeiten. Bekanntlich wenn eine geeignete Flüssigkeit in ein starkes homogenes elektr. Feld gesetzt wird, benimmt sie sich gegenüber durchgehendem Lichtes wie ein einachsiger Kristall. Die Lage der Hauptebene ist durch die Richtung der Feldlinien gegeben.

B. 1

Bitte das erste Bild!

Wenn man nun eine Kerrzelle Z, aus einem in Nitrobenzol eingetauchten Kondensator K bestehend, zwischen zwei gekreuzten Nicolschen Prismen einlegt, und zwar so dass die Kraftlinien mit den Polarisationsebenen der Nicols Winkel von 45° bilden, wird folgendes passieren: Wenn das Feld = 0 ist wird vom Kohlenbogen B kein Licht hindurch gehen weil $N_1$ und $N_2$ gekreuzt sind. Wenn man an Z eine gewisse Spannung anlegt kann man sich dass durch $N_1$ durchgehende linearpolarisiertes Licht

Hay dos cosas interesantes en esta primera página: una es que medía tiempos de nanosegundos (¡en 1925!), y otra es que pide que le pasen la imagen (*Bitte das erste Bild!*): usaban proyectores, como nosotros, no tan sofisticados como los que 100 años después nos dio, precisamente, la revolución cuántica que ellos hicieron.

Antes de publicar la tesis había que aprobar el examen final. Dice Gaviola:

*«Eran en realidad cuatro exámenes: Física Teórica, Física Experimental, Matemática y Filosofía, tomados por von Laue, Nernst, [Hans] Reichenbach y Kohler. Obtuve la calificación magna cum laude.»*

El trabajo se publicó en Annalen der Physick en 1926.

El 6 de junio de 1926 fue la ceremonia de graduación, en el despacho del Rector, a donde había que concurrir en traje de etiqueta, poner rodilla en tierra y recibir un espaldarazo con un diploma simbólico. Dice Gaviola:

*«La ceremonia me produjo profunda impresión; con el espaldarazo del Rector había sido armado caballero andante de la física. Pronto empecé a librar combates singulares contra la farsa, la corrupción, el fraude y el atraso en la física, primero en Estados Unidos y después en la Argentina. Por supuesto, en la mayoría de los casos salí “descalabrado”.»*

3. ***Die Abklingungszeiten der Fluoreszenz von Farbstofflösungen;***
***von Enrique Gaviola***

Zur Untersuchung der zeitlichen Veränderung des elektrischen Kerreffektes in Flüssigkeiten bei rasch abnehmender Feldstärke verwandten Abraham und Lemoine[1]) als erste eine sehr empfindliche Anordnung.

Die Anordnung war die folgende (Fig. 1).

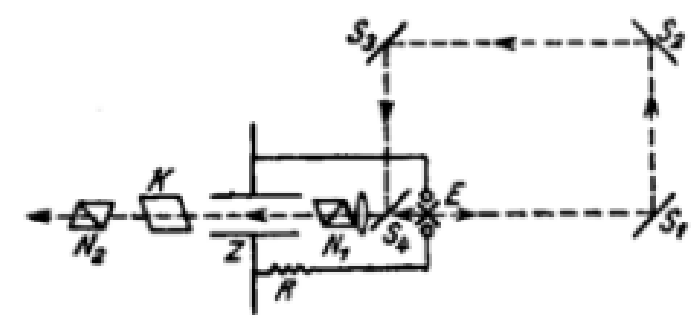

Fig. 1

Die Kondensatorplatten $Z$ waren in einer mit Schwefelkohlenstoff gefüllten Zelle eingetaucht und mit den Polen eines Hochspannungstransformators verbunden. Die Zelle befand sich zwischen zwei gekreuzten Nikols $N_1$ und $N_2$ so, daß die Kraftlinien des elektrischen Feldes gegen die Polarisationsebenen Winkel von $45^0$ bildeten. Der Kondensator $Z$ wurde durch die Funkenstrecke $E$ und den Widerstand $R$ entladen. $E$ diente als Lichtquelle. Das Licht vom Funken $E$ konnte entweder die Zelle $Z$ direkt erreichen oder nach Spiegelung an $S_1$, $S_2$, $S_3$ und $S_4$. Mittels eines Kalkspates, dessen Hauptebene parallel der Polarisationsebene von $N_1$ lag, und durch Drehung des Nikols $N_2$ wurde die beim Durchgang durch $Z$ erfolgte Phasenverschiebung $\alpha$ der beiden Komponenten des Lichtes $E$ gemessen. Abraham und Lemoine fanden nun, daß, wenn das Licht von $E$ auf dem kürzesten Weg durch die Zelle ging, die Phasenverschiebung am größten war, und daß sie abnahm, wenn der vom Lichte zurückgelegte Weg (durch Entfernen der

1) Compt. rend. **129.** S. 206. 1899.

**Annalen der Physik. IV. Folge. 81.** 44

El diploma está en latín, y lleva el sello de la universidad y la firma solamente del decano de la Facultad de Filosofía, pequeñísima y en lápiz. El texto dice:

QUE SEA FELIZ Y PROPICIO

De la UNIVERSIDAD LITERARIA FEDERICO GUILLERMO de BERLÍN, el Rector Magnífico, HEINRICH TRIEPEL, Doctor en Derecho y Ciencias Políticas en esta Universidad, Profesor Público Ordinario, Consejero Íntimo de Justicia, Caballero de la Orden del Águila Roja en Cuarta Clase, Comendador de otras Órdenes, por decreto de la Muy Honorable Facultad de Filosofía y bajo la autoridad legítima del Promotor, JULIUS PETERSEN, Doctor en Filosofía, Profesor Público Ordinario en esta Universidad, Miembro Ordinario de la Academia Prusiana de Ciencias, Senador de la Academia Prusiana de las Artes, Caballero de la Cruz de Hierro en Segunda Clase, Decano actual de la Facultad de Filosofía, al ilustrísimo y doctísimo señor ENRIQUE GAVIOLA, argentino, habiendo sustentado con gran elogio (*suma cum laude*) el examen de Filosofía y presentado una disertación muy loable, cuyo título es: *Die Abklingungszeiten der Fluoreszenz von Farbstofflösungen* (Los tiempos de decaimiento de la fluorescencia de soluciones de colorantes), la cual fue aprobada por la autoridad de la Facultad, los grados y honores de DOCTOR EN FILOSOFÍA y MAESTRO EN ARTES LIBERALES, confieren, el 21 de diciembre del año 1926, de forma legítima y solemne; y mediante el presente diploma, verificado con el sello oficial de la Facultad de Filosofía, se declara públicamente este otorgamiento.

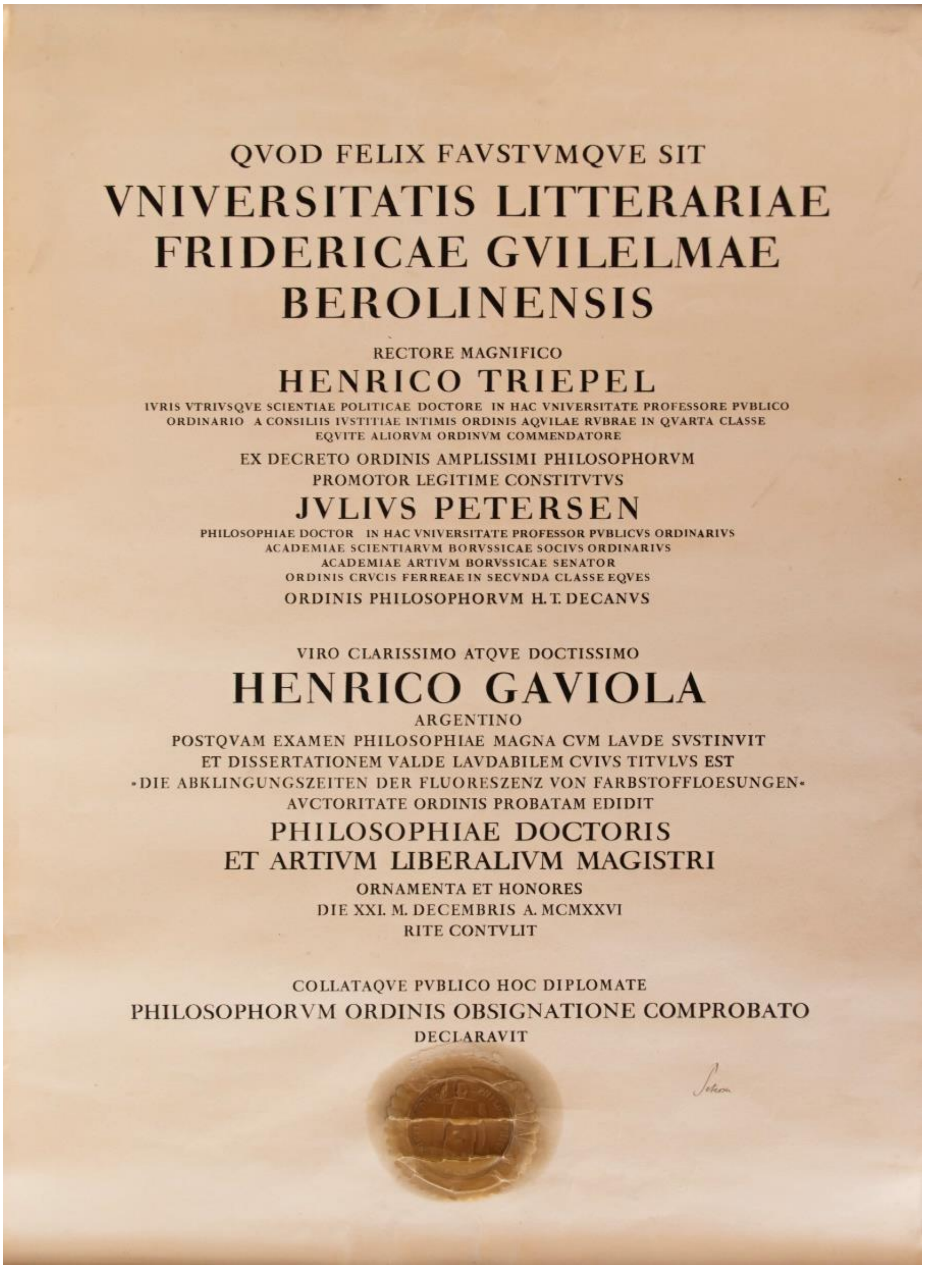

QVOD FELIX FAVSTVMQVE SIT

VNIVERSITATIS LITTERARIAE
FRIDERICAE GVILELMAE
BEROLINENSIS

RECTORE MAGNIFICO
HENRICO TRIEPEL
IVRIS VTRIVSQVE SCIENTIAE POLITICAE DOCTORE IN HAC VNIVERSITATE PROFESSORE PVBLICO
ORDINARIO A CONSILIIS IVSTITIAE INTIMIS ORDINIS AQVILAE RVBRAE IN QVARTA CLASSE
EQVITE ALIORVM ORDINVM COMMENDATORE
EX DECRETO ORDINIS AMPLISSIMI PHILOSOPHORVM
PROMOTOR LEGITIME CONSTITVTVS
JVLIVS PETERSEN
PHILOSOPHIAE DOCTOR IN HAC VNIVERSITATE PROFESSOR PVBLICVS ORDINARIVS
ACADEMIAE SCIENTIARVM BORVSSICAE SOCIVS ORDINARIVS
ACADEMIAE ARTIVM BORVSSICAE SENATOR
ORDINIS CRVCIS FERREAE IN SECVNDA CLASSE EQVES
ORDINIS PHILOSOPHORVM H. T. DECANVS

VIRO CLARISSIMO ATQVE DOCTISSIMO
HENRICO GAVIOLA
ARGENTINO
POSTQVAM EXAMEN PHILOSOPHIAE MAGNA CVM LAVDE SVSTINVIT
ET DISSERTATIONEM VALDE LAVDABILEM CVIVS TITVLVS EST
»DIE ABKLINGUNGSZEITEN DER FLUORESZENZ VON FARBSTOFFLOESUNGEN«
AVCTORITATE ORDINIS PROBATAM EDIDIT
PHILOSOPHIAE DOCTORIS
ET ARTIVM LIBERALIVM MAGISTRI
ORNAMENTA ET HONORES
DIE XXI. M. DECEMBRIS A. MCMXXVI
RITE CONTVLIT

COLLATAQVE PVBLICO HOC DIPLOMATE
PHILOSOPHORVM ORDINIS OBSIGNATIONE COMPROBATO
DECLARAVIT

Después de doctorarse, Gaviola hizo lo que hoy llamaríamos postdocs. El primero fue con Robert Wood, un destacadísimo físico experimental de la Johns Hopkins University, en Baltimore. Gaviola se anotó para una beca de la Fundación Rockefeller, y aunque logró las calificaciones necesarias, no le dieron la beca porque era sudamericano, y no estaba previsto que alguien que no fuese norteamericano o europeo la obtuviese. Cuando se lo comunicaron, Gaviola fue a la casa de Einstein, pasó por supuesto el filtro de visitantes que ejercía su esposa, y se lo contó. Según Gaviola, fue la única vez que lo vio furioso. Einstein se sentó de inmediato a escribir una carta de protesta, y le preguntó a Gaviola: *«¿Escribo en alemán o en inglés?»*, a lo que Gaviola contestó *«Ud. es Einstein, escriba en alemán, que se la hagan traducir»*. Einstein mandó la carta, y a Gaviola le dieron la beca.

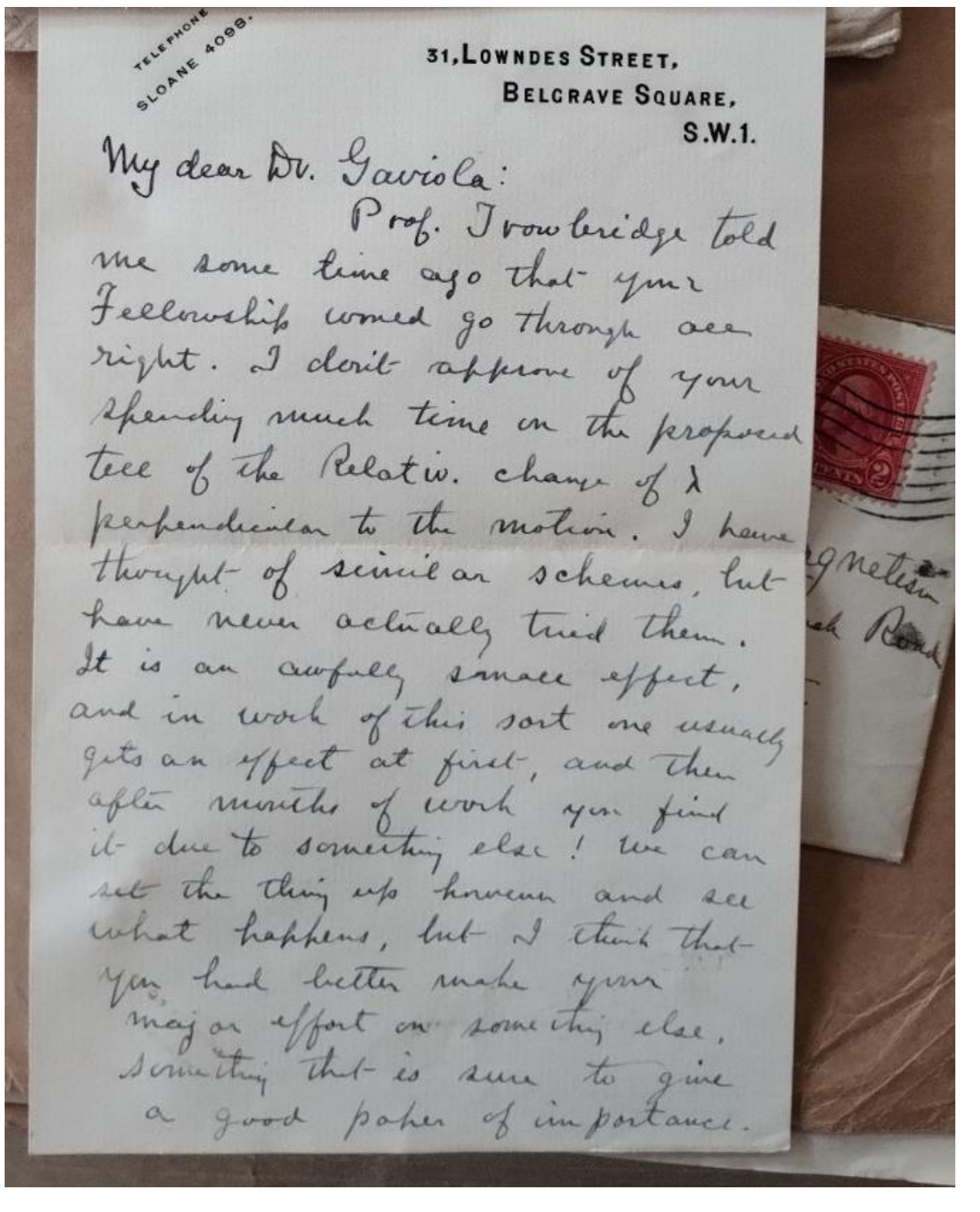

TELEPHONE SLOANE 4098.

31, LOWNDES STREET, BELGRAVE SQUARE, S.W.1.

My dear Dr. Gaviola:

Prof. Trowbridge told me some time ago that your Fellowship would go through all right. I don't approve of your spending much time on the proposed test of the Relativ. change of λ perpendicular to the motion. I have thought of similar schemes, but have never actually tried them. It is an awfully small effect, and in work of this sort one usually gets an effect at first, and then after months of work you find it due to something else! We can set the thing up however and see what happens, but I think that you had better make your major effort on something else, something that is sure to give a good paper of importance.

Esta es una carta de Wood a Gaviola, en la que le confirma que la beca saldría. Está enviada a París, donde Gaviola estaba trabajando con Jean Perrin (Nobel 1926). Perrin había hecho el trabajo experimental para verificar las predicciones de Einstein del movimiento browniano, resolviendo la disputa de un siglo sobre la existencia de los átomos. Fue también el primero en sugerir que la energía de las estrella venía de la fusión de hidrógeno en helio.

Gaviola, por sugerencia de Einstein, había propuesto intentar observar el efecto Doppler relativista, algo que era muy difícil. En la carta, Wood discute la propuesta, y le recomienda que mejor hagan otra cosa, más fácilmente publicable. Si Wood hubiera conocido previamente a Gaviola, tal vez le habría asignado ese trabajo, Gaviola lo habría hecho sin dudas, y se habría ganado el premio Nobel de Física.

El texto dice:

> *«Prof. Trowbridge* [Augustus Trowbridge, quien en 1927 era director de la Oficina Europea del International Education Board de la Fundación Rockefeller] *told me some time ago that your Fellowship would go through all right. I don't approve of your spending much time on the proposed test of the Relativ. change of λ perpendicular to the motion. I have thought of similar schemes, but have never actually tried them.*
>
> *«It is an awfully square effect, and in work of this sort one usually gets an effect at first, and then after months of work, you find it due to something else. One can set the thing up however and see what happens, but I think that you had better make your major effort on something else, something that is sure to give a good paper of importance.»*

También encontré un tarjeta postal de Gans, en la que lo felicita por la beca conseguida, y le agradece los *reprints* que le mandó, y dice que le mandará los suyos a Baltimore, y que siga escribiéndole.

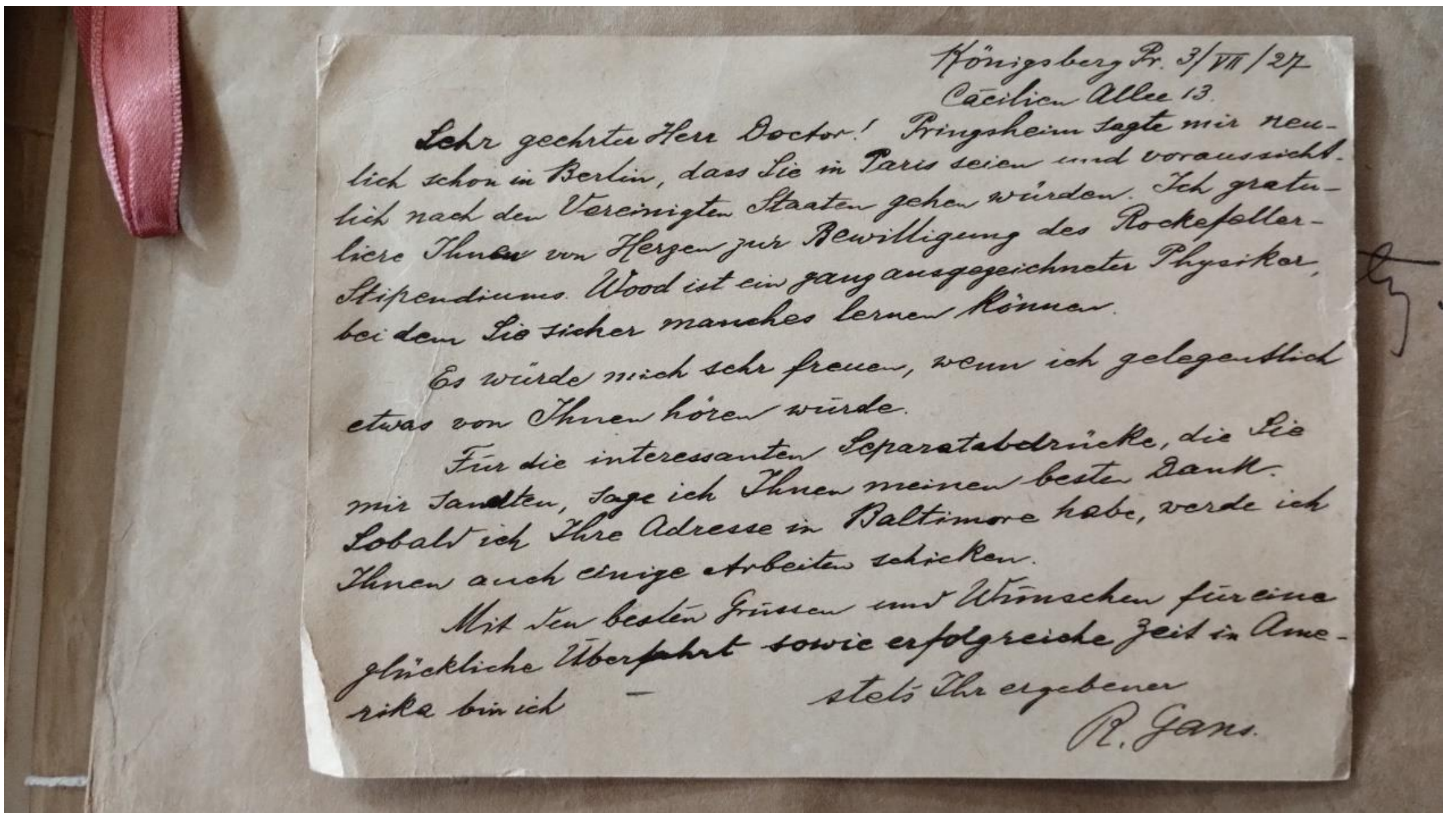

Königsberg Pr. 3/VII/27
Cäcilien Allee 13.

Sehr geehrter Herr Doctor! Pringsheim sagte mir neulich schon in Berlin, dass Sie in Paris seien und voraussichtlich nach den Vereinigten Staaten gehen würden. Ich gratuliere Ihnen von Herzen zur Bewilligung des Rockefeller-Stipendiums. Wood ist ein ganz ausgezeichneter Physiker, bei dem Sie sicher manches lernen können.

Es würde mich sehr freuen, wenn ich gelegentlich etwas von Ihnen hören würde.

Für die interessanten Separatabdrücke, die Sie mir sandten, sage ich Ihnen meinen besten Dank. Sobald ich Ihre Adresse in Baltimore habe, werde ich Ihnen auch einige Arbeiten schicken.

Mit den besten Grüssen und Wünschen für eine glückliche Überfahrt sowie erfolgreiche Zeit in Amerika bin ich

stets Ihr ergebener

R. Gans

El texto está en alemán, como toda la correspondencia con Gans posterior a aquella primera carta de 1922. Dice:

*Königsberg, Prusia, 3 de agosto de 1927*
*Cäcilienallee 13*

*Querido Doctor,*

*Pringsheim me dijo recientemente in Berlín que Ud. está en París y que probablemente irá a los Estados Unidos. Lo felicito de corazón por obtener la Beca Rockefeller.*

*Wood es realmente un físico excelente, de quien aprenderá mucho.*

*Me encatará recibir noticias suyas de vez en cuando.*

*Gracias por los interesantes reprints que me mandó. Apenas tenga su dirección en Baltimore le mandaré algunos de mis propios trabajos. Con mis mejores saludos, deseos de un viaje seguro y una estadía exitosa en América, quedo*

*siempre su devoto*

*R. Gans*

Gaviola estuvo un año con Wood, y publicaron *siete* trabajos, mayormente en temas de espectroscopía del mercurio, de los cuales encontré manuscritos a máquina y de puño y letra.

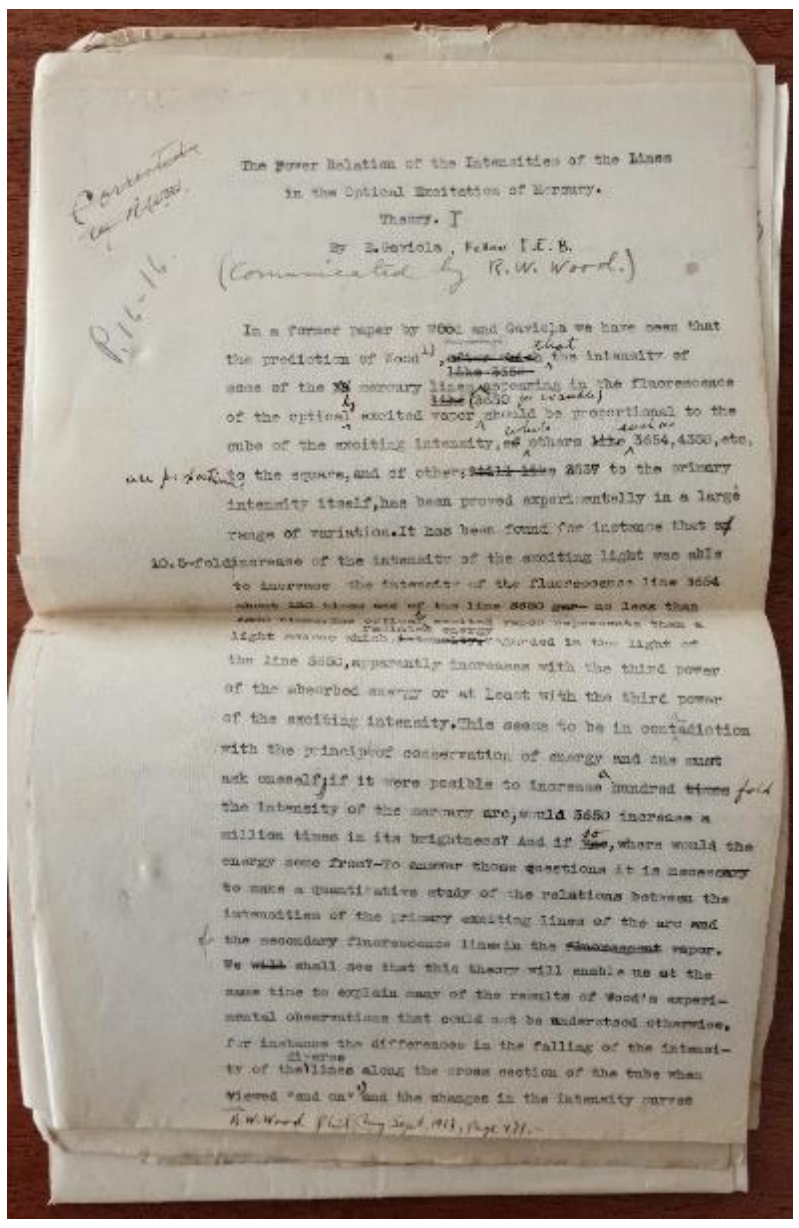

The Power Relation of the Intensities of the Lines
in the Optical Excitation of Mercury.
Theory. I
By E. Gaviola, Fellow I.E.B.
(Communicated by R.W. Wood.)

In a former paper by Wood and Gaviola we have seen that the prediction of Wood[1], that the intensity of some of the mercury lines appearing in the fluorescence of the optical excited vapor (3650 for example) should be proportional to the cube of the exciting intensity, of others 3654, 4358, etc. to the square, and of others 2537 to the primary intensity itself, has been proved experimentally in a large range of variation. It has been found for instance that a 10.5-fold increase of the intensity of the exciting light was able to increase the intensity of the fluorescence line 3654 about 100 times and of the line 3650 no less than ... light energy which ... the line 3650, apparently increases with the third power of the absorbed energy or at least with the third power of the exciting intensity. This seems to be in contradiction with the principle of conservation of energy and one must ask oneself, if it were possible to increase hundred fold the intensity of the mercury arc, would 3650 increase a million times in its brightness? And if so, where would the energy come from? To answer those questions it is necessary to make a quantitative study of the relations between the intensities of the primary exciting lines of the arc and the secondary fluorescence lines in the vapor. We shall see that this theory will enable us at the same time to explain many of the results of Wood's experimental observations that could not be understood otherwise, for instance the differences in the falling of the intensity of the diverse lines along the cross section of the tube when viewed "end on" and the changes in the intensity curves

[1] R.W. Wood Phil. Mag. Sept. 1913, Page 771.

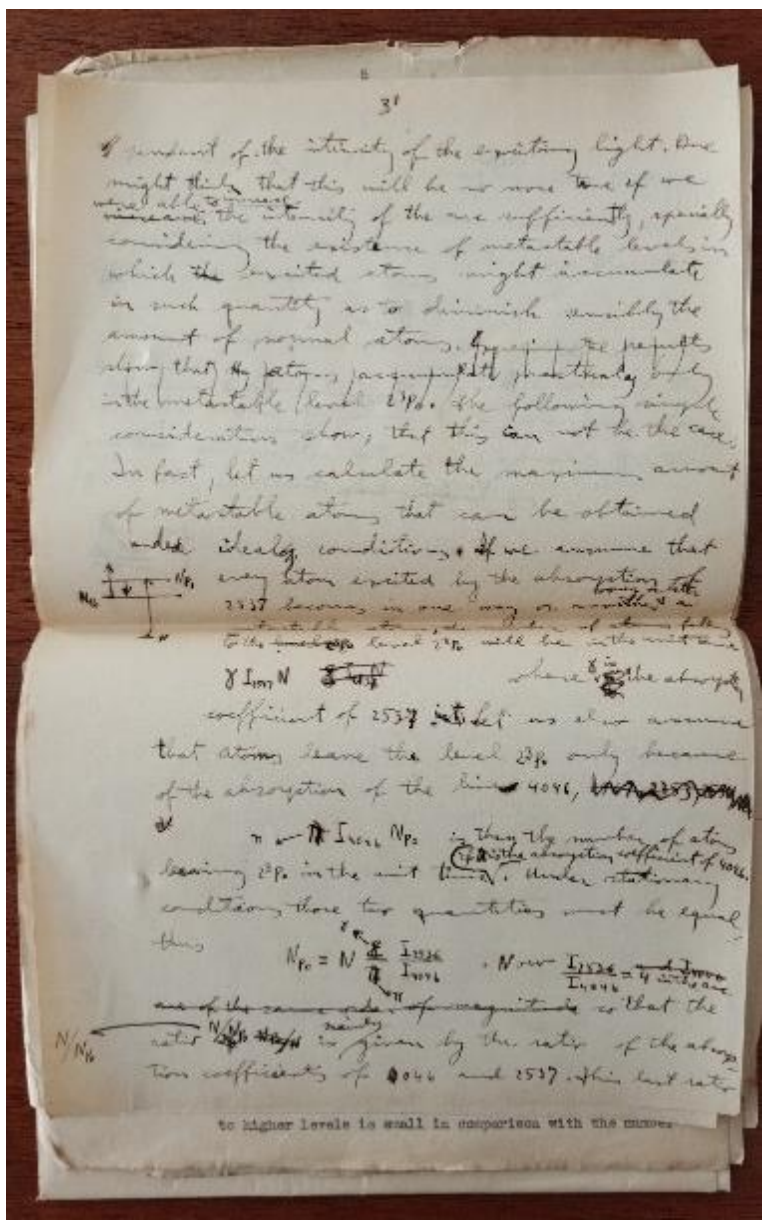

Después de Baltimore, Gaviola estuvo dos años más en Estados Unidos, en la Carnegie Institution, en Washington DC, donde trabajó mucho y en diversas áreas, y logró un prestigio internacional interesante. Volvió ilusionado a La Plata, la encontró envuelta en incomprensibles rencillas internas, se cansó y en seis meses se volvió a ir a Berlín.

En 1930 volvió de nuevo, a Exactas de la UBA, donde estuvo 4 años. Su carta de renuncia, motivada porque no aceptaban sus cambios propuestos al doctorado (y a toda la Facultad), tiene 40 páginas. Se fue "de vacaciones" varios meses a Europa, pero en realidad estuvo visitando laboratorios y trabajando, y lo invitaban a quedarse, porque era muy reconocido internacionalmente. Estuvo un tiempo en Madrid, al final del cual le ofrecieron ir a trabajar con Linus Pauling (Nobel de Química 1954, por su revolucionaria explicación del origen cuántico del enlace químico, y también Nobel de la Paz 1962). Pero Félix Aguilar lo propuso para dirigir el Observatorio de Córdoba, que enfrentaba la paralización de la construcción del "gran espejo" para el nuevo telescopio (proyecto iniciado 30 años antes). En lugar de hacerse cargo inmediatamente, como no se había formado como óptico, aprovechó la beca que le habían ofrecido para trabajar con Pauling, para irse a CALTECH y el Observatorio de Mt Wilson a trabajar con John Strong. Allí desarrolló una técnica revolucionaria para aluminizar y conformar grandes espejos, que cambió la historia de la construcción de grandes telescopios. El gigante de 5 m de Monte Palomar, por ejemplo, estaba paralizado por problemas técnicos, y finalmente pudo ser terminado. Aquí lo vemos junto a Strong al terminar el aluminizado del espejo de 100 pulgadas, en una famosa *selfie* tomada por Gaviola mismo.

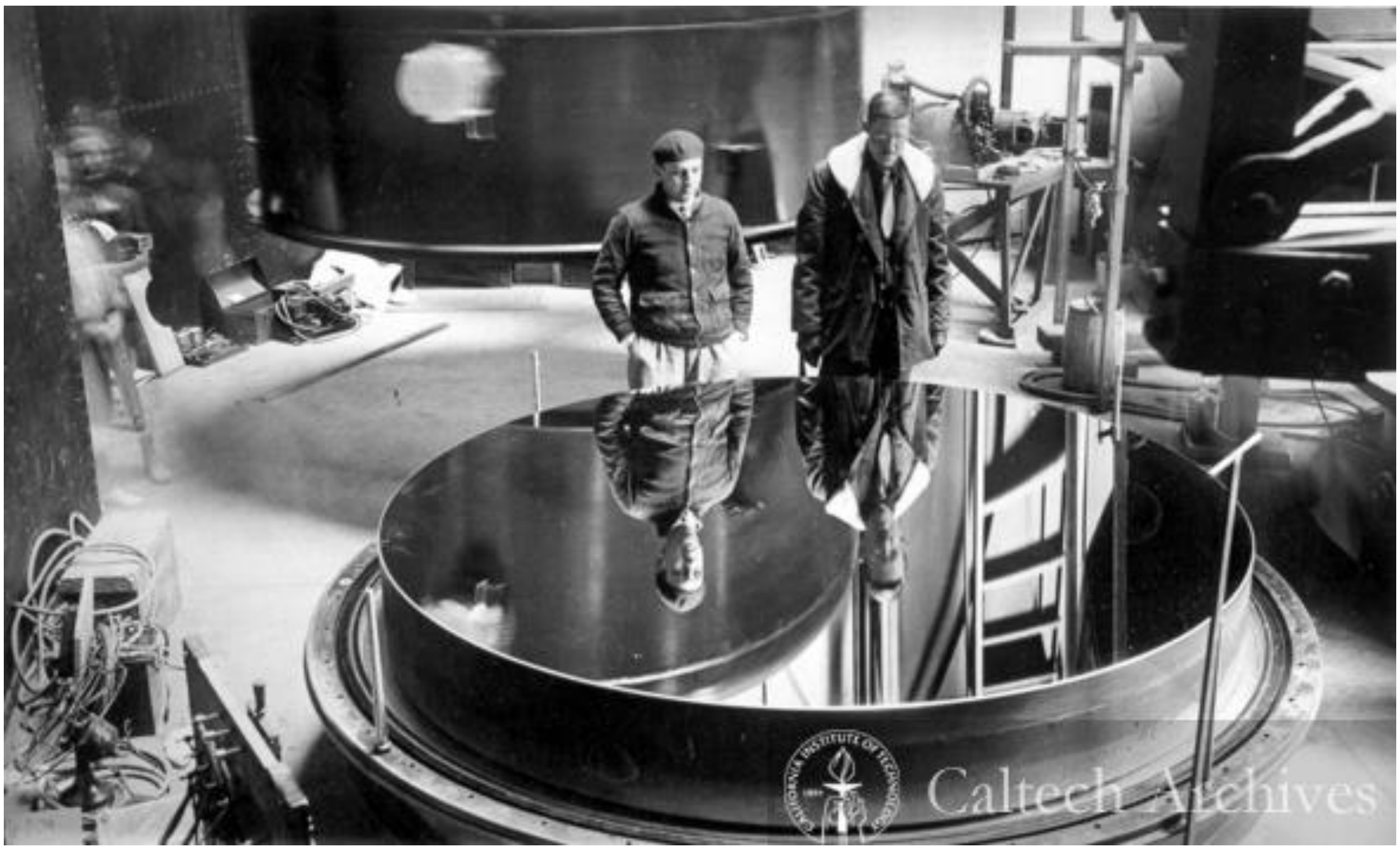

El espejo de Córdoba, finalmente, se mandó a Pittsburg para ser terminado, y hacia el final Gaviola viajó para certificarlo. Hubo un montón de demoras, pero finalmente se terminó el telescopio (es el que está en el Observatorio de Bosque Alegre), y fue una revolución para la astronomía argentina. En esta foto (de Córdoba Estelar, de Minniti Morgan y Paolantonio), lo vemos de pie frente al espejo finalizado, antes del aluminizado.

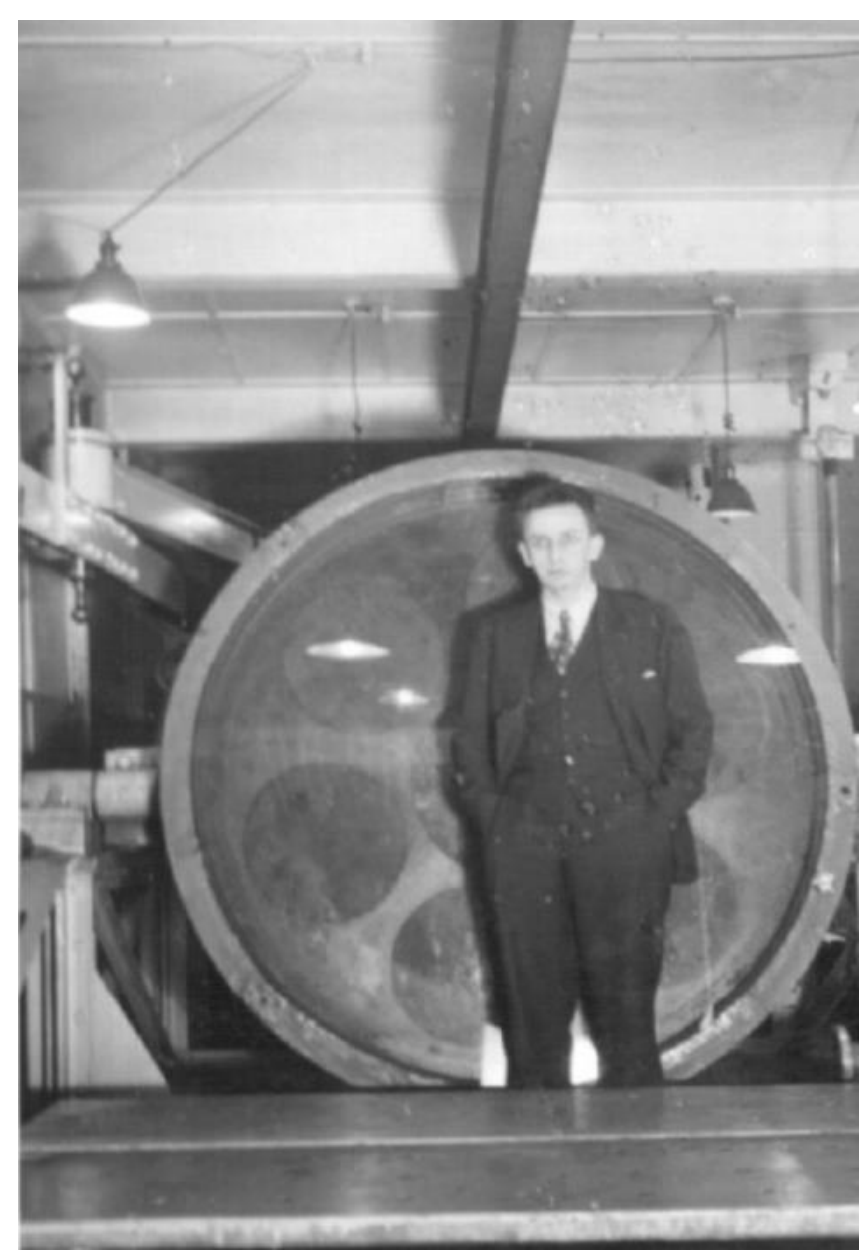

Llevó años incorporar a Gaviola al Observatorio, donde acabaría siendo Director. Gaviola quería transformar el observatorio en un centro de astrofísica, y fue crucial el trabajo que hizo con Ricardo Platzeck: el desarrollo de un extraordinario espectrógrafo, que ganó fama mundial. Tenemos en el archivo planos del espectrógrafo.

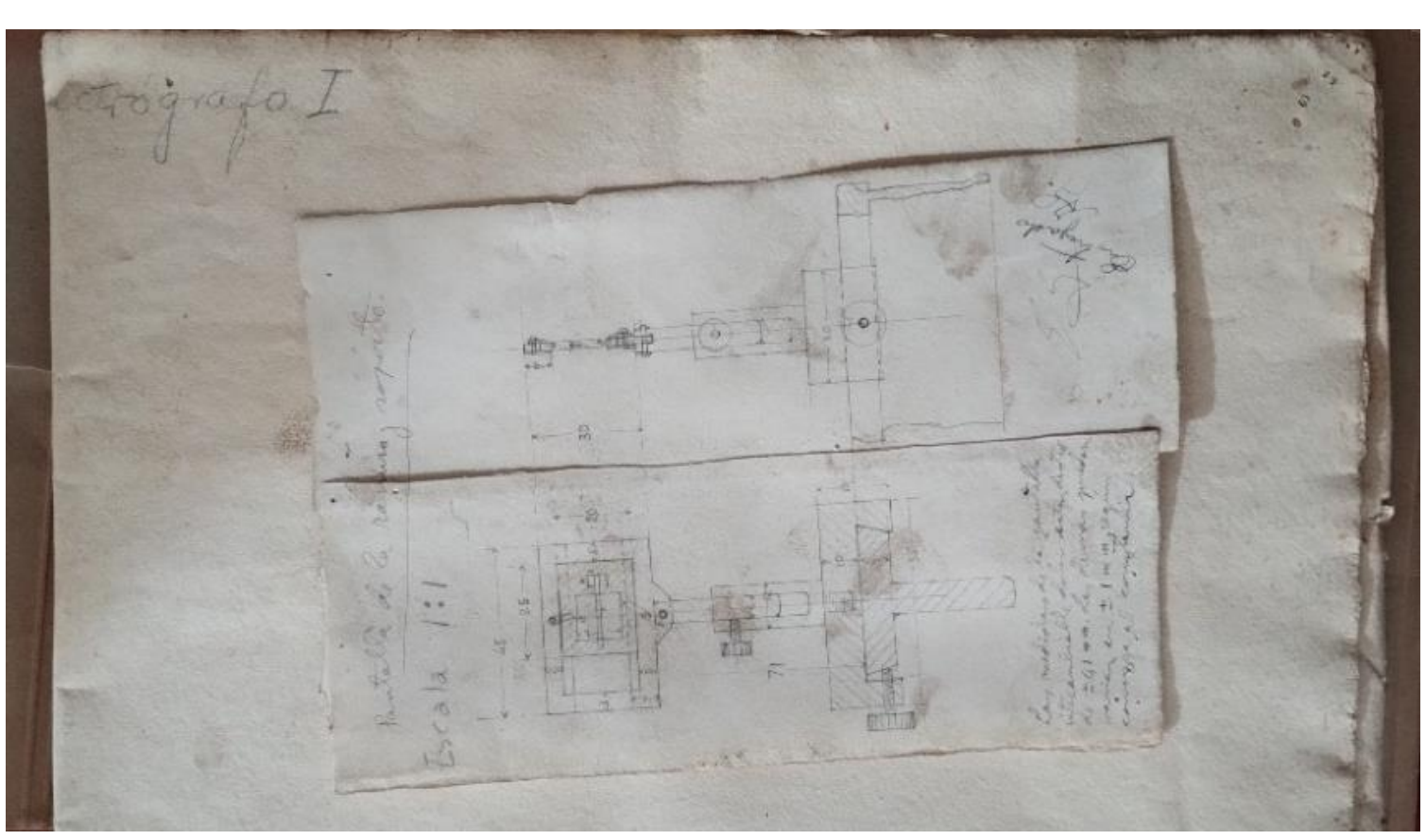

Entre los muchos documentos interesantes sobre el trabajo de astronomía, está este cuaderno llamado Espectrógrafo Mediano, que comienza en marzo de 1943 con una descripción de cómo hay que poner el portafilm, y cómo hay que enfocar.

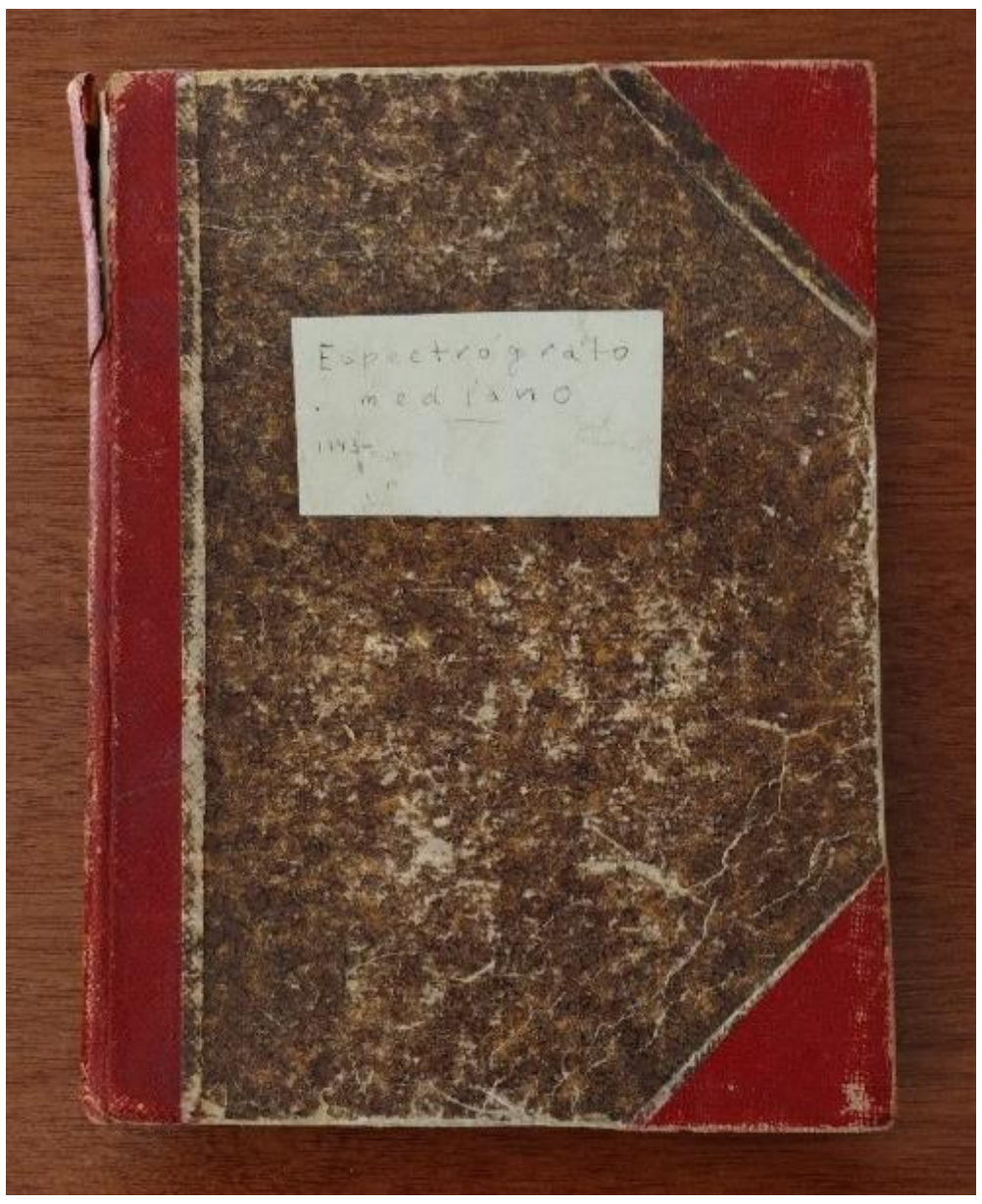

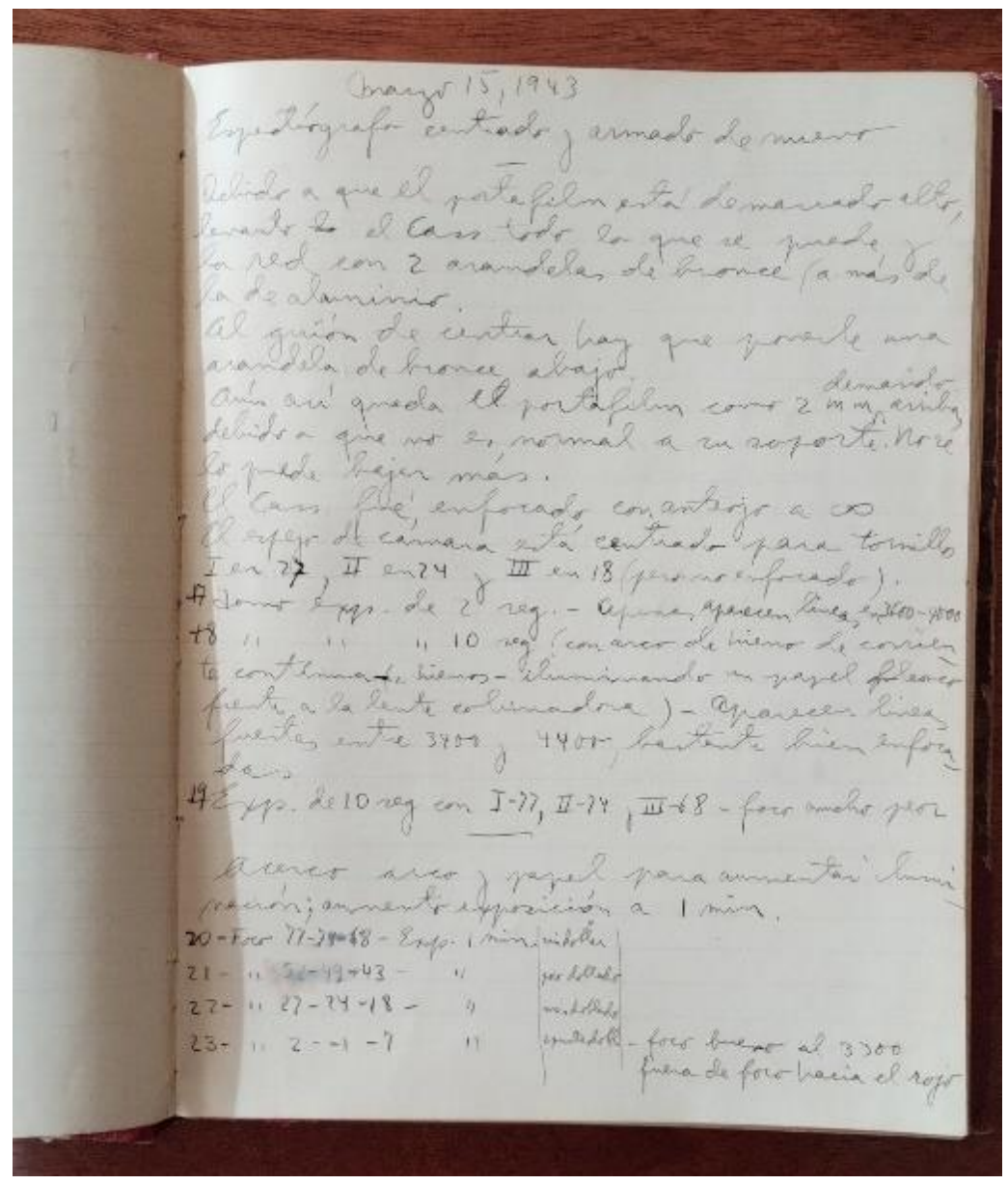

Es un típico cuaderno de laboratorio, con dibujos, indicaciones de cómo usar los instrumentos, pruebas de medición, espectros que alguna vez estuvieron pegados... Hay observaciones seguramente valiosas, como este espectro de la Nova Puppis.

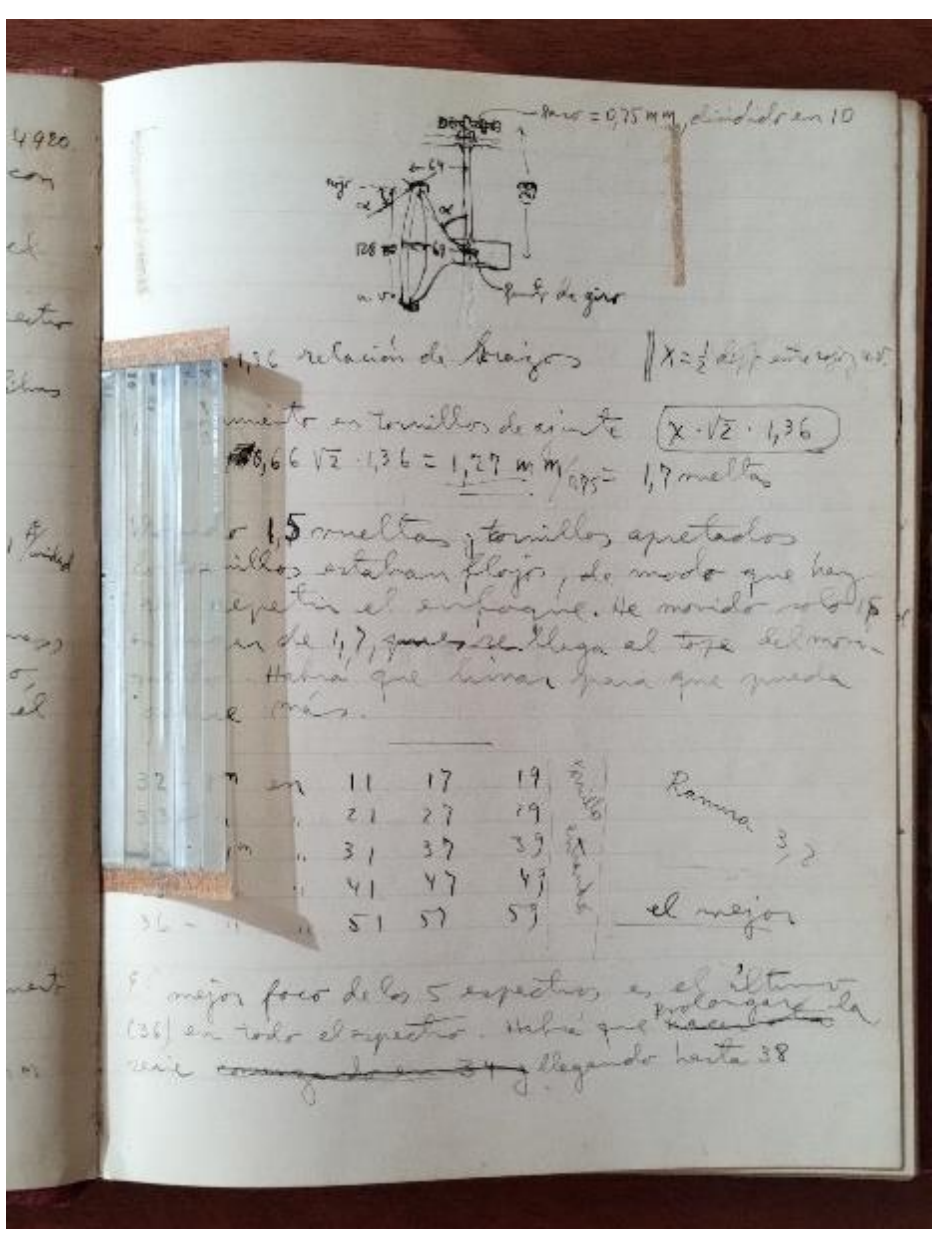

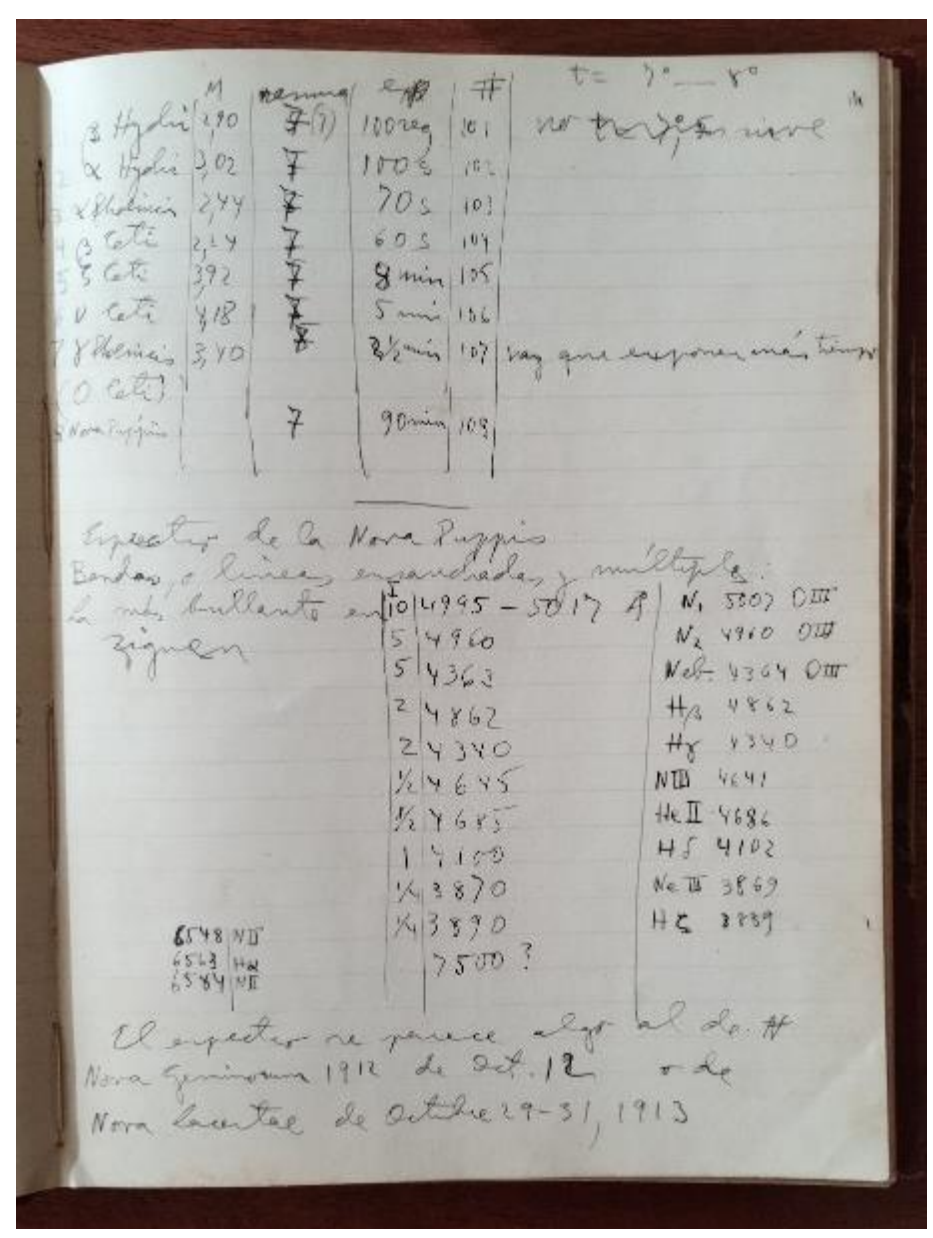

Es interesante ver cómo aumentan las observaciones de Eta Carinae, que se convertiría en su estrella favorita, la frustración de los días nublados y de las exposiciones mal guiadas, la lucha para no confundir dónde están los puntos cardinales.

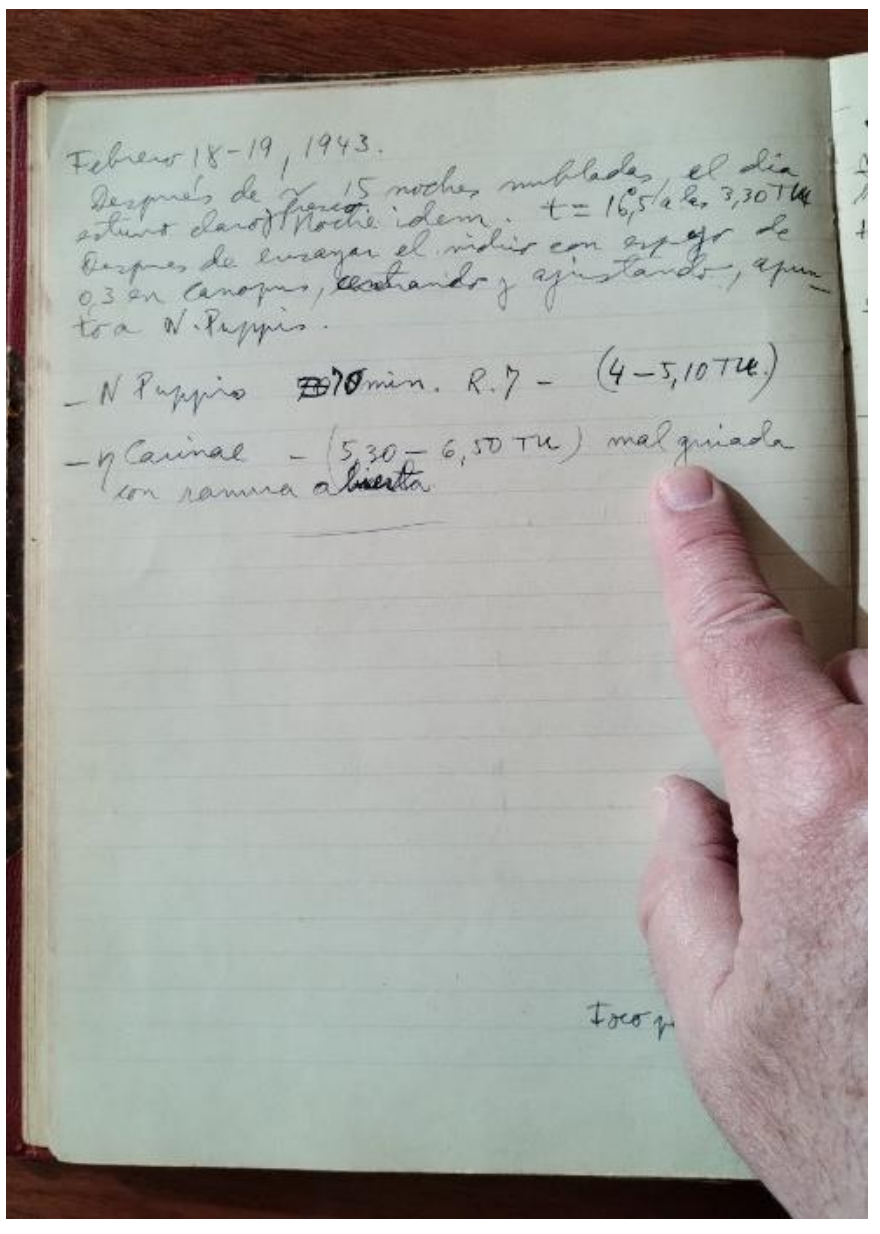

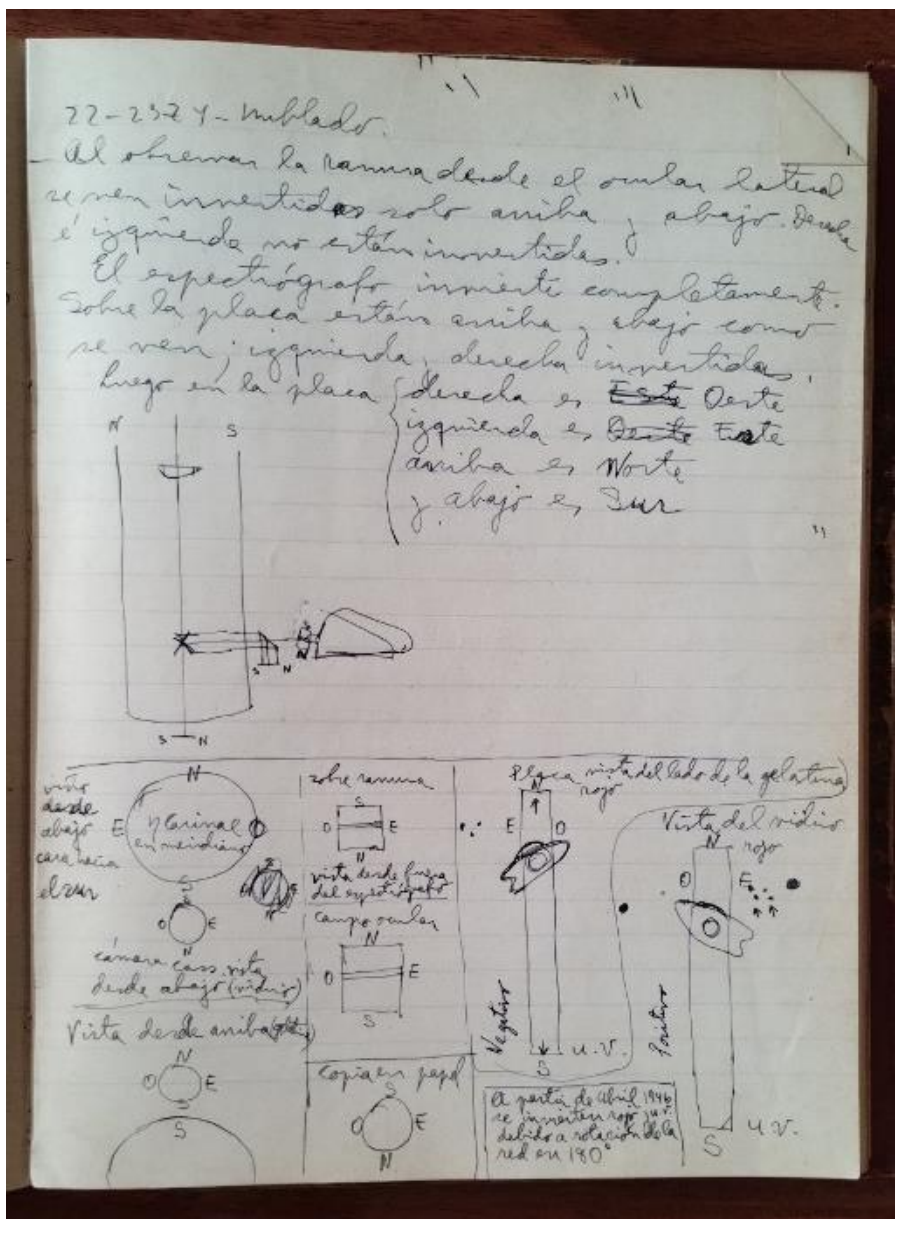

Los días de observación hay indicaciones del tiempo. También encontramos recordatorios, como el que al pie de la siguiente imagen dice *"llevar a Bosque tul negro para trabajar en colmenas"* (abajo, izquierda).

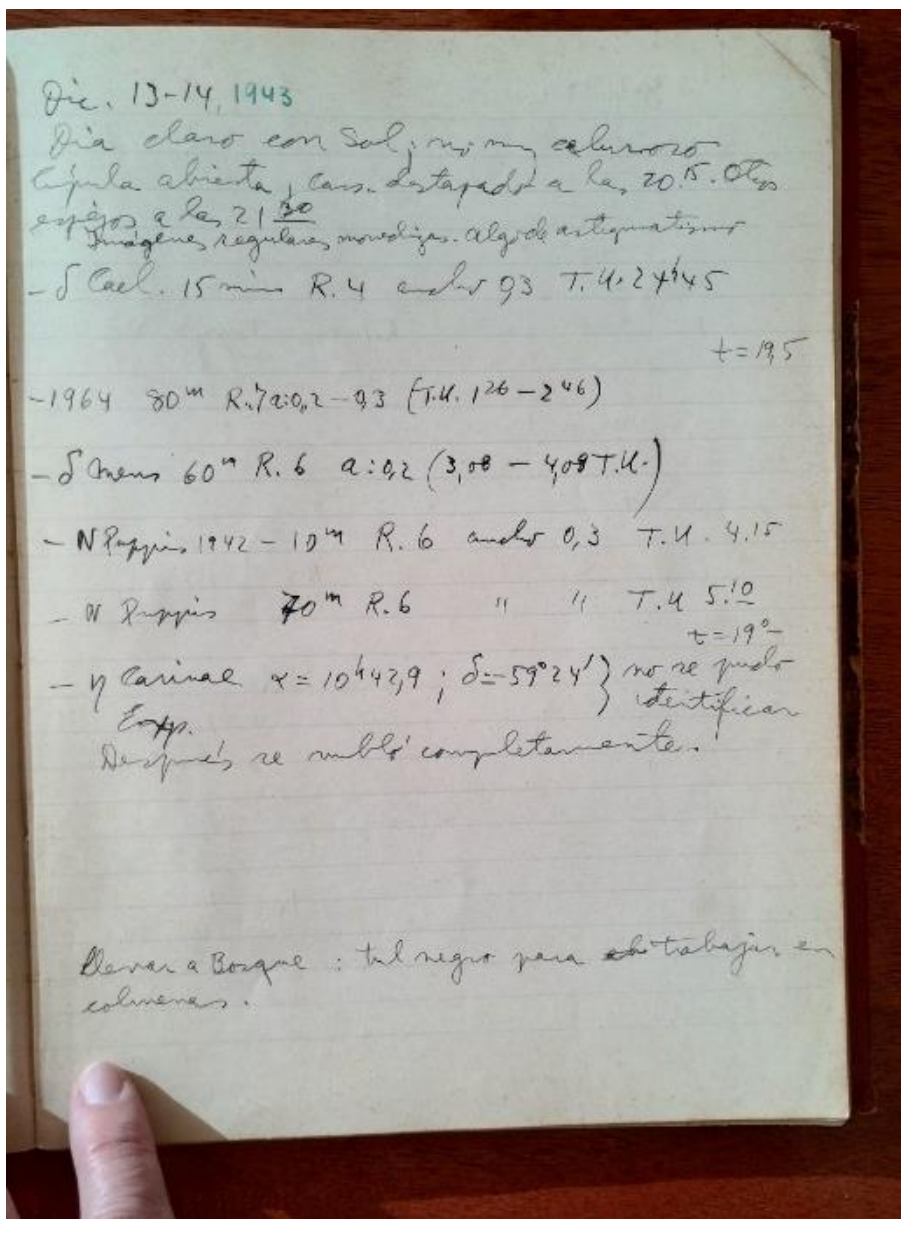

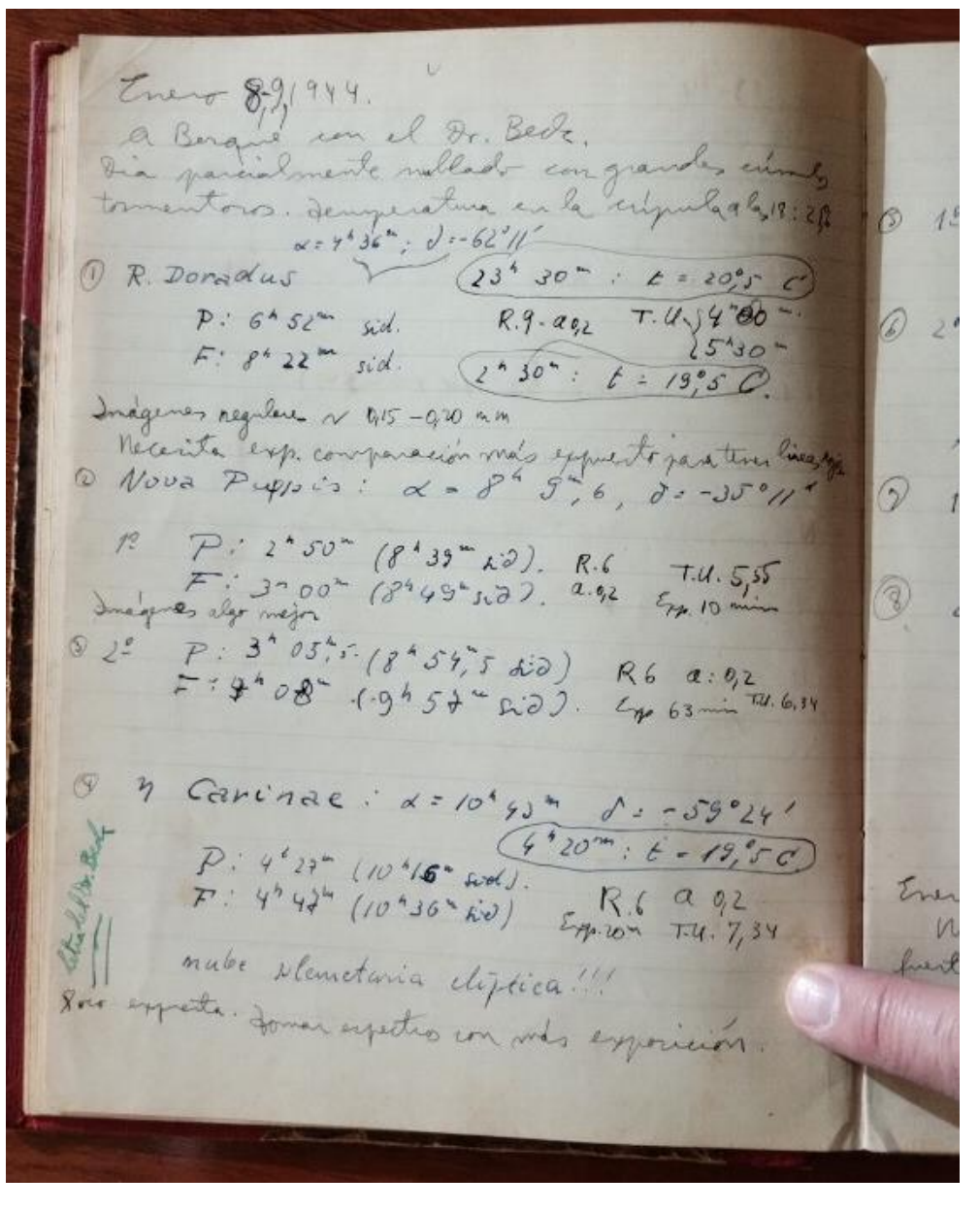

El 8 de enero del 44 fue a Bosque con Guido Beck (imagen de arriba a la derecha). Beck era un físico extraordinario, a quien Gaviola ayudó a escapar de Europa, y revolucionó la física teórica en Argentina primero y en Brasil después. Ese día hay una observación de Eta Carinae, con una notita marcada con triple signo de exclamación y señalada en verde como *"Letra del Dr. Beck"*. A continuación muestro un zoom del texto:

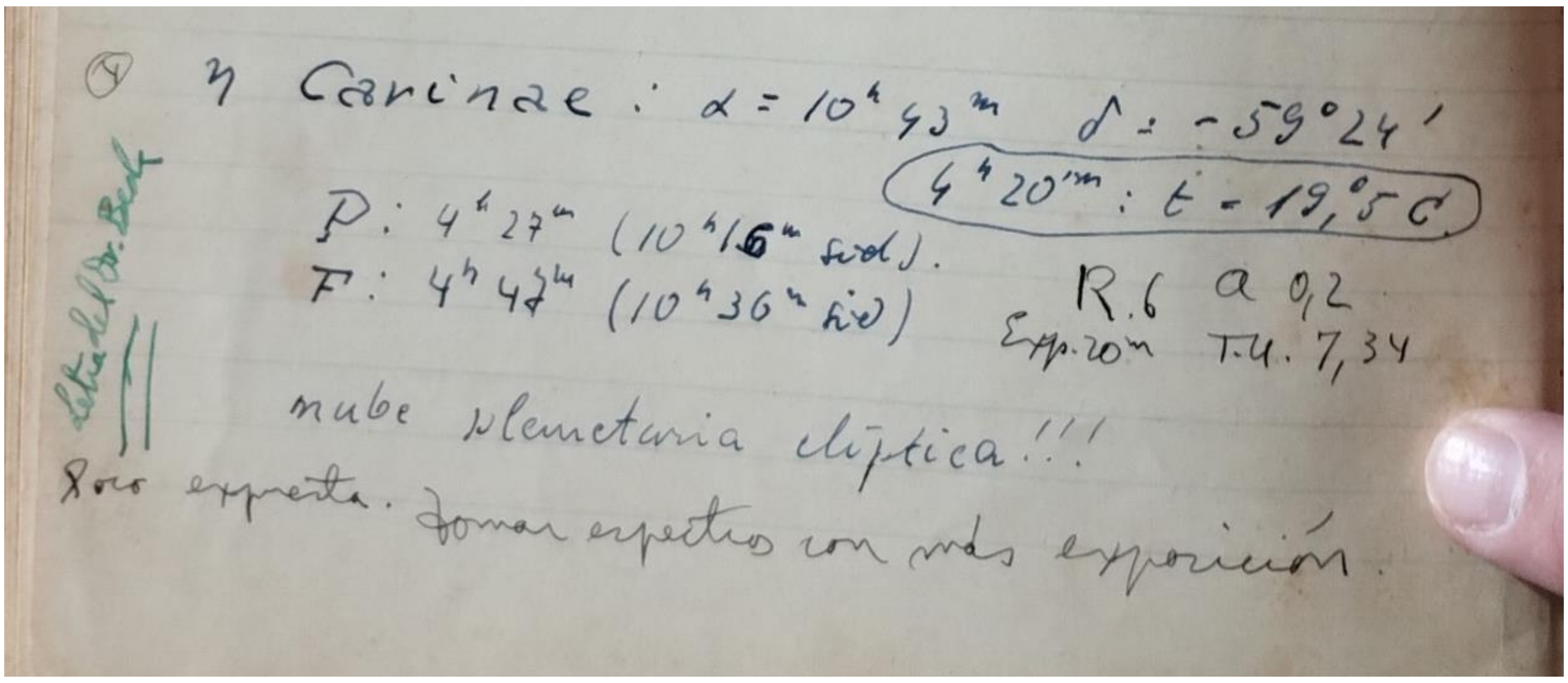

Ese día, entonces, descubrieron la famosa nebulosa alrededor de Eta Carinae, así que el descubridor de la nebulosa que hoy llamamos *Homúnculo de Gaviola* es Guido Beck. Esta rara nebulosa se convirtió en una obsesión para Gaviola. Encontré en el archivo un par de placas de vidrio con imágenes positivas, donde se aprecia la (hoy) familiar formita.

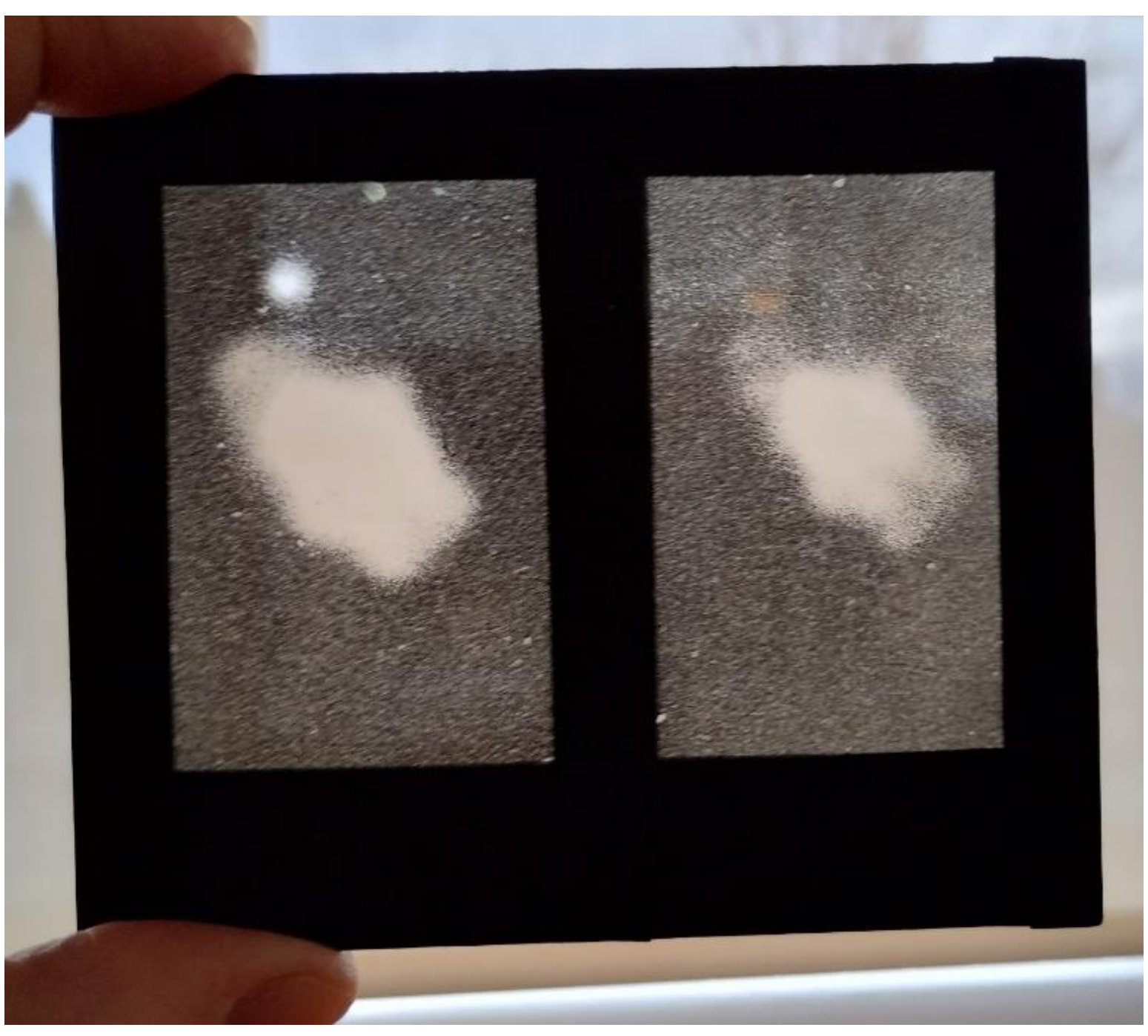

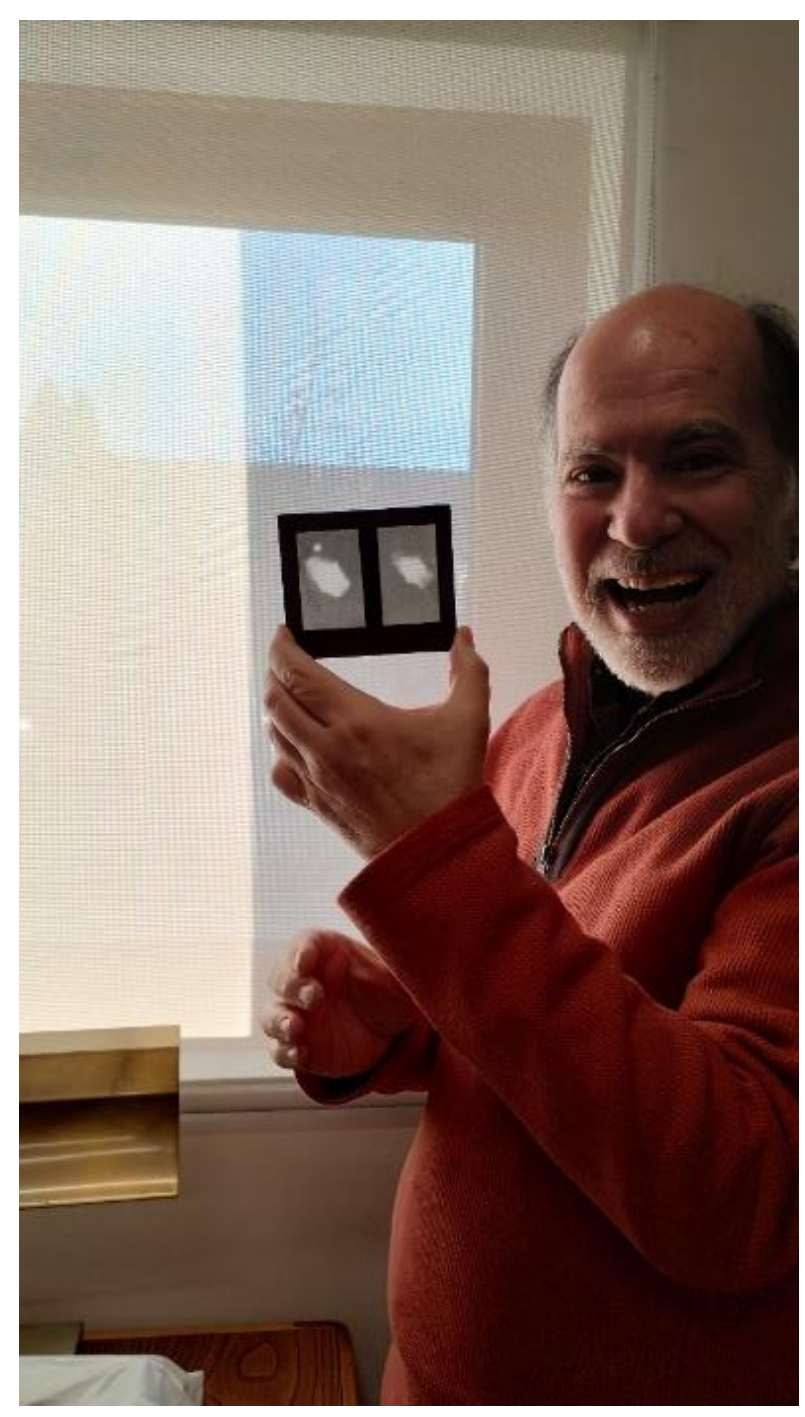

En la Reunión de la AFA de 1945 en Córdoba, Gaviola la describe como un “muñeco” (tercer párrafo).

ENRIQUE GAVIOLA (Córdoba): *El espectro de Eta Carinae.*

El espectro de Eta Carinae ha sido fotografiado con el gran reflector de la Estación Astrofísica y el espectrógrafo a red de reflexión cubriendo la zona entre 3076 y 6678. Varios cientos de líneas en emisión han sido medidas. Aparecen prominentes la serie de Balmer, algunas líneas del He y Na y líneas prohibidas y permitidas del hierro ionizado. $H\alpha$ tiene una fuerte componente desplazada 1000 km/seg. hacia el rojo que no aparece en el resto de la serie.

En absorción aparecen los términos superiores de la serie de Balmer desplazados hacia el violeta y las líneas amarillas del sodio.

Fotografías directas obtenidas con 31,50 metros de distancia focal equivalente e imágenes inferiores al segundo de arco muestran que Eta Carinae es una nebulosa con un punto de máxima intensidad sin forma estelar. El conjunto tiene la forma de un muñeco de 12 segundos de alto y 8 de ancho. En el mismo se distinguen la nebulosa central en forma de coma y 10 nubecillas periféricas.

El espectro de la cabeza es diferente del del cuerpo. Son prominentes $H\alpha$ desplazada 1000 km/seg. hacia el rojo y las líneas no clasificadas del nebulio 3869 y 4069 con poco desplazamiento. Otras líneas del nebulio (las más fuertes generalmente) no aparecen.

Gaviola eligió una de sus mejores fotos para un artículo genérico que publicó en Nature en ocasión del 75o aniversario del OAC. Allí no le da un nombre. El otro astrónomo que estaba estudiando la estrella era David Thackeray, de Cambridge, en Pretoria, pero publicó apenas un dibujo recién en el 49, describiéndola cono forma de "cuña" o "barba".

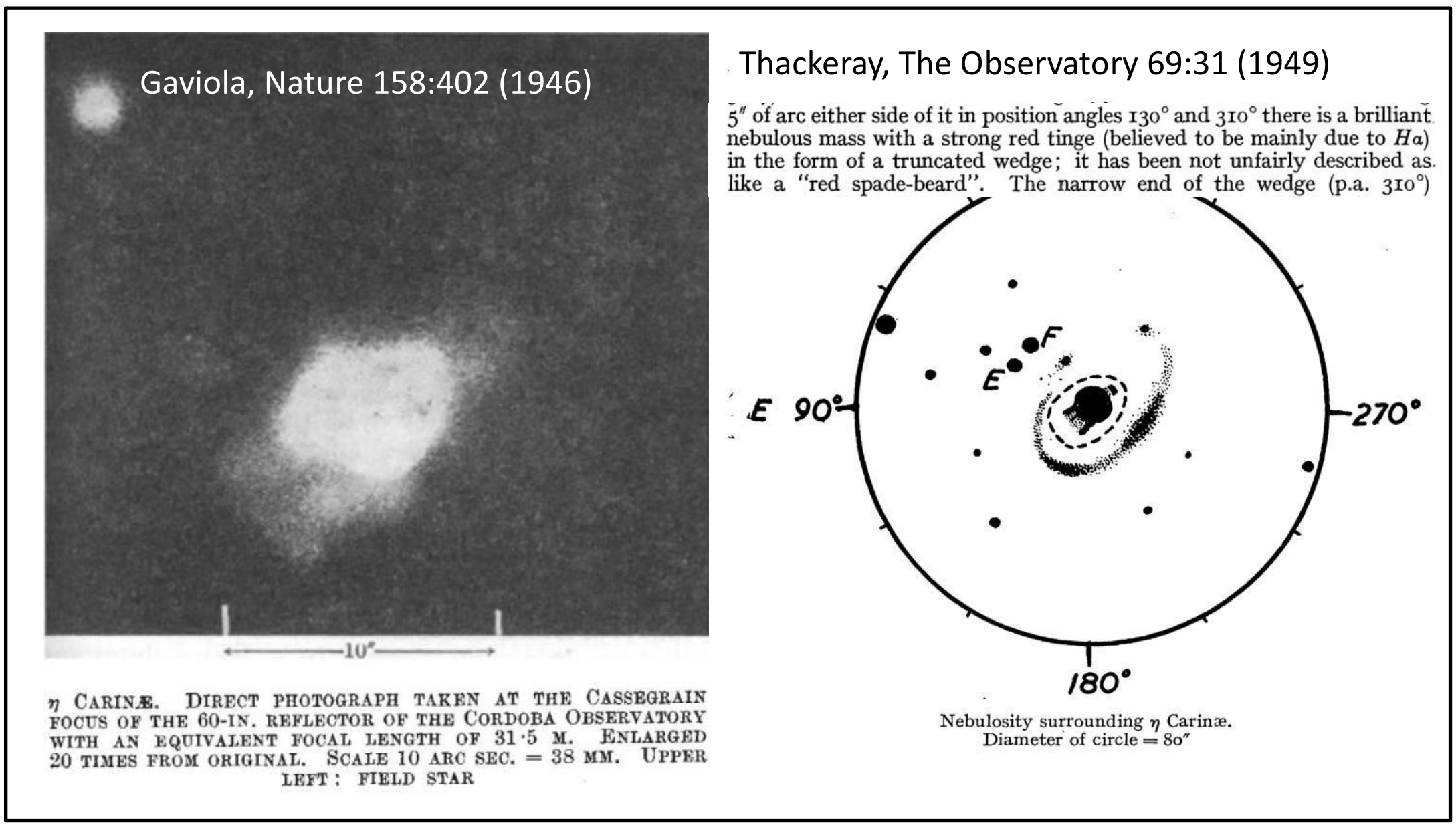

Gaviola publicó dos trabajos cruciales sobre Eta Carinae, uno sobre la nebulosa y otro sobre su espectro. De los dos tenemos los manuscritos y toda la correspondencia con la revista, hasta la cuenta de cuánto iba a costar, y los pedidos diversos para conseguir los dólares.

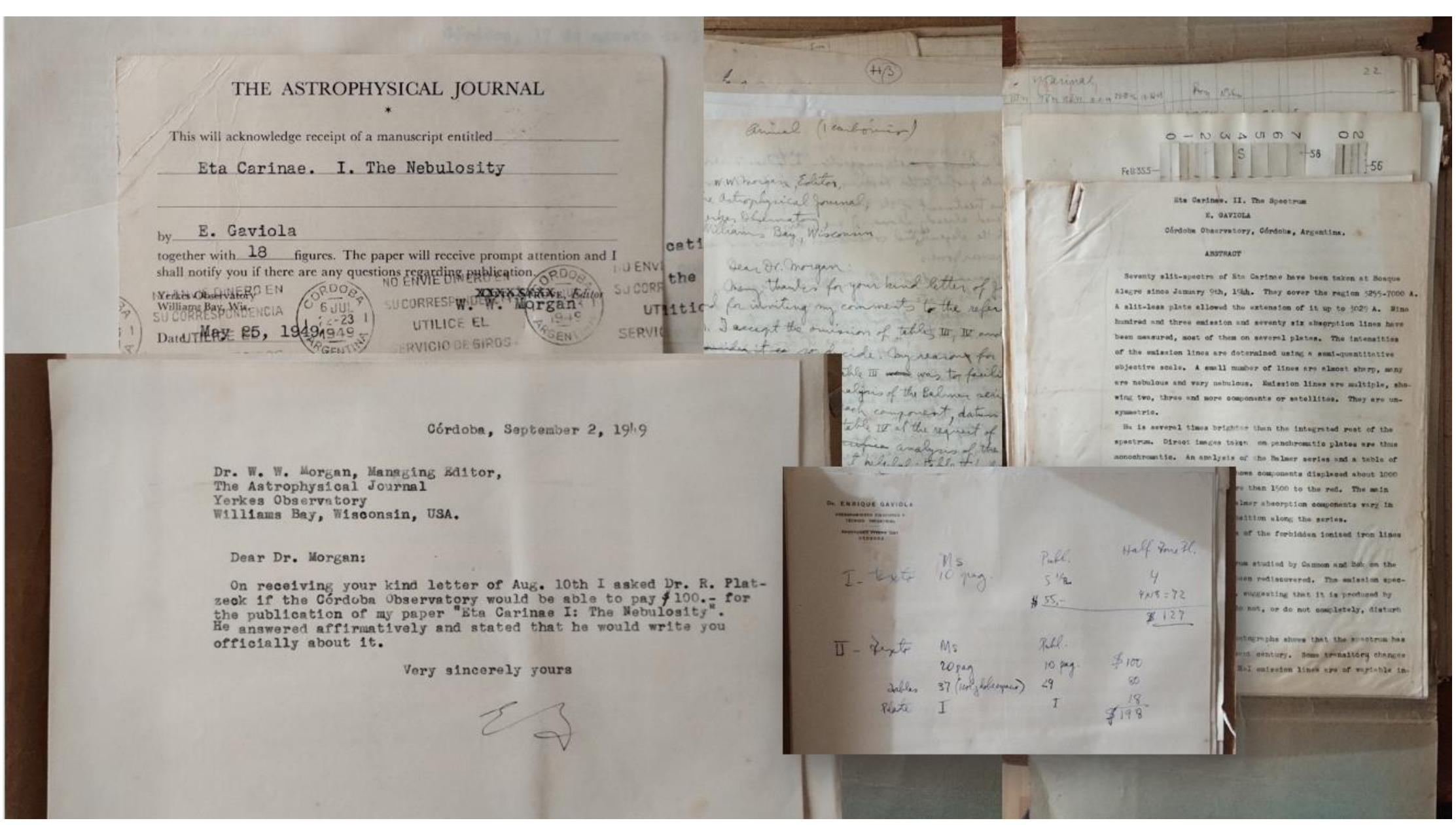

Hay manuscritos mecanografiados (como el que vemos en la imagen de arriba, a la derecha), y otros de puño y letra, donde aparece la palabra *"homunculus"*, seguramente por primera vez (cuarta línea, imagen de abajo a la derecha).

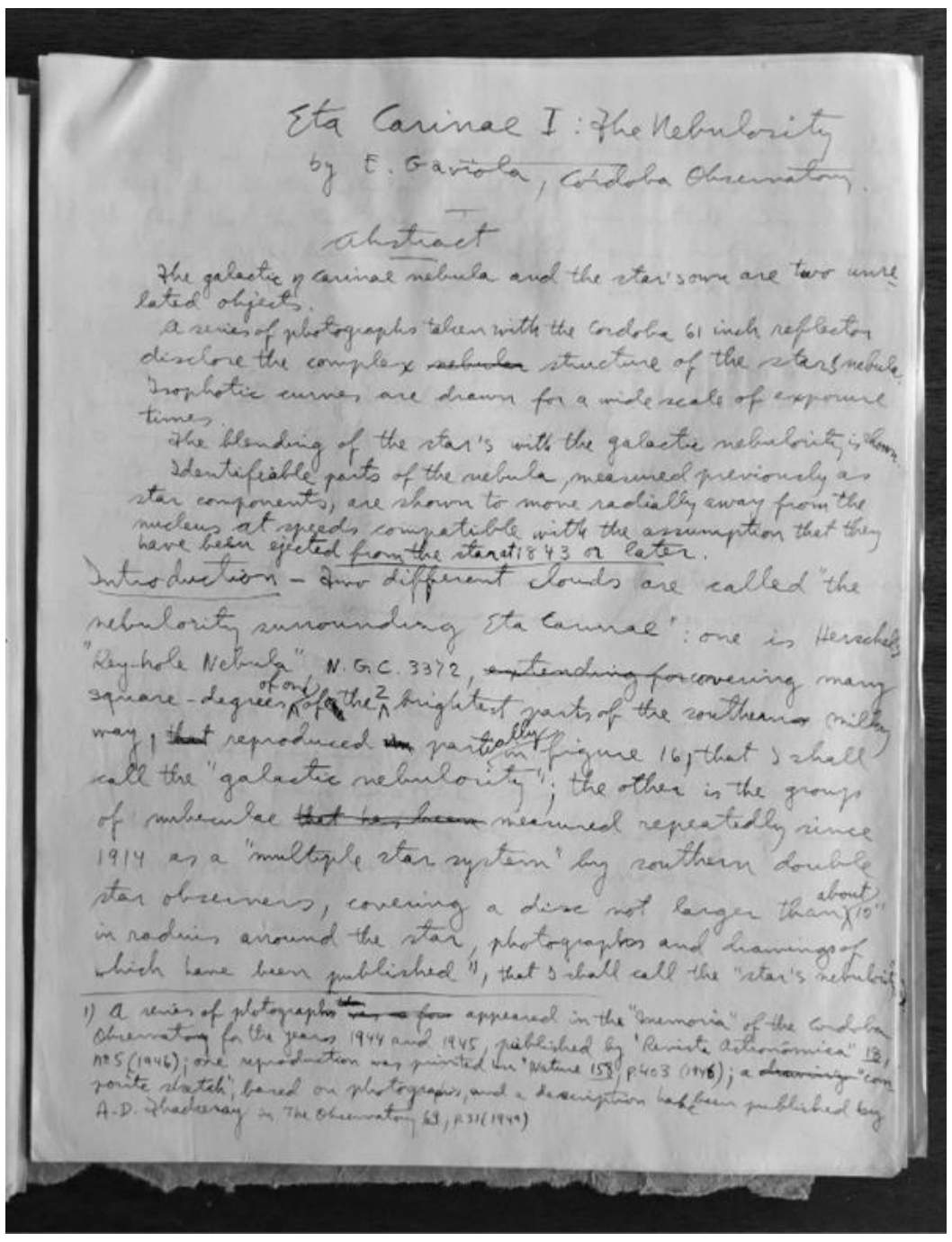

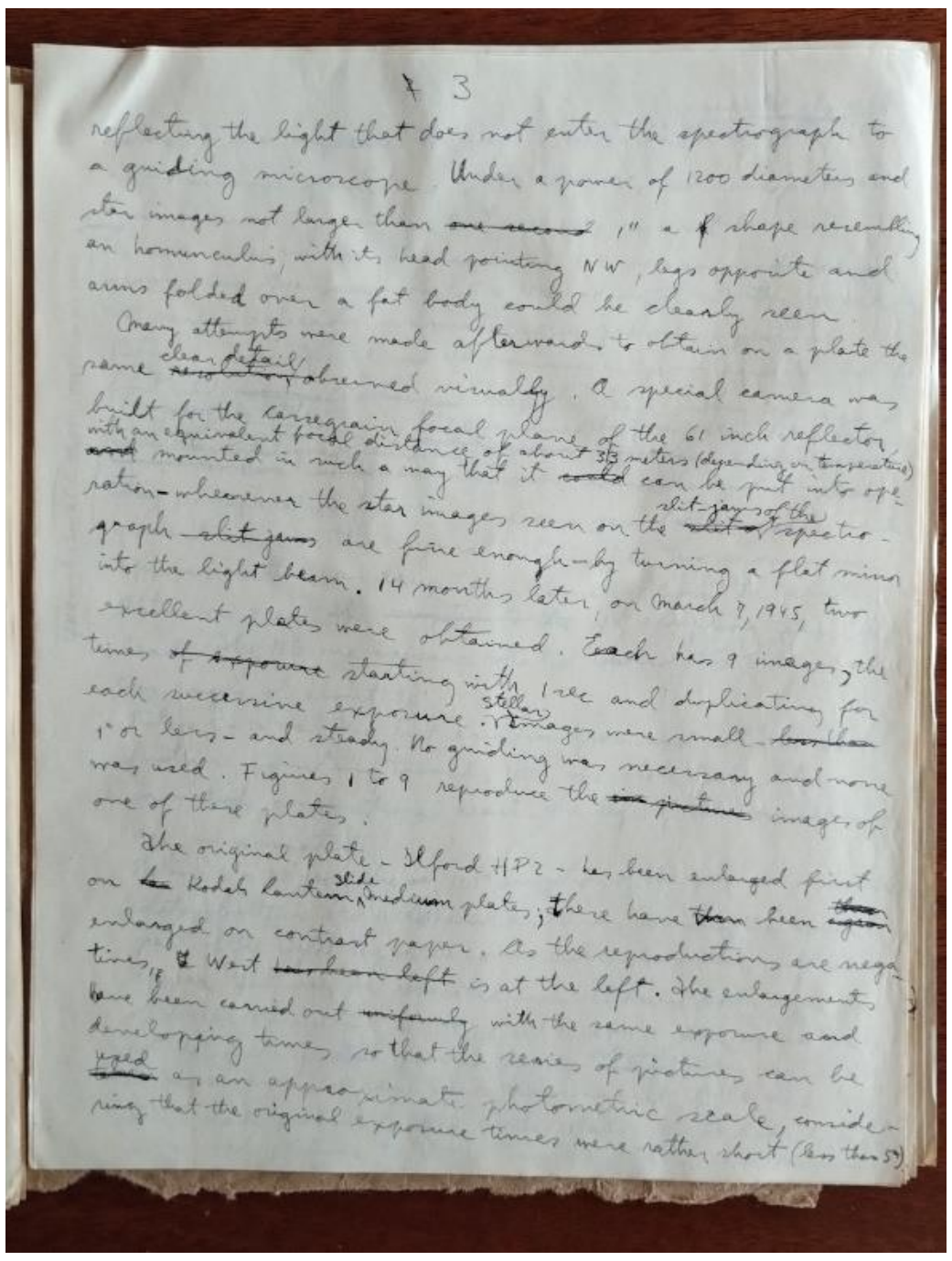

Las tomas definitivas del Homúnculo, las que aparecieron en las publicaciones, están registradas en el cuaderno en marzo 6/7, descriptas como "buenas, quietas, ⌀0.2 mm":

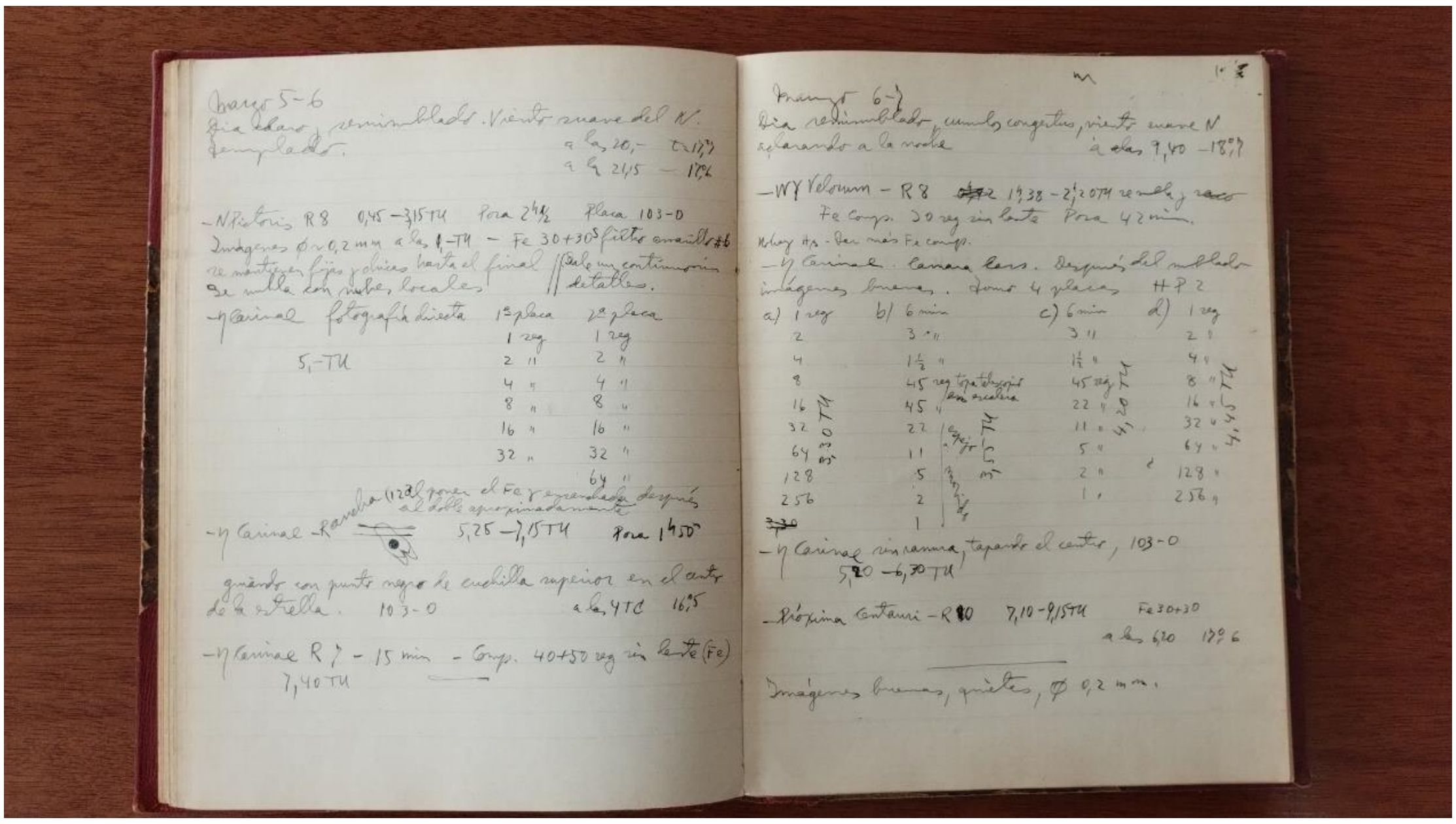

Esas fotos, así como los espectros hechos con el mundialmente famoso espectrógrafo de reflexión que había diseñado y construido con Platzeck, tal como aparecieron en las publicaciones, se conservan en el archivo:

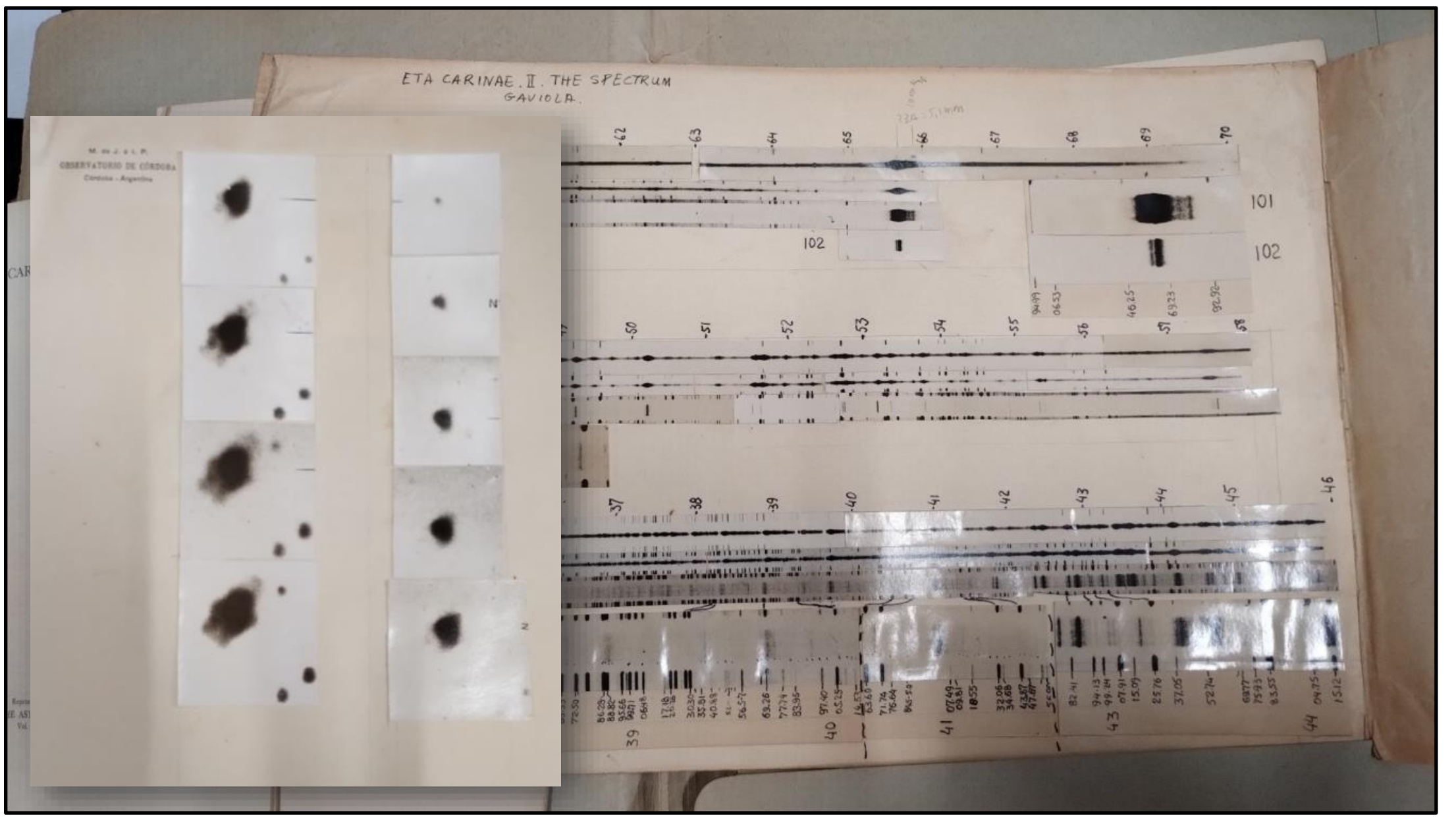

¿Qué es el Homúnculo, y por qué es tan notable? Cerca de la Cruz del Sur hay un rombo de cúmulos estelares alrededor de la gran Nebulosa de Carina. Es la región del cielo más rica en grandes cúmulos y nebulosidad. La Nebulosa de Carina es una maravilla: es tan brillante, y emplazada en medio de tantos ricos cúmulos estelares, que nada se compara con ella, en ningún lugar del cielo. Es más grande y más brillante que la famosa Nebulosa de Orión, si bien tal vez es menos conspicua por encontrarse justo en medio de la Vía Láctea.

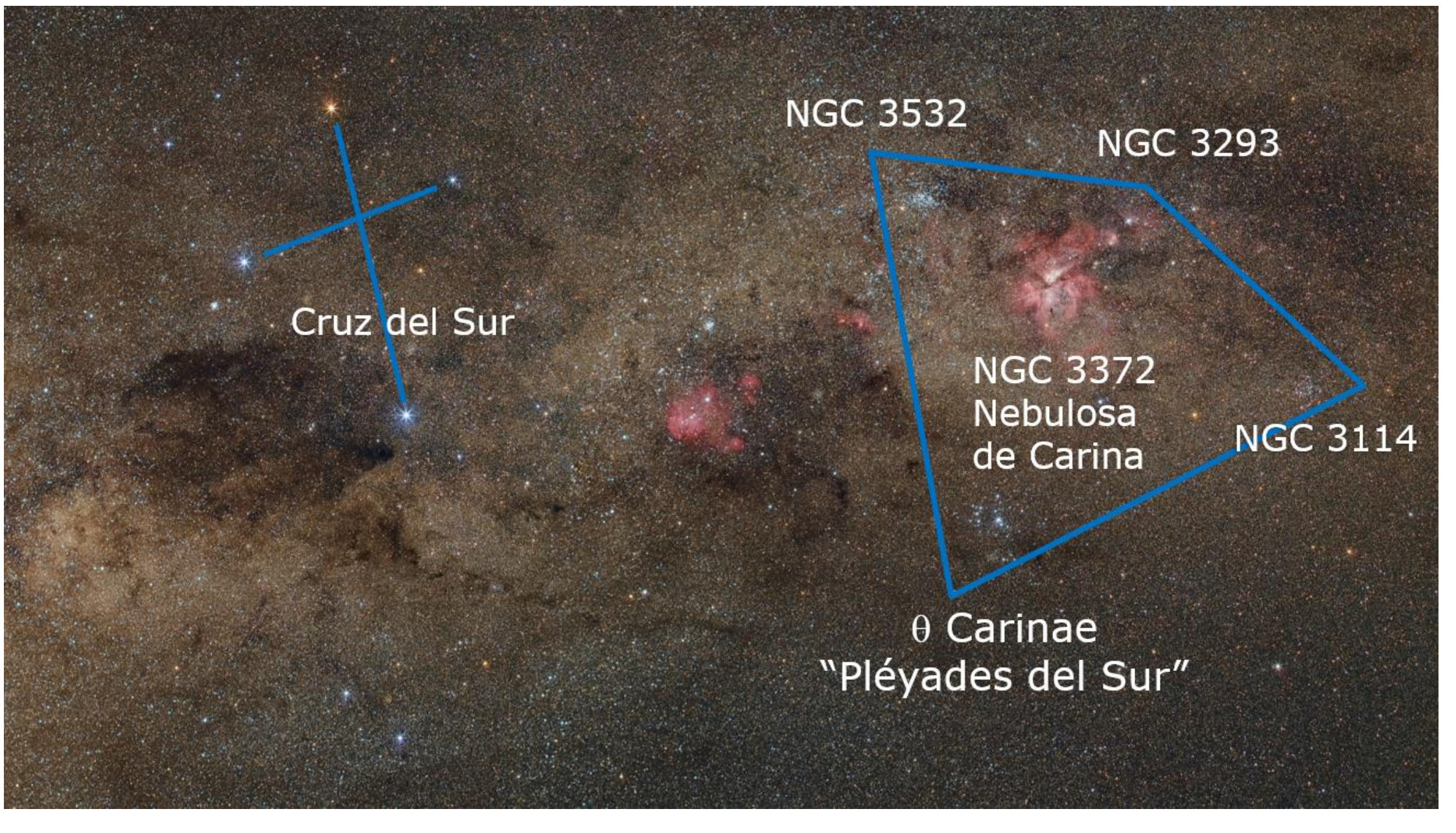

La nebulosa es una gigantesca región de formación estelar, tal vez la más grande de nuestra galaxia. Tiene una estructura intrincada de gas brillante y filamentos oscuros. La estrella Eta Carinae se encuentra en la parte más brillante de la nebulosa, y ya era una estrella famosa, notable en particular por su variabilidad.

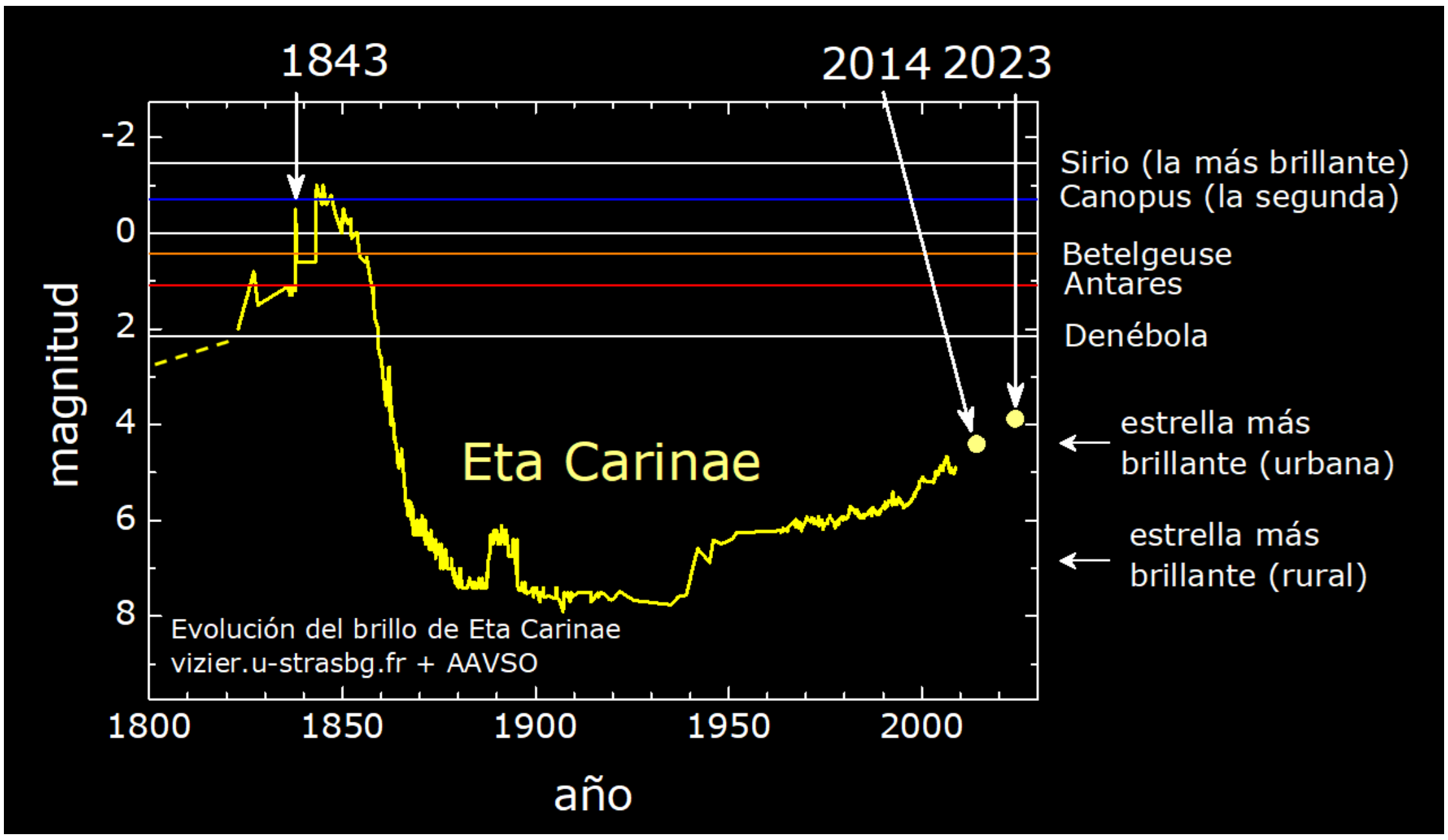

La figura muestra una curva de luz correspondiente al último par de siglos. Durante el siglo XIX fue aumentando de brillo hasta que, de golpe, en 1843, sufrió lo que se llama su Gran Erupción. Se convirtió en la segunda estrella más brillante del cielo nocturno, más brillante que Canopus, y luego se fue apagando hasta hacerse invisible. Hoy sabemos que durante este episodio eruptivo la estrella expulsó lo que hoy es el Homúnculo. A mediados de la década de 1940, justo cuando Gaviola jugó un rol fundamental en su observación, Eta Carinae empezó a aumentar de brillo nuevamente. Sigue haciéndolo, y hoy es de nuevo una estrella visible a simple vista.

Si hacemos un zoom en la región más brillante de la nebulosa, vemos varios cúmulos estelares en su interior. Una región oscura llamada Nebulosa de la Cerradura ocupa la parte central. A su izquierda, formando parte de uno de los cúmulos estelares, se encuentra el objeto (señalado por una flecha), designado Eta Carinae como si fuera una estrella. Pero no es solamente una estrella, tiene efectivamente la forma nebulosa elíptica que llamó la atención de Beck el mismo día que la fotografiaron.

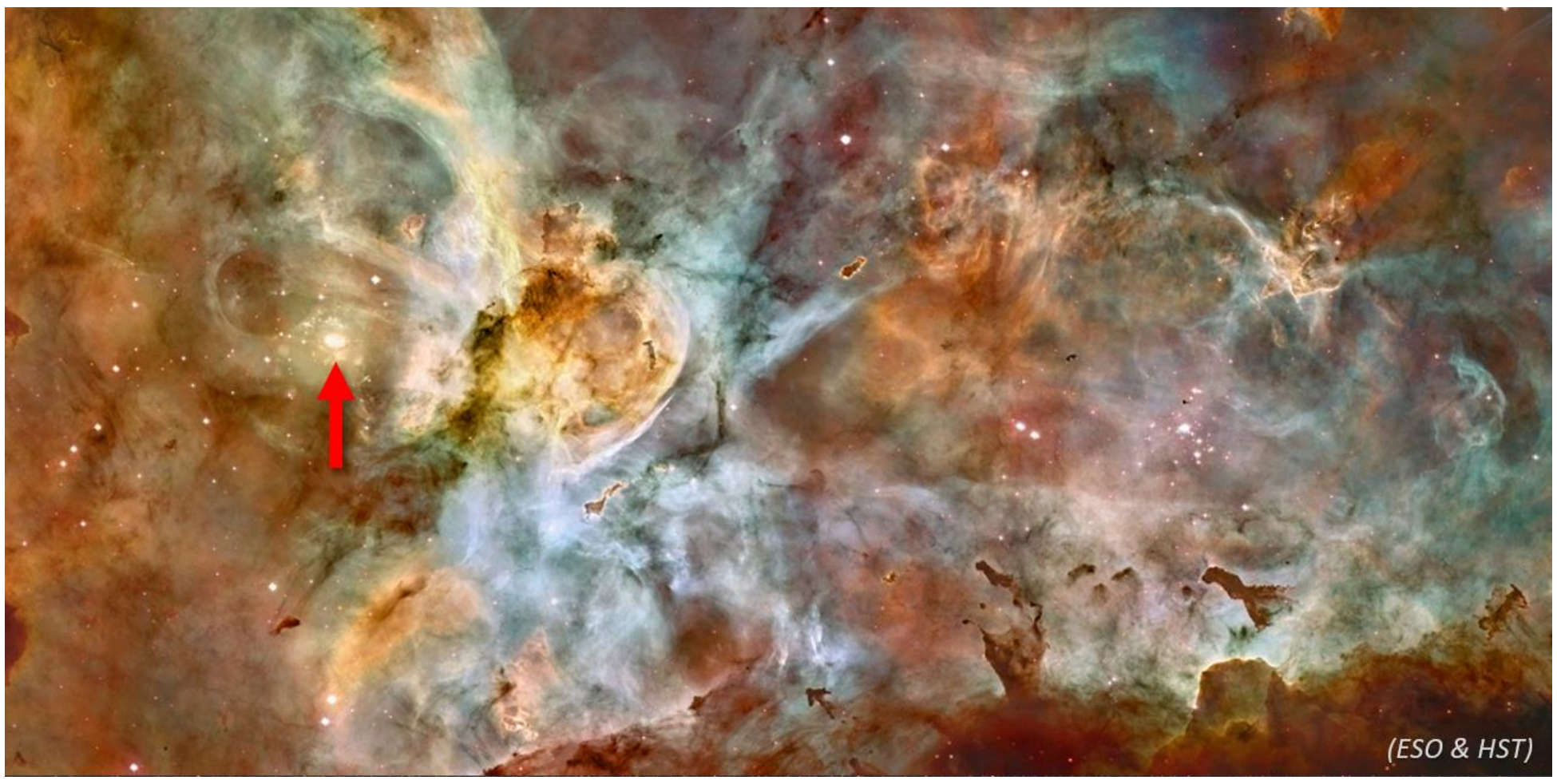

El Homúnculo es una nebulosa muy peculiar, un objeto único. En las increíbles imágenes modernas podemos ver su complejidad. Pero es incluso posible fotografiarla con un telescopio modesto, desde el balcón de mi casa. Esta es la nebulosa que le llamó la atención a Beck.

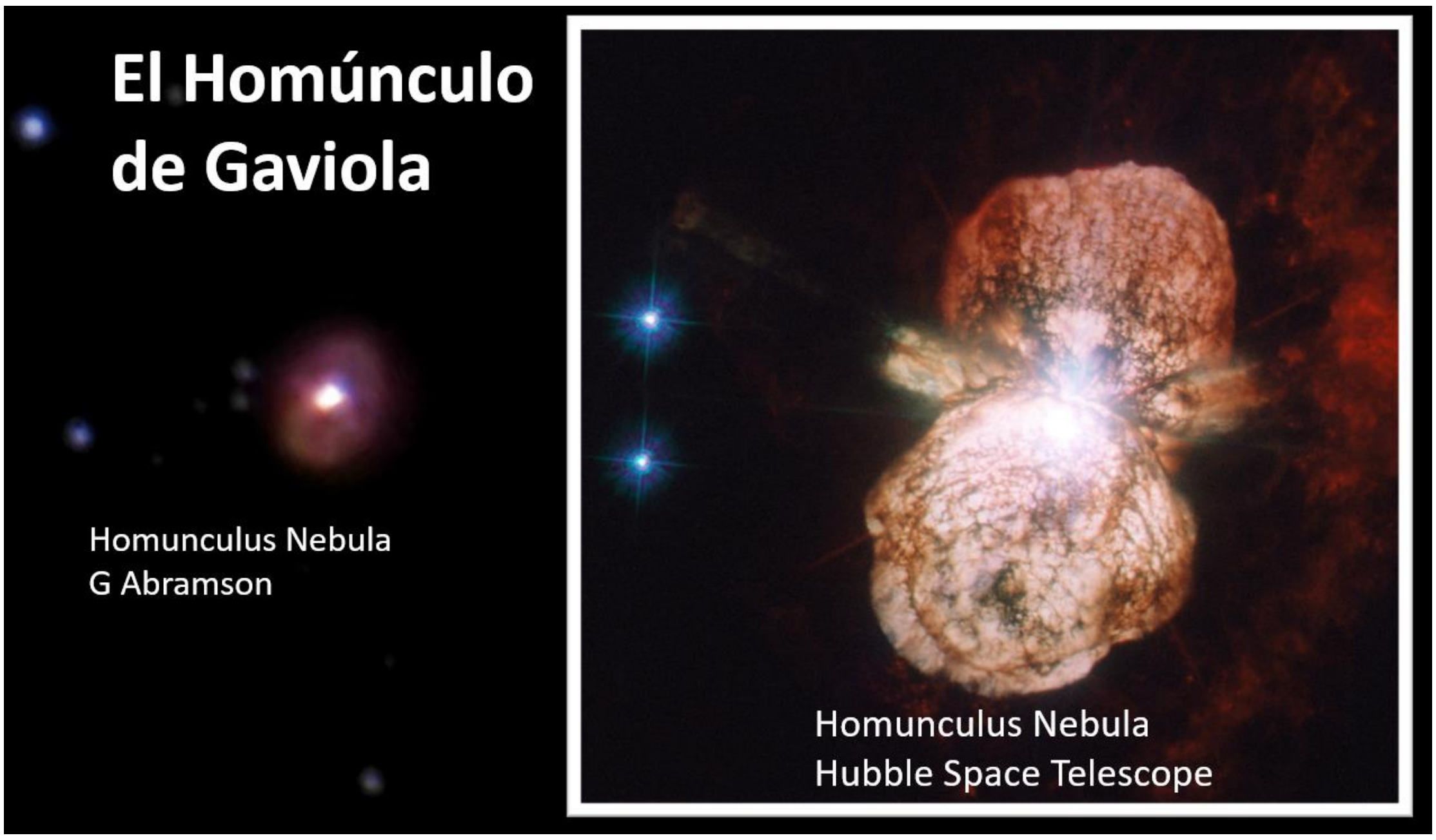

Para Gaviola fue fundamental el estudio que hizo del espectro del Homúnculo. Tenemos muchos de los espectros en placas de vidrio, en film y hasta en película de cine.

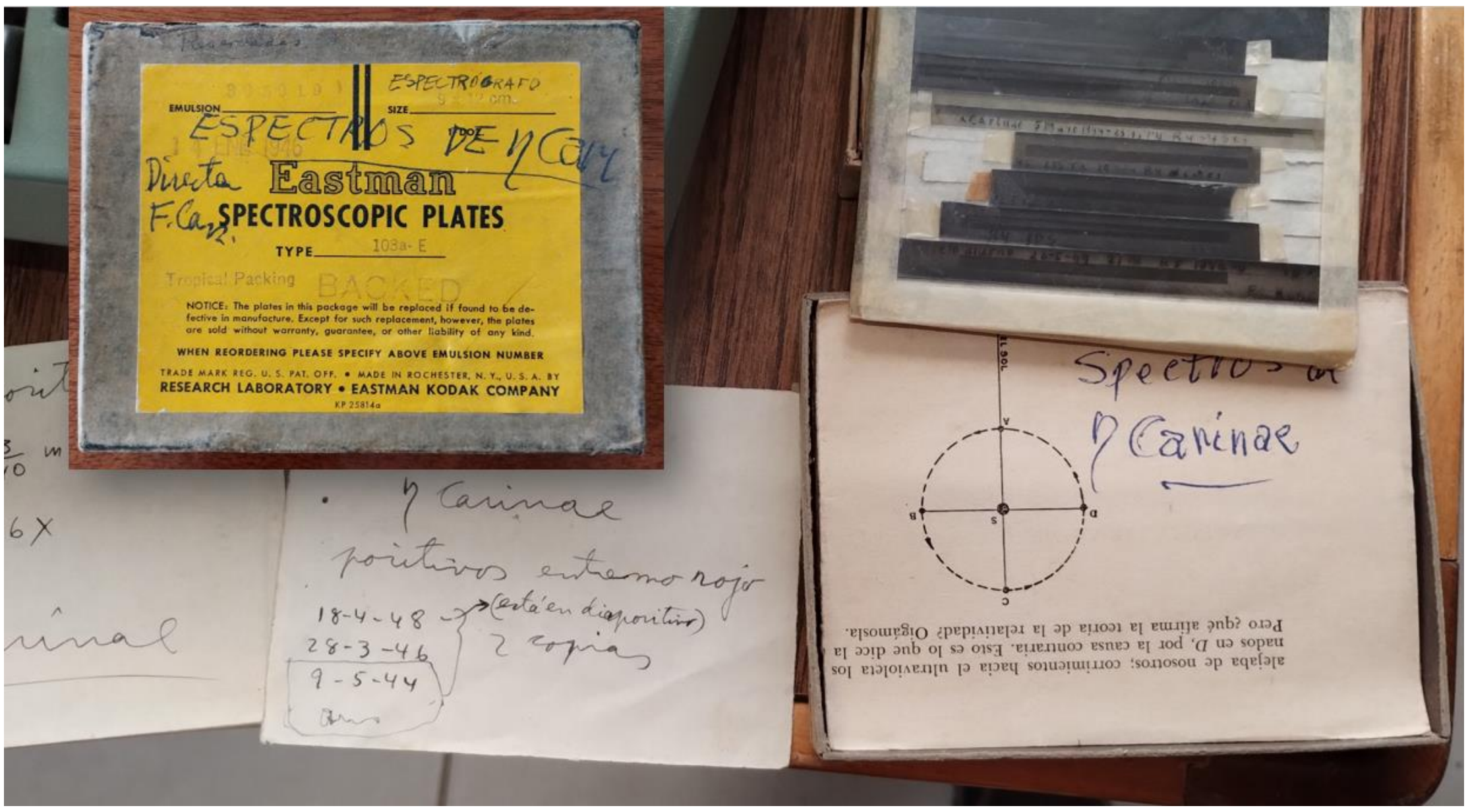

Este es uno de los espectros de vidrio, y tal vez este es el tipo de rotura que se producía porque el plano focal del espectrógrafo era cóncavo. Con Platzeck, desarrollaron una técnica de secado del vidrio antes de usarlo, para reducir las roturas.

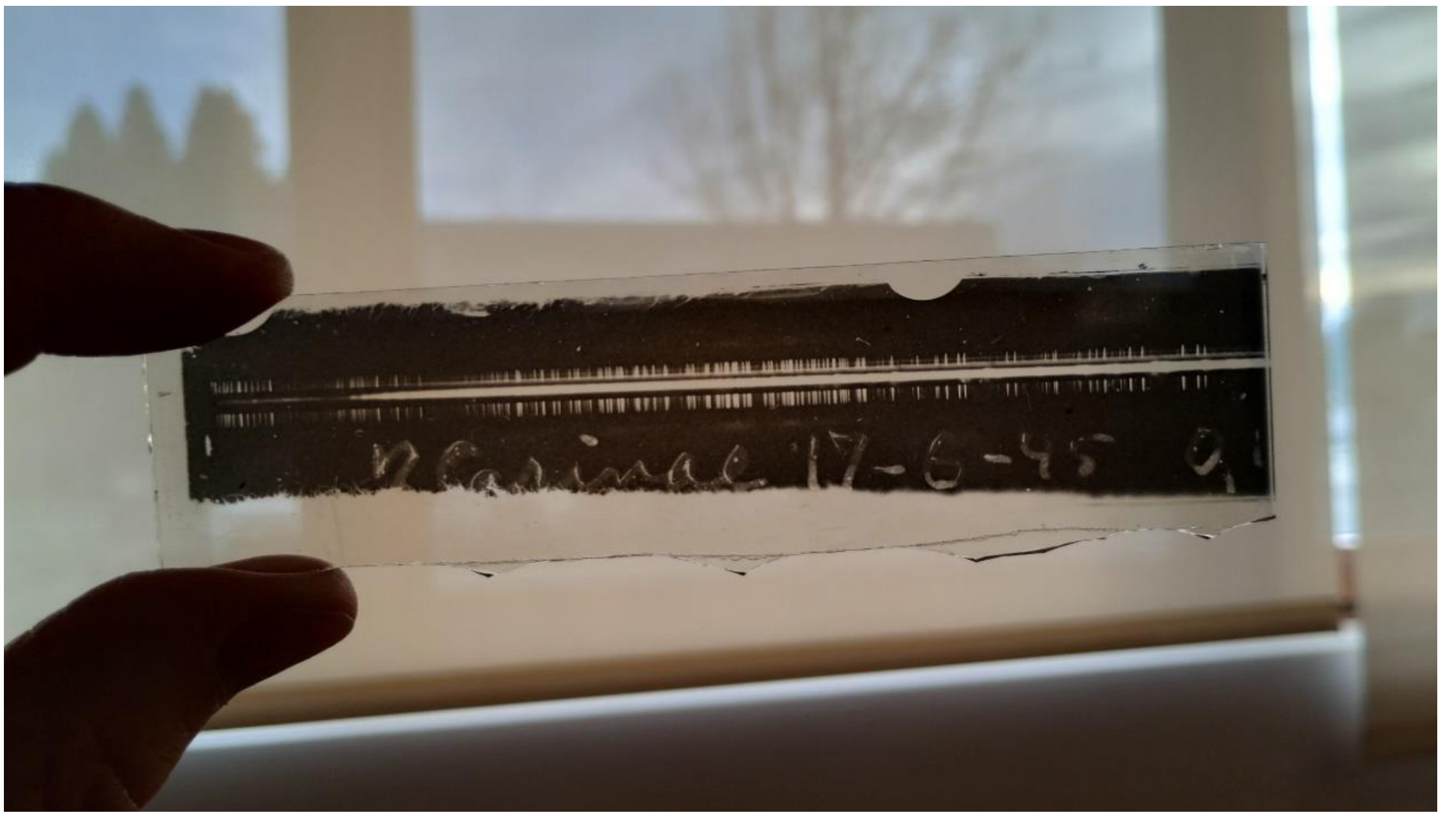

Y también tenemos hojas y hojas de tablas de líneas espectrales. En la imagen de abajo, a la izquierda está la lista de todos los espectros. La tabla de la derecha tiene las velocidades calculadas (ver el encabezamiento marcado en rojo). Esas velocidades fueron cruciales para determinar la dinámica del Homúnculo, y establecer que se había originado en la Gran Erupción de 1843.

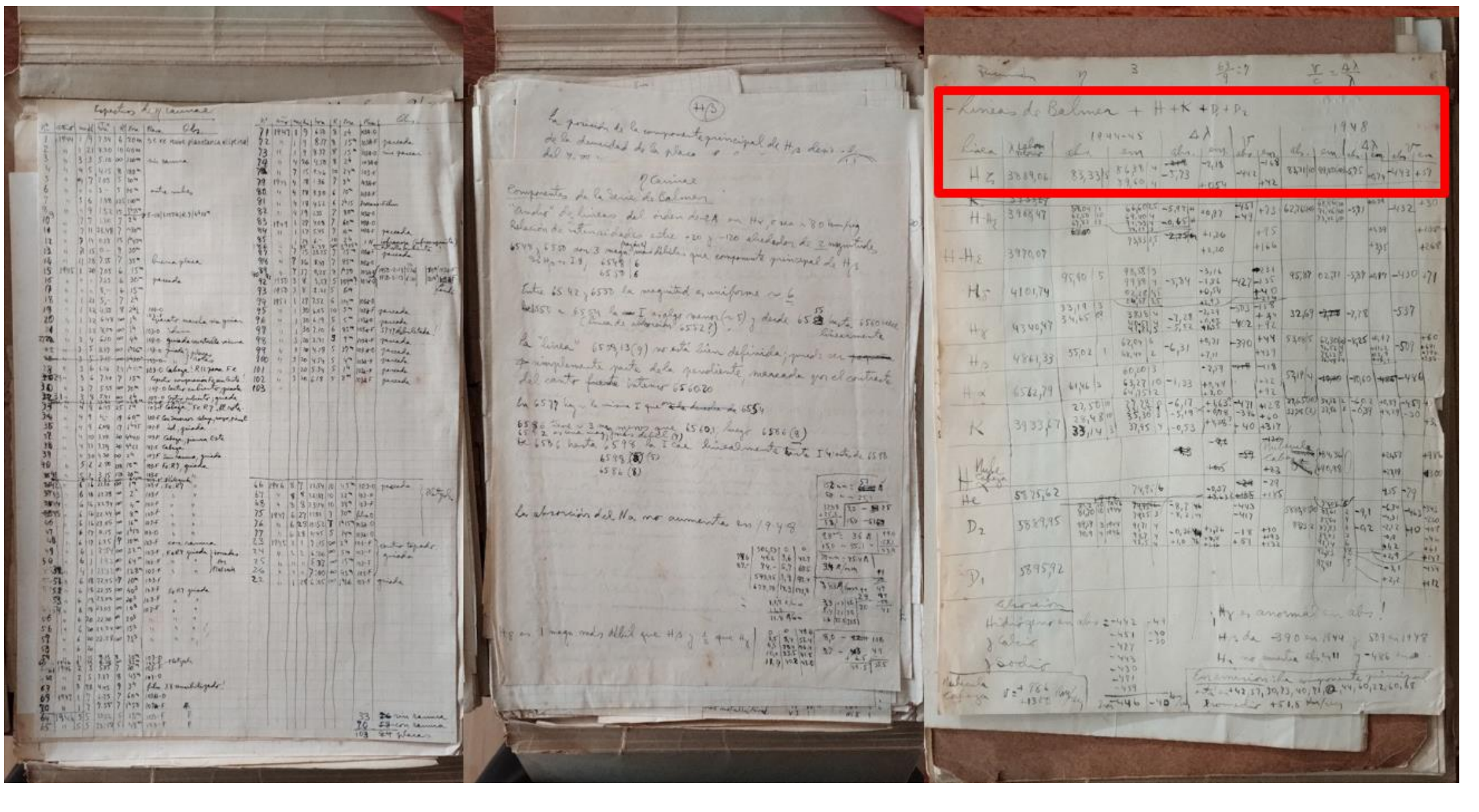

El hecho de que el Homúnculo sea tan peculiar se debe a que es tan joven. Gaviola fue el primero en estudiar esta nebulosa fotográficamente a fondo, y medir la velocidad con la que se está expandiendo.

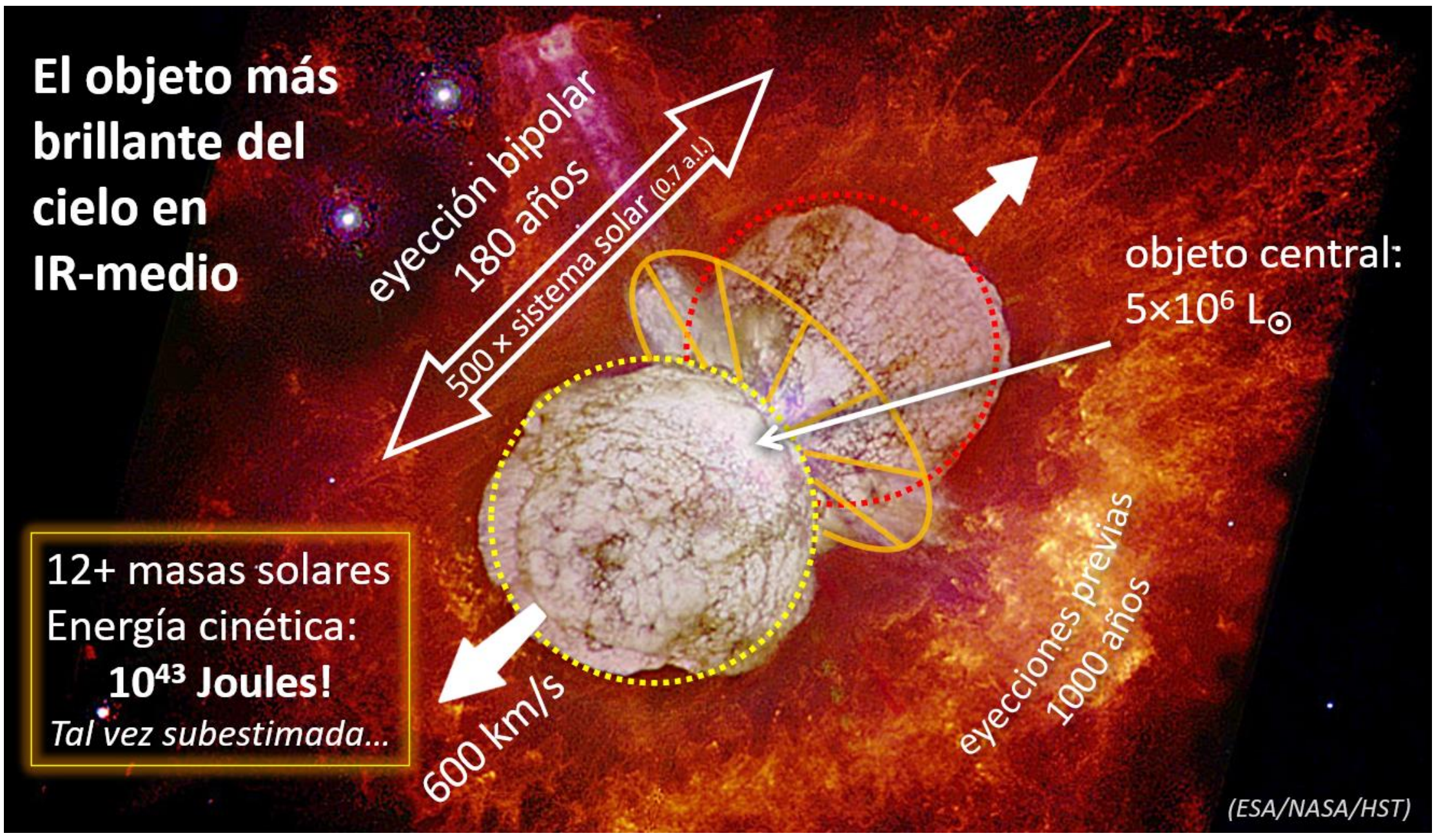

Sabemos que su masa es de entre 12 y 20 masas solares, compuesto de gas y polvo, y que es muy denso. Se encuentra expandiéndose libremente a unos 600 km/s (medidos por primera vez por Gaviola). Existen regiones en el interior con velocidades aún más increíbles, de hasta 3000 km/s. Sólo la energía cinética puede calcularse en $10^{43}$ Joules. Es una energía inmensa, comparable a la de una explosión de supernova, pero la estrella sobrevivió.

Tiene una falda ecuatorial mucho menos densa, rápida y joven, y lo rodea material más antiguo, que ya está frenándose, muy parecido a los restos de una supernova. Una onda rápida de 1843 ya está impactando contra este medio, brillando en rayos X, con tanta energía como el Homúnculo entero. Es el objeto más brillante del cielo en infrarrojo de 5 a 20 micrones, con una temperatura de unos 140 K. Funciona, de hecho, como un calorímetro del objeto interior, que resulta con un brillo de millones de soles. Curiosamente, refleja la luz de la estrella interior desde muchas direcciones, y fuera del Sol es la única estrella para la cual podemos observar algo así. Hay un Pequeño Homúnculo dentro, más lento, más joven, eyectado en la "pequeña erupción" de 1890.

Eta Carinae misma (la estrella) es muy difícil de estudiar, tan oscurecida está por el Homúnculo. Es muy brillante en rayos X, y los telescopios espaciales dedicados finalmente permitieron un importante descubrimiento: además de la variabilidad irregular, había una muy regular. Esto correspondía a la presencia de una compañera en órbita, con un período de 5 años y medio y una órbita muy elongada, similar en forma y tamaño a la de nuestro cometa Halley. Ya se han estudiado varios periastros en detalle, con varios descubrimientos increíbles. Por ejemplo, este "evento espectroscópico", como lo llaman. Es una observación de una línea prohibida del helio. Muestra más de 300 luminosidades solares en una única línea espectroscópica.

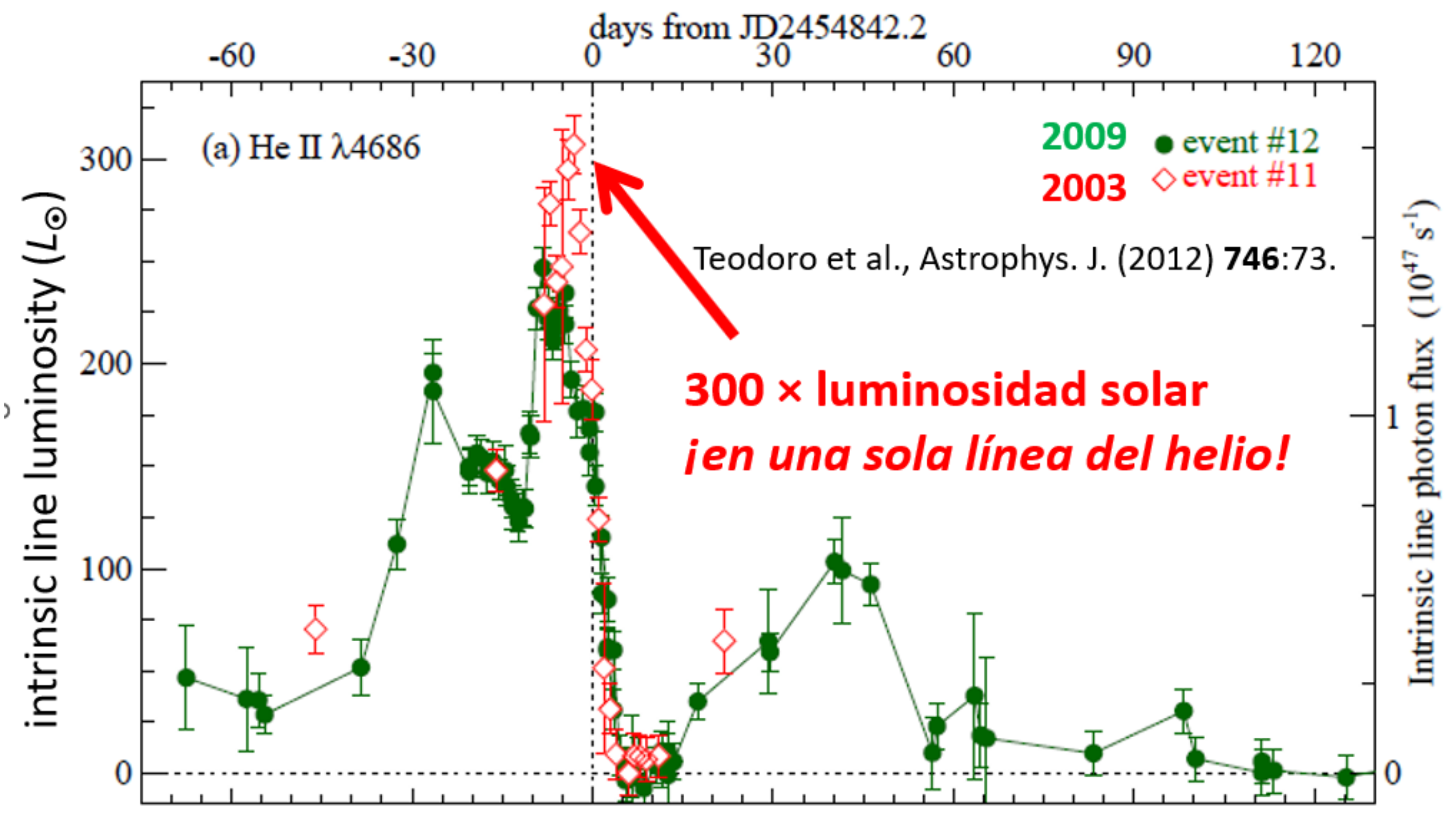

Estos eventos de rayos X han sido modelados con gran éxito como la colisión de los vientos de ambas estrellas. La secundaria no se ve directamente, pero su viento es más rápido que el de la primaria, y repite en cada órbita una loca espiral de turbulencia. Todavía no es del todo obvio cuál es la orientación de sus órbitas en el espacio, pero otros parámetros ya han sido determinados con precisión. Se acepta en general que la estrella primaria es de 90 masas solares, y que la secundaria es de 30, un monstruo sólo empequeñecido por Eta Car A. Considerando las sucesivas erupciones, se conjetura que la primaria nació con más de 150 masas solares, lo cual la hace una de las estrellas más pesadas de la era actual del universo.

A partir de diversos episodios de aumento y disminución de brillo y de líneas espectrales, se ha inferido una compleja estructura interna con cáscaras y nubes compactas, que han ido evolucionando bastante rápidamente. Se estima que el "ocultador" actual (tan pequeño que funciona como un coronógrafo), está disipándose desde los 2000, a 0.11 mag/año, y que desaparecerá en la próxima década, a más tardar en 2036, cuando la estrella alcance V=2.5.

Estas estrellas hiper masivas no duran mucho. El mecanismo exacto de su final no está del todo claro. Podría ser una supernova de colapso de núcleo, como las que se esperan para Antares o Betelgeuse, o podría ser otra inestabilidad. En todo caso, es seguro que los combustibles nucleares se acabarán rápidamente, y que Eta explotará en un futuro cercano. Entendido en sentido astronómico, por supuesto. Puede ser en el próximo millón de años, o puede ser este fin de semana. Cuando lo haga, brillará en nuestro cielo con magnitud –7, más brillante incluso que el planeta Venus en su máximo brillo.

Cuando esto ocurra las capas de elementos cada vez más pesados que se habrán ido acumulando como capas de cebolla serán expulsados al espacio interestelar. Junto con otros que se sintetizarán en el momento de la explosión. La explosión se propagará a través de la Nebulosa de Carina, disparando la formación de nuevas estrellas.

Estrellas como Eta Carinae, aun siendo tan raras, juegan un rol crucial en la evolución química y dinámica de la galaxia. Fue en estrellas como Eta Carinae donde se forjaron los átomos (salvo el hidrógeno) de nuestros cuerpos, de nuestras cosas, de nuestro mundo. Y fue probablemente la explosión de una estrella como Eta Carinae la que, hace 5 mil millones de años, puso en movimiento esos átomos que acabarían formando el sistema solar, la Tierra, y a nosotros mismos.

Regresemos a Gaviola, que se fue y volvió y se volvió a ir del Observatorio de Córdoba. Trabajó en muchos lados, en la cristalería Rigolleau, en General Electric, y a fines de los 50 estuvo en la Universidad de Córdoba, en el IMAF (actual FAMAF). De esos años es este cuadernito, donde hay listas de alumnos de sus clases. Se reconocen personalidades conocidas

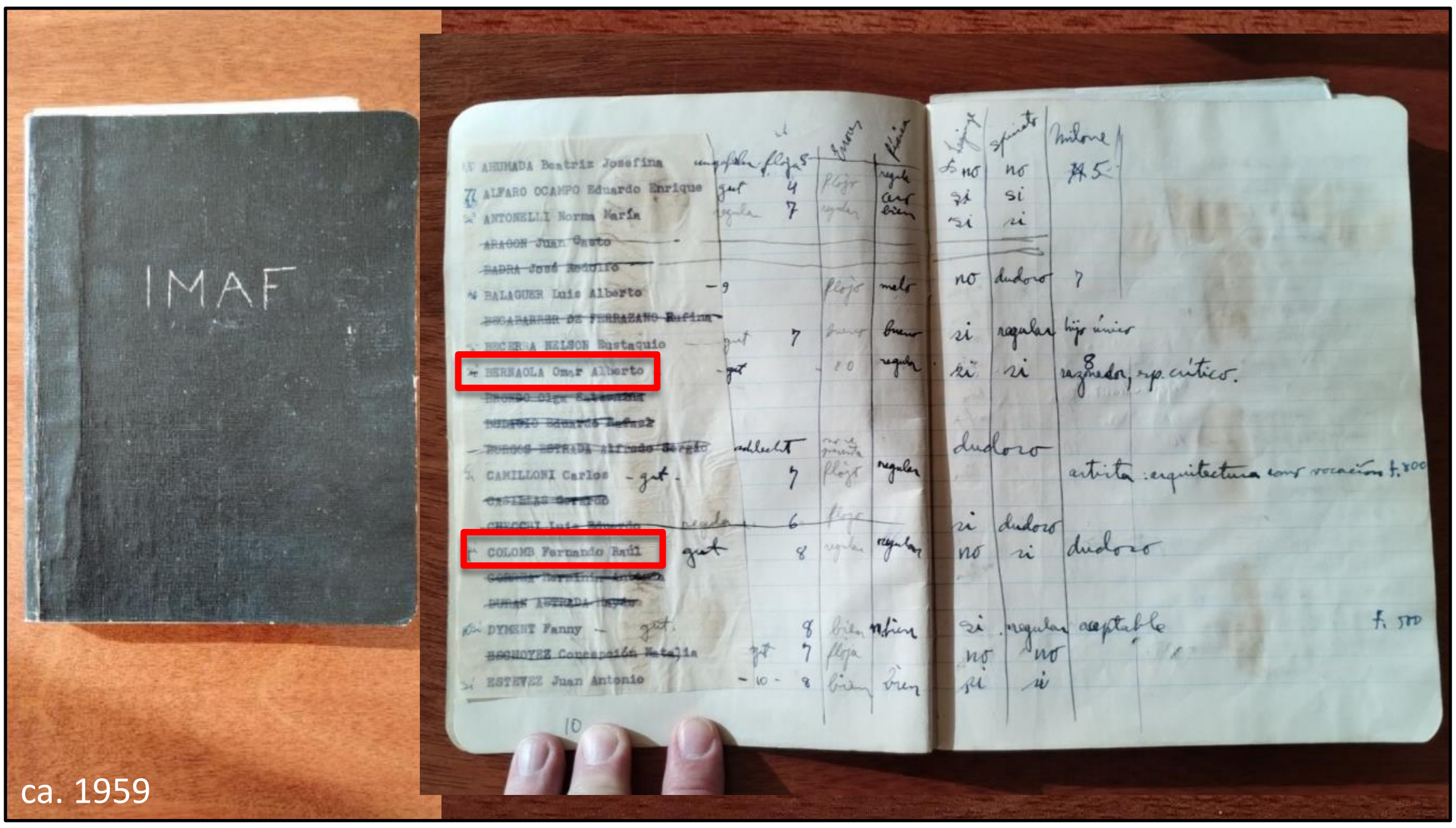

Marqué a Omar Bernaola, por ejemplo, físico de la CNEA y biógrafo de Gaviola. Está anotado como *"razonador, esp. crítico"*. Marqué también a Fernando Colomb, quien en cambio está marcado como *"dudoso"*, pero llegó a ser el fundador del Instituto Argentino de Radioastronomía, presidente de la AAA, del CASLEO, e Investigador Principal de los satélites SAC-C y SAC-D/Aquarius.

Hay más, mezclados con temas de clase y notas de exámenes. Seguro que hay más gente conocida. Y en el reverso de las mismas páginas hay anotaciones sobre su método de la cáustica para medir superficies ópticas. Tiene aspecto de ser material preparado para una clase, más que trabajo de laboratorio.

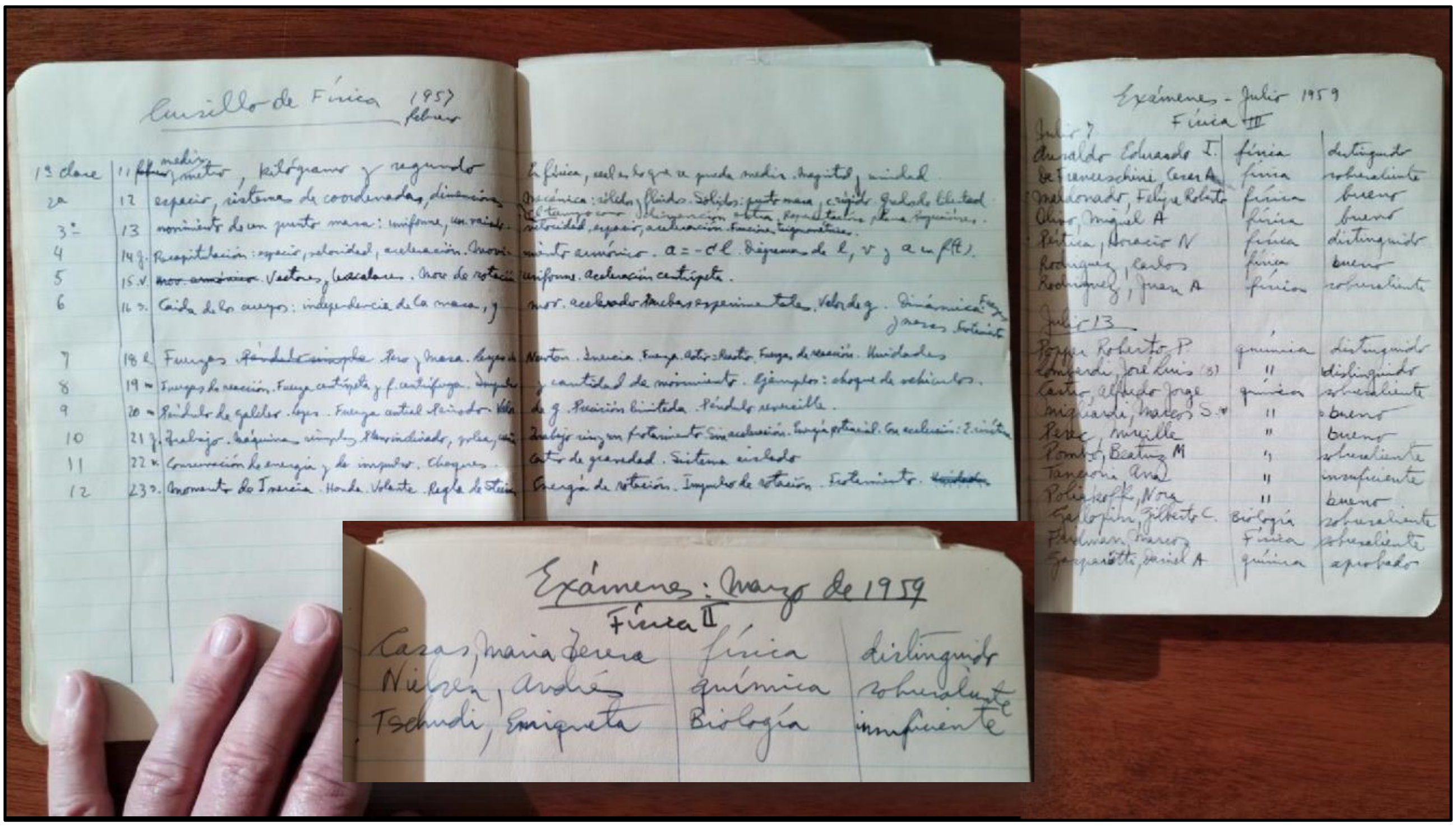

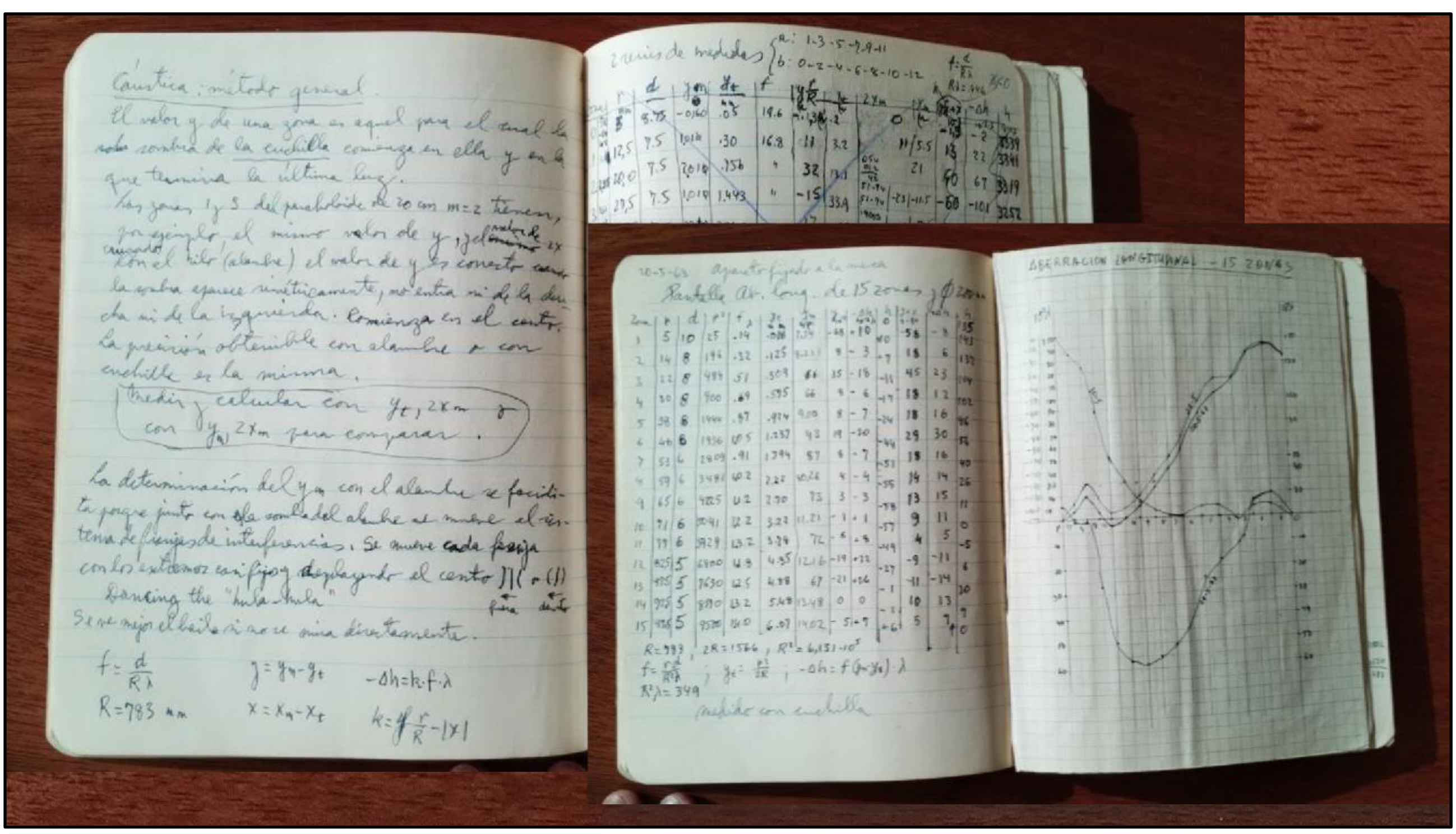

De esos años hay también unas notas manuscritas de Termodinámica, y conferencias sobre Eta Carinae.

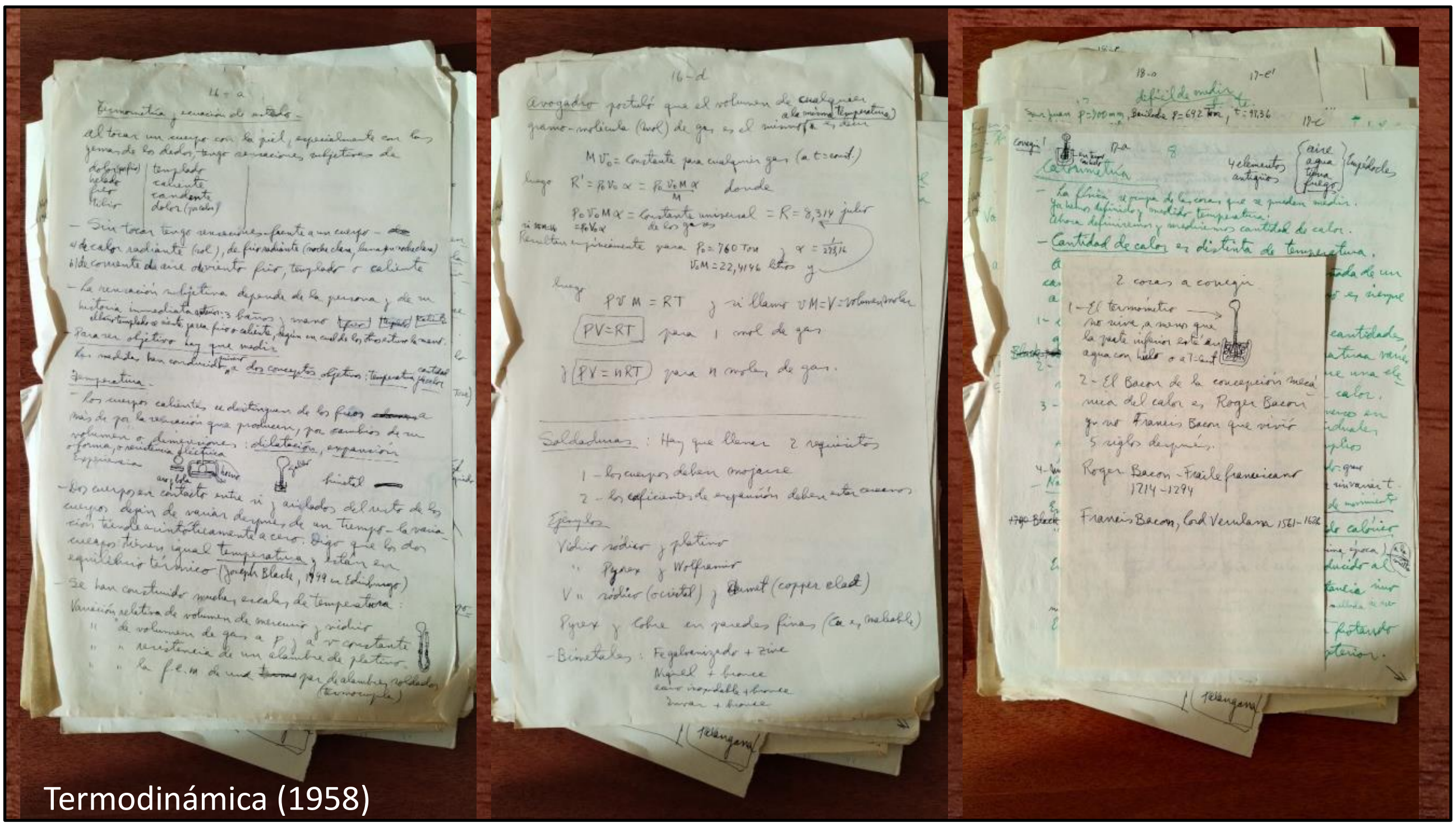

Termodinámica (1958)

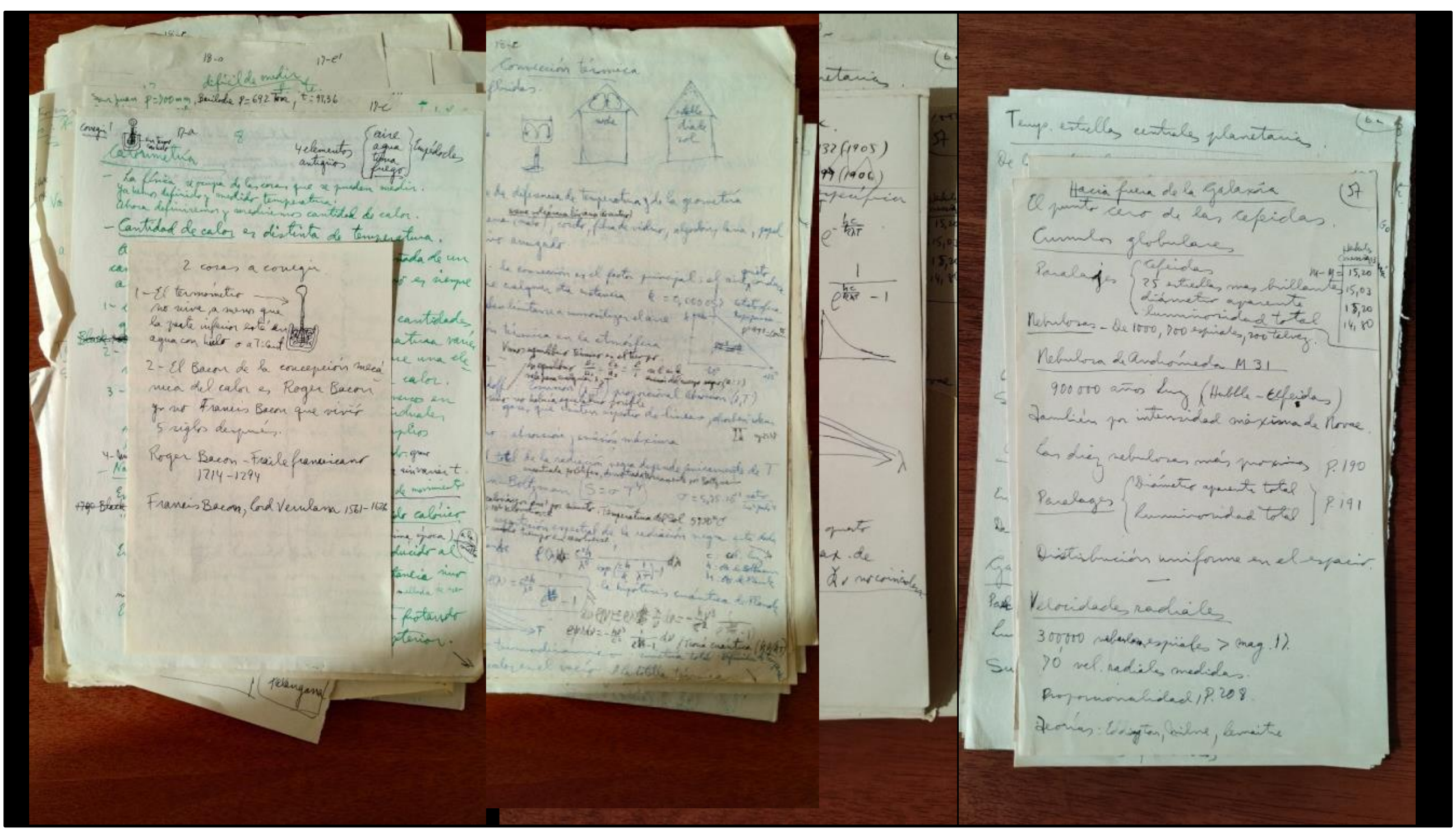

En la Reunión de la AFA en La Plata, en 1953, Gaviola se reunió con Ernesto Galloni y José Balseiro, para discutir las posibilidades que había dejado la interrupción del Proyecto de Richter. Poco después presentaron a las autoridades de la CNEA la propuesta de la creación de un Instituto de Física, en varias reuniones y también por escrito. Al principio todo parecía marchar bien, pero en septiembre, en una reunión, para Gaviola se pudrió todo. La CNEA quería hacerlo, pero suprimiendo los primeros 2 años (algo que Gaviola interpretó como un “curso de perfeccionamiento”, en lugar de una carrera universitaria). El hecho de que las autoridades fueran militares no ayudó a la interacción con Gaviola. Se levantó y se fue. La sucesión de eventos está relatada en este documento, que tituló Cronología de la Prehistoria, que se conserva junto a este papelito, donde garabateó sus impresiones de ese día, y también una carta muy indignada a Iraolagoitía, que no sé si llegó a enviar.

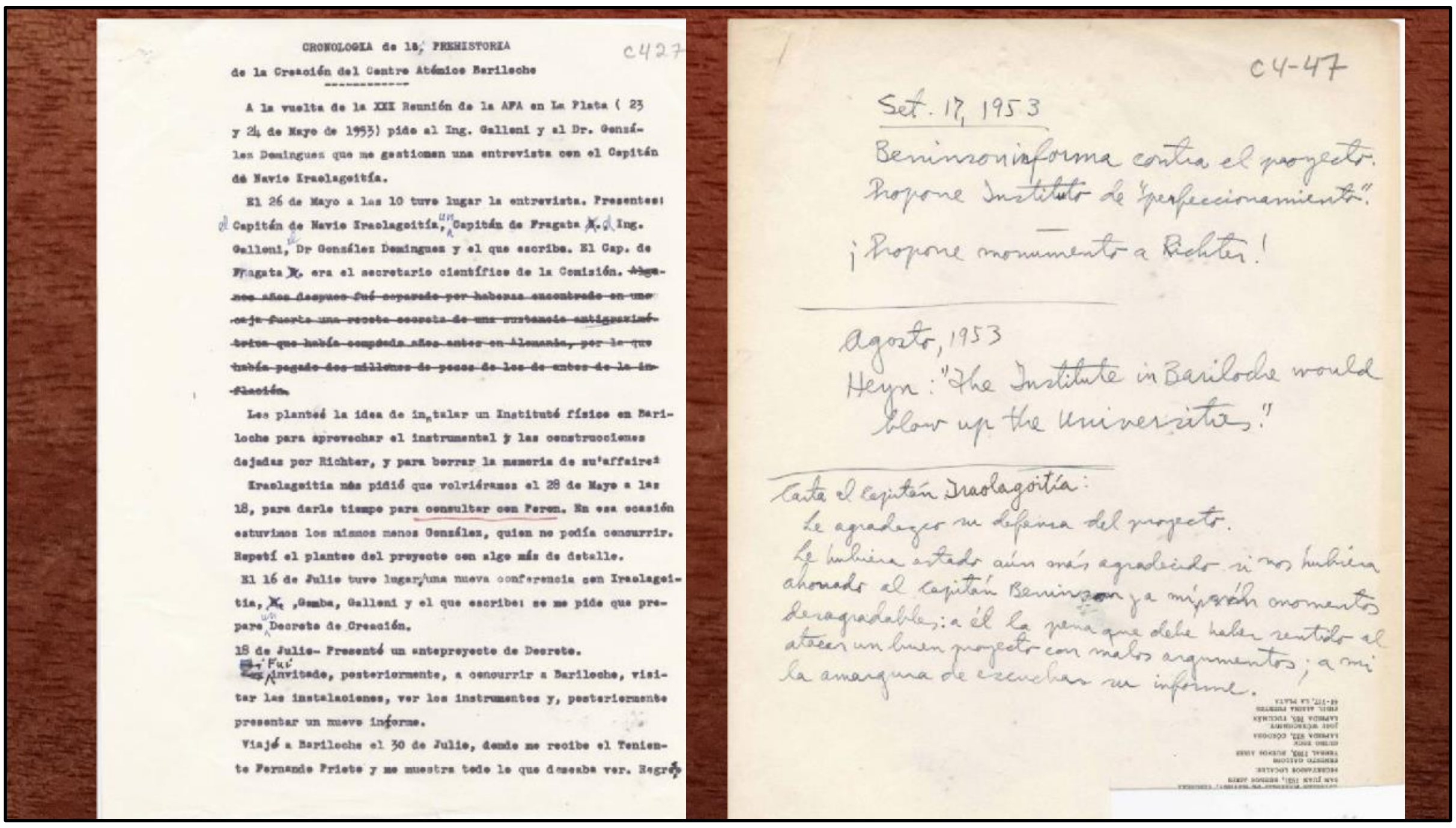

c427

CRONOLOGIA de la PREHISTORIA
de la Creación del Centro Atómico Bariloche

A la vuelta de la XXI Reunión de la AFA en La Plata ( 23 y 24 de Mayo de 1953) pido al Ing. Galloni y al Dr. González Dominguez que me gestionen una entrevista con el Capitán de Navio Iraolagoitía.

El 26 de Mayo a las 10 tuvo lugar la entrevista. Presentes: el Capitán de Navio Iraolagoitía, un Capitán de Fragata X., el Ing. Galloni, el Dr González Dominguez y el que escribe. El Cap. de Fragata X. era el secretario científico de la Comisión. ~~Algunos años despues fué separado por haberse encontrado en una caja fuerte una receta secreta de una sustancia [illegible] que había comprado años antes en Alemania, por lo que había pagado dos millones de pesos de los de antes de la inflación.~~

Les planteé la idea de instalar un Instituto físico en Bariloche para aprovechar el instrumental y las construcciones dejadas por Richter, y para borrar la memoria de su 'affaire'.

Iraolagoitia nos pidió que volviéramos el 28 de Mayo a las 18, para darle tiempo para consultar con Peron. En esa ocasión estuvimos los mismos menos González, quien no podía concurrir. Repetí el planteo del proyecto con algo más de detalle.

El 16 de Julio tuvo lugar una nueva conferencia con Iraolagoitia, X., Gamba, Galloni y el que escribe: se me pide que prepare un Decreto de Creación.

18 de Julio- Presenté un anteproyecto de Decreto.

Fui invitado, posteriormente, a concurrir a Bariloche, visitar las instalaciones, ver los instrumentos y, posteriormente presentar un nuevo informe.

Viajé a Bariloche el 30 de Julio, donde me recibe el Teniente Fernando Prieto y me muestra todo lo que deseaba ver. Regresé

C4-47

Set. 17, 1953

Beninson informa contra el proyecto.
Propone Instituto de "perfeccionamiento".

¡Propone monumento a Richter!

Agosto, 1953
Heyn: "The Institute in Bariloche would blow up the universities!"

Carta al Capitán Iraolagoitía:
Le agradezco su defensa del proyecto.
Le hubiera estado aún más agradecido si nos hubiera ahorrado al capitán Beninson y a mí esos momentos desagradables: a él la pena que debe haber sentido al atacar un buen proyecto con malos argumentos; a mí la amargura de escuchar su informe.

Balseiro, mucho más conciliador, siguió reuniéndose, organizó dos escuelas de verano, y finalmente en agosto de 1955 comenzaron las clases en el Instituto de Física, que se convirtió en el trabajo de su vida. Aquí los vemos juntos en 1961, durante una visita de Gaviola a Bariloche.

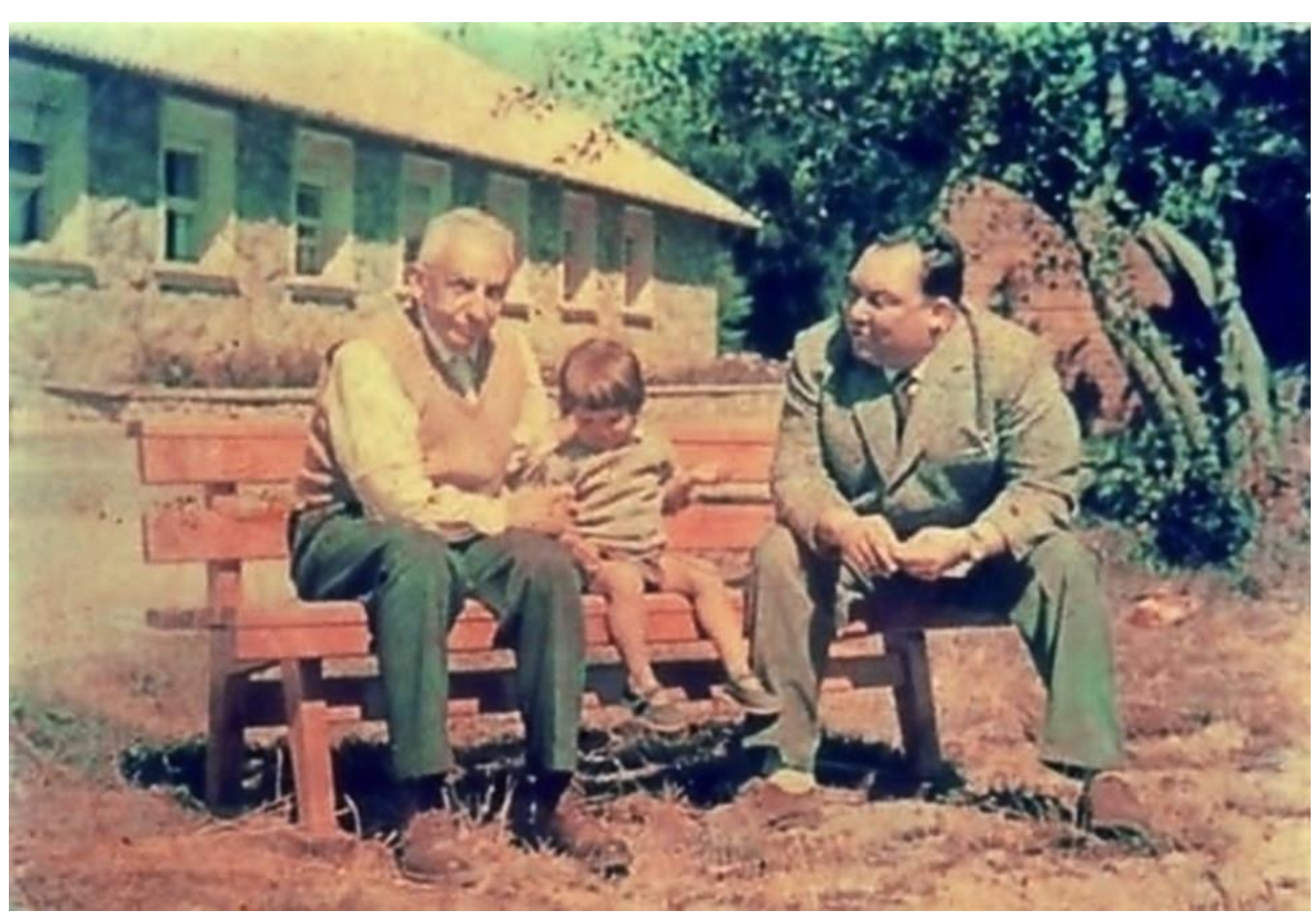

En marzo de 1962 murió Balseiro, muy joven. En un homenaje a socios fallecidos de la AFA, Gaviola dijo sobre él:

*«Balseiro fue el conquistador intrépido que en 1955 fue al desierto cultural de Bariloche y en siete años creó, organizó, orientó y le insufló su espíritu a la mejor escuela de Física de Latinoamérica. Desde Emil Bose no se había visto un milagro semejante.»*

En 1963, el nuevo director del Instituto, Carlos Mallman, convocó a Gaviola para que se uniera al plantel docente. Gaviola, con gran generosidad (tenía 63, y se había peleado con la CNEA) acepta.

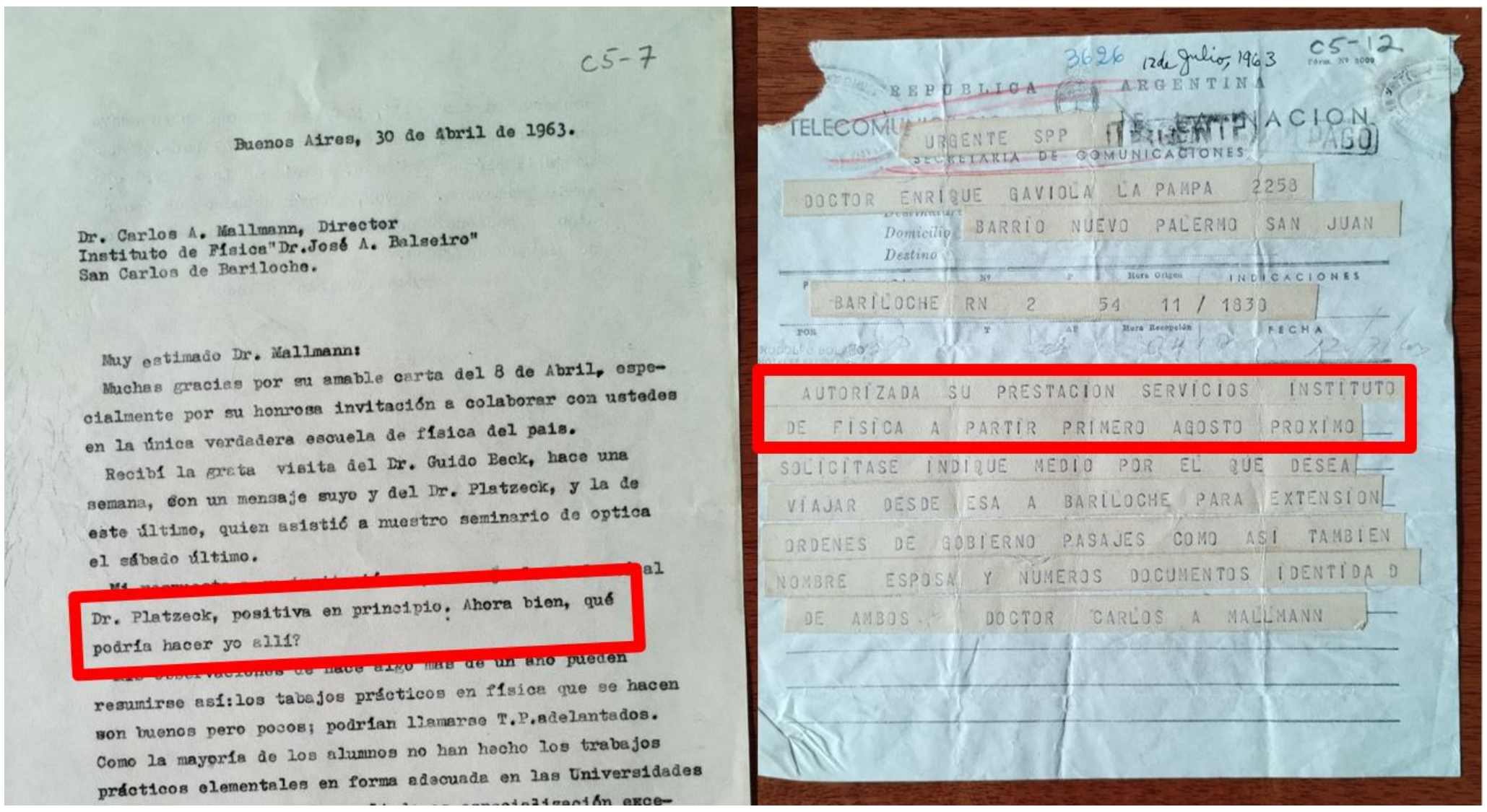

C5-7

Buenos Aires, 30 de Abril de 1963.

Dr. Carlos A. Mallmann, Director
Instituto de Física"Dr.José A. Balseiro"
San Carlos de Bariloche.

Muy estimado Dr. Mallmann:

Muchas gracias por su amable carta del 8 de Abril, especialmente por su honrosa invitación a colaborar con ustedes en la única verdadera escuela de física del país.

Recibí la grata visita del Dr. Guido Beck, hace una semana, con un mensaje suyo y del Dr. Platzeck, y la de este último, quien asistió a nuestro seminario de optica el sábado último.

 al Dr. Platzeck, positiva en principio. Ahora bien, qué podría hacer yo allí?

 de hace algo más de un año pueden resumirse así:los tabajos prácticos en física que se hacen son buenos pero pocos; podrían llamarse T.P.adelantados. Como la mayoría de los alumnos no han hecho los trabajos prácticos elementales en forma adecuada en las Universidades 

3626 12 de Julio, 1963 C5-12

REPÚBLICA ARGENTINA
TELECOMUNICACIONES
SECRETARÍA DE COMUNICACIONES

URGENTE SPP

DOCTOR ENRIQUE GAVIOLA LA PAMPA 2258
BARRIO NUEVO PALERMO SAN JUAN

BARILOCHE RN 2 54 11 / 1830

AUTORIZADA SU PRESTACION SERVICIOS INSTITUTO DE FISICA A PARTIR PRIMERO AGOSTO PROXIMO SOLICITASE INDIQUE MEDIO POR EL QUE DESEA VIAJAR DESDE ESA A BARILOCHE PARA EXTENSION ORDENES DE GOBIERNO PASAJES COMO ASI TAMBIEN NOMBRE ESPOSA Y NUMEROS DOCUMENTOS IDENTIDAD DE AMBOS. DOCTOR CARLOS A MALLMANN

Varios cursos tuvieron la gran oportunidad de interactuar con él, aprender de él. Produjo una gran impresión, que todavía se siente en algunos ámbitos del Balseiro. De sus clases de Física Experimental hay una cantidad de apuntes sueltos. Se ve que usó apuntes previos de Balseiro.

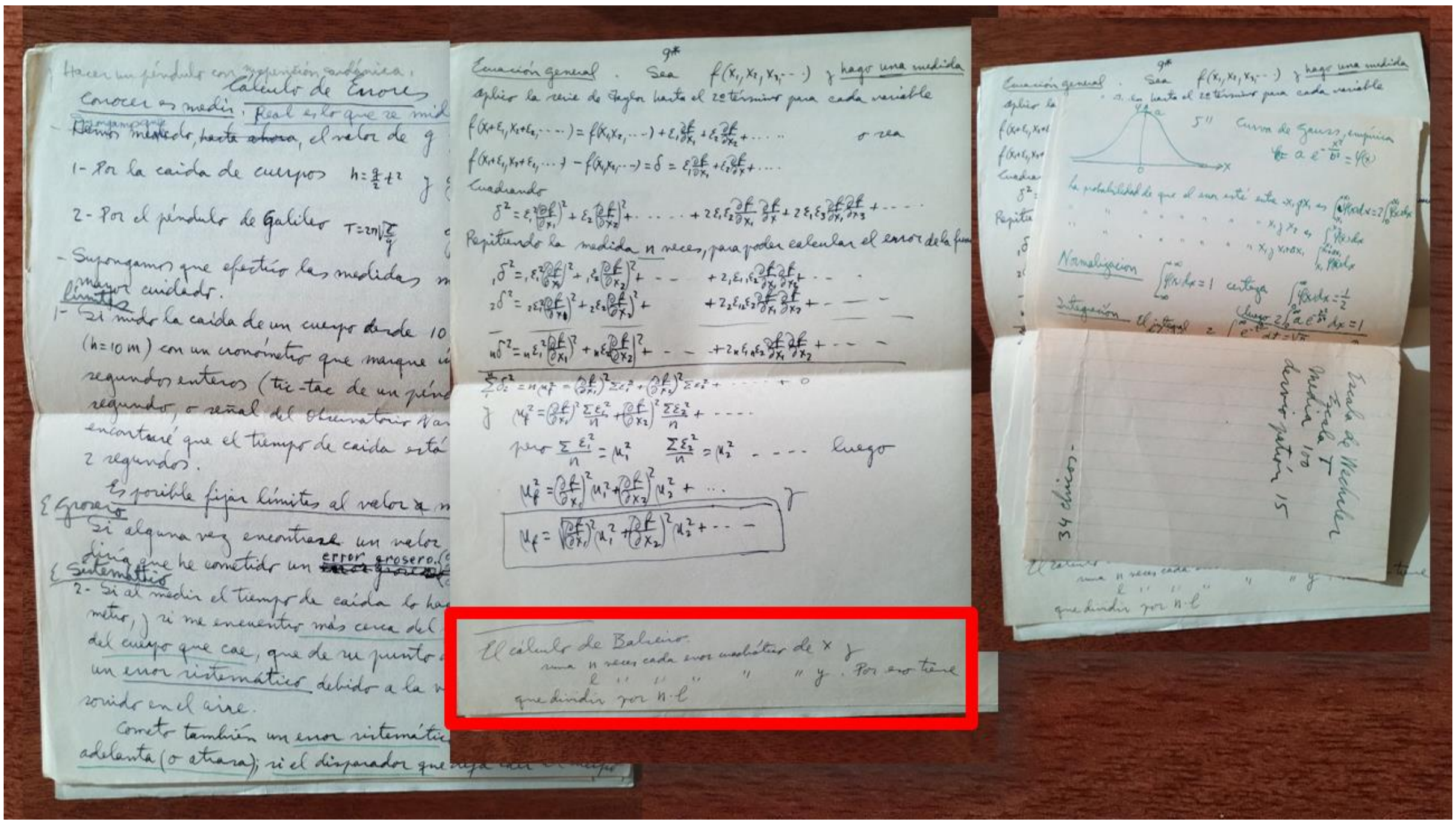

Para ir terminando les traje su última batalla quijotesca. Los que conocen el Instituto Balseiro saben que es un lugar hermoso, lleno de árboles. Lo que no todos saben es que prácticamente todos esos árboles los plantó Gaviola. El Centro Atómico Bariloche era un descampado. Para cambiarle el aspecto de cuartel (que es lo que había sido), Gaviola consiguió que le donaran 3000 retoños, de numerosas especies, y los plantó el mismo.

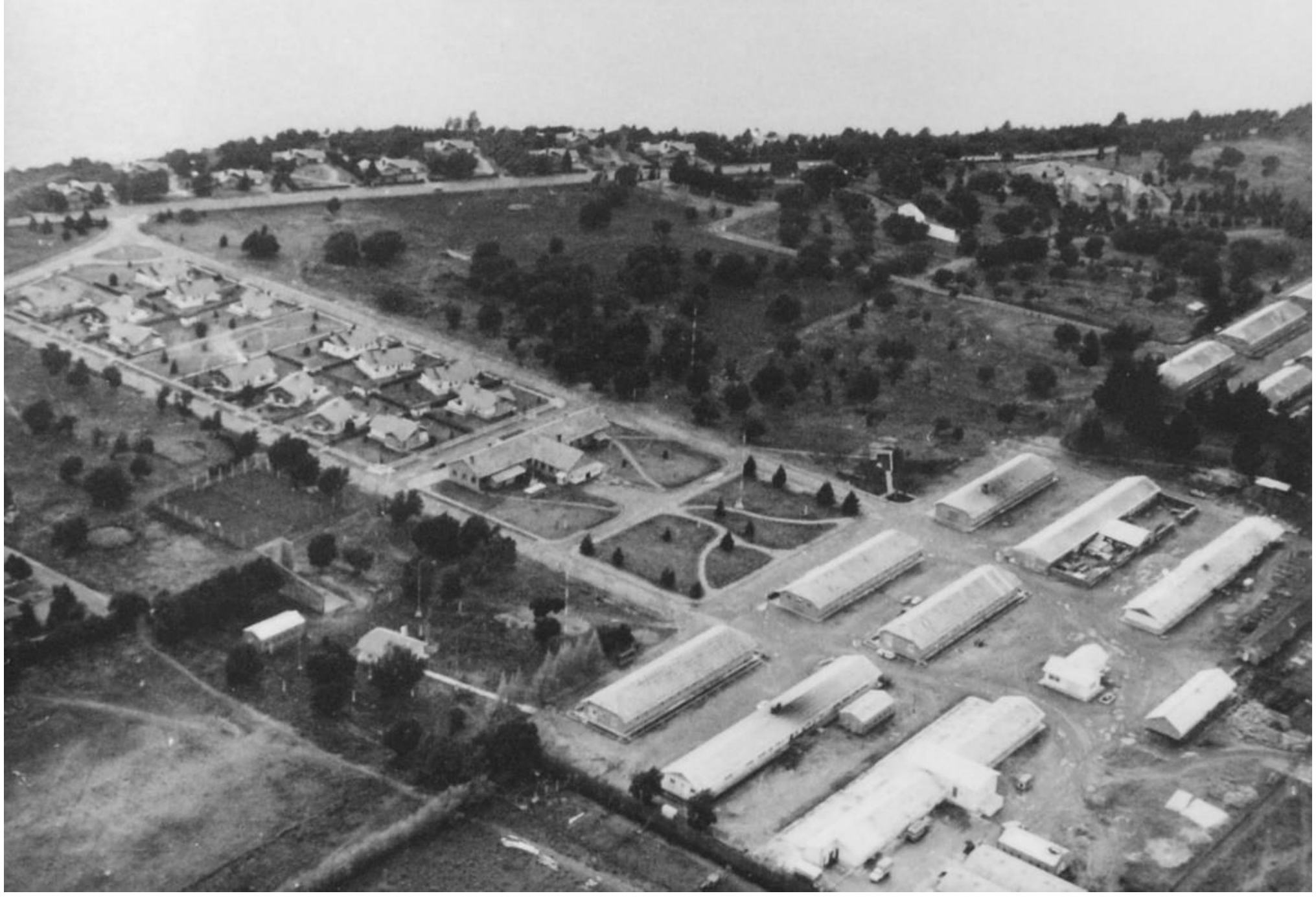

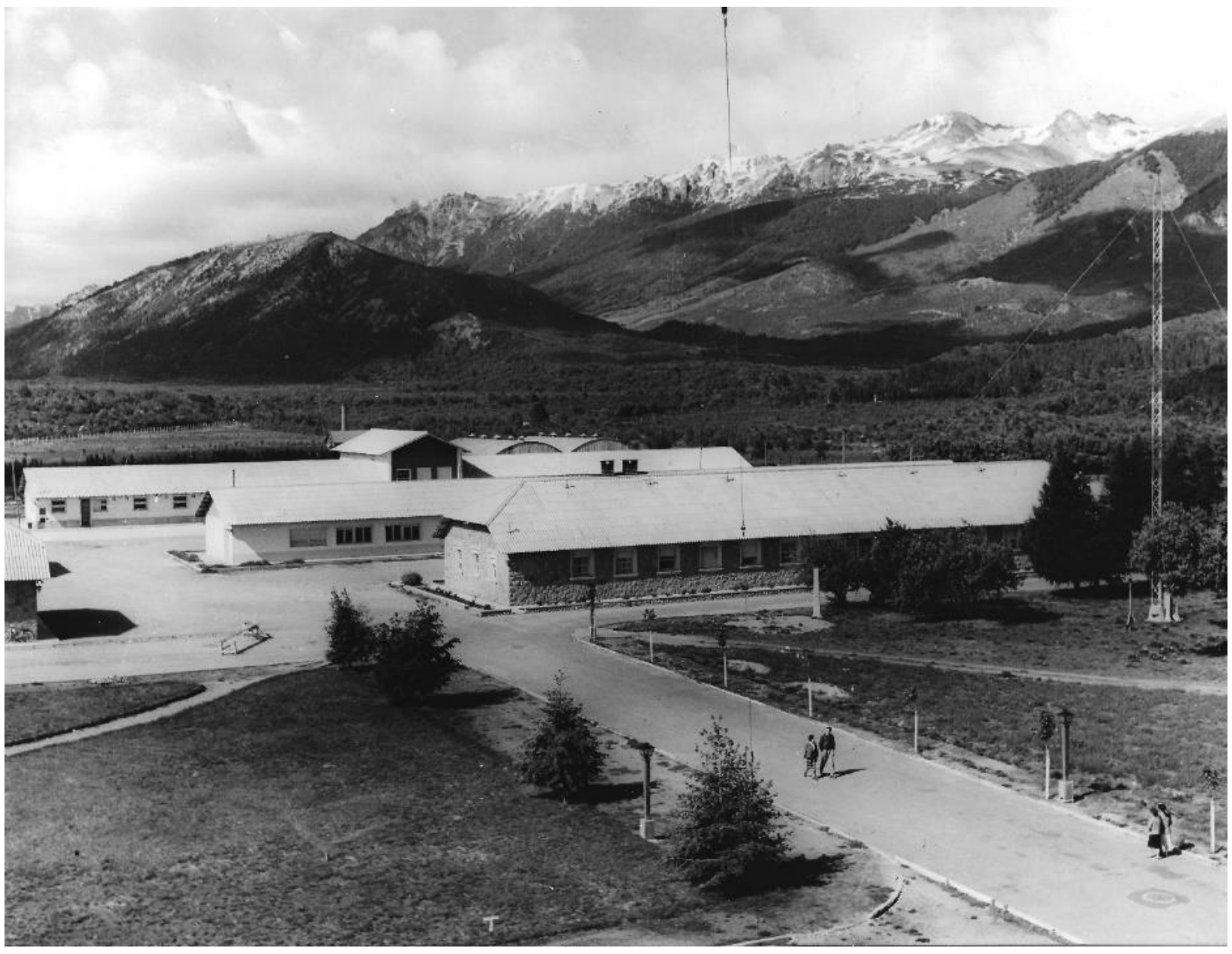

Un día, tal vez ya estaba retirado, o fue su último año, el Director hizo un plan de pavimentar algunas de las calles de tierra. ¡Y para eso iban a talar varios de los árboles de Gaviola! Gaviola declaró la *Guerra de la Clorofila*. Se conservan las cartas con sus argumentos. Es una guerra que en parte ganó, porque vi los planos de la propuesta, y unos cuantos árboles que pude identificar todavía existen.

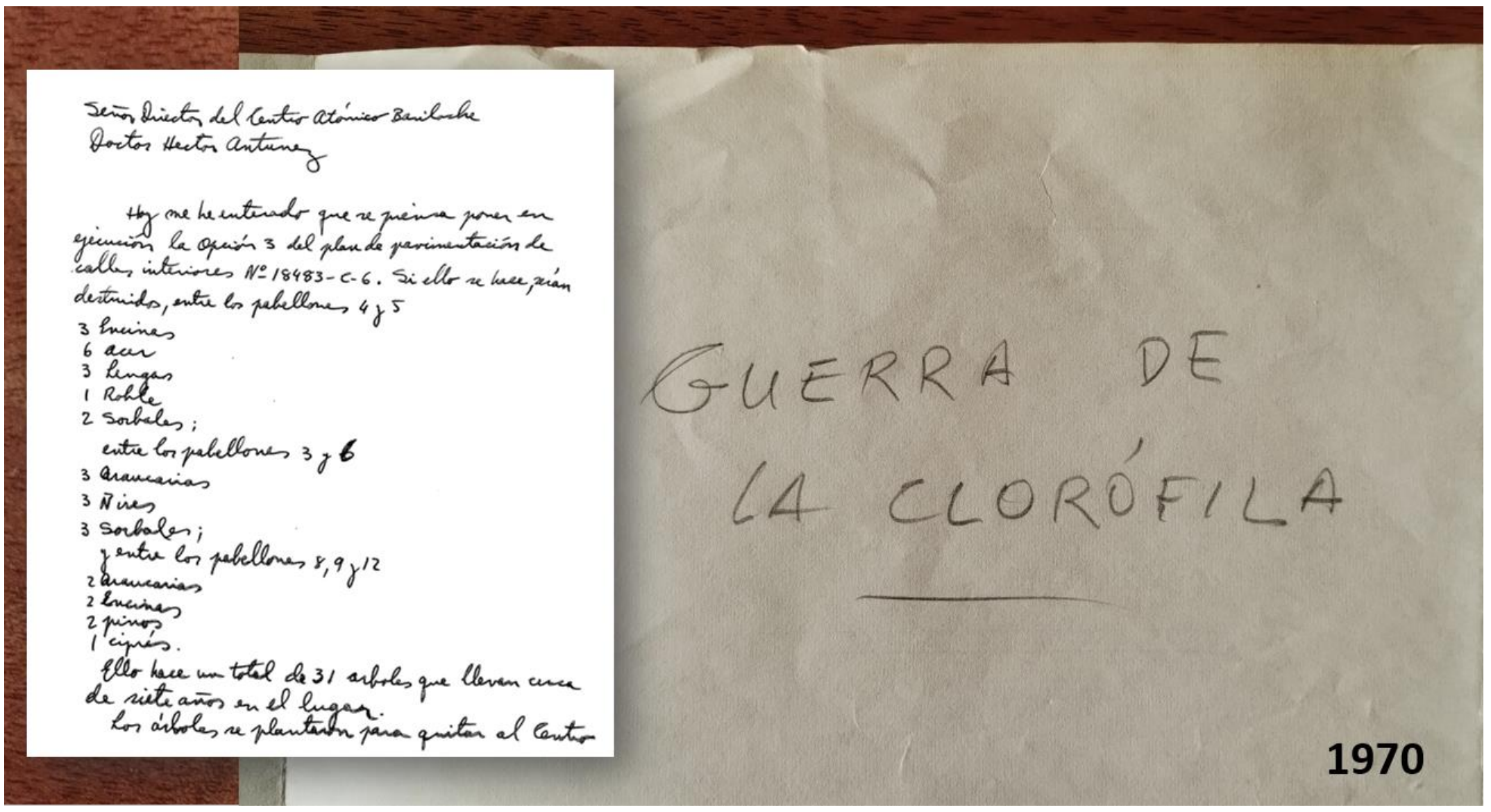

Gaviola estaba muy orgulloso de su trabajo en Bariloche. En un discurso en la DAIA, en 1965, en ocasión de recibir un premio de la colectividad judía, dijo que:

*«En el semi exilio de Bariloche estoy efectuando la labor más fecunda de mi vida; estoy ayudando a formar una decena de físicos por año.»*

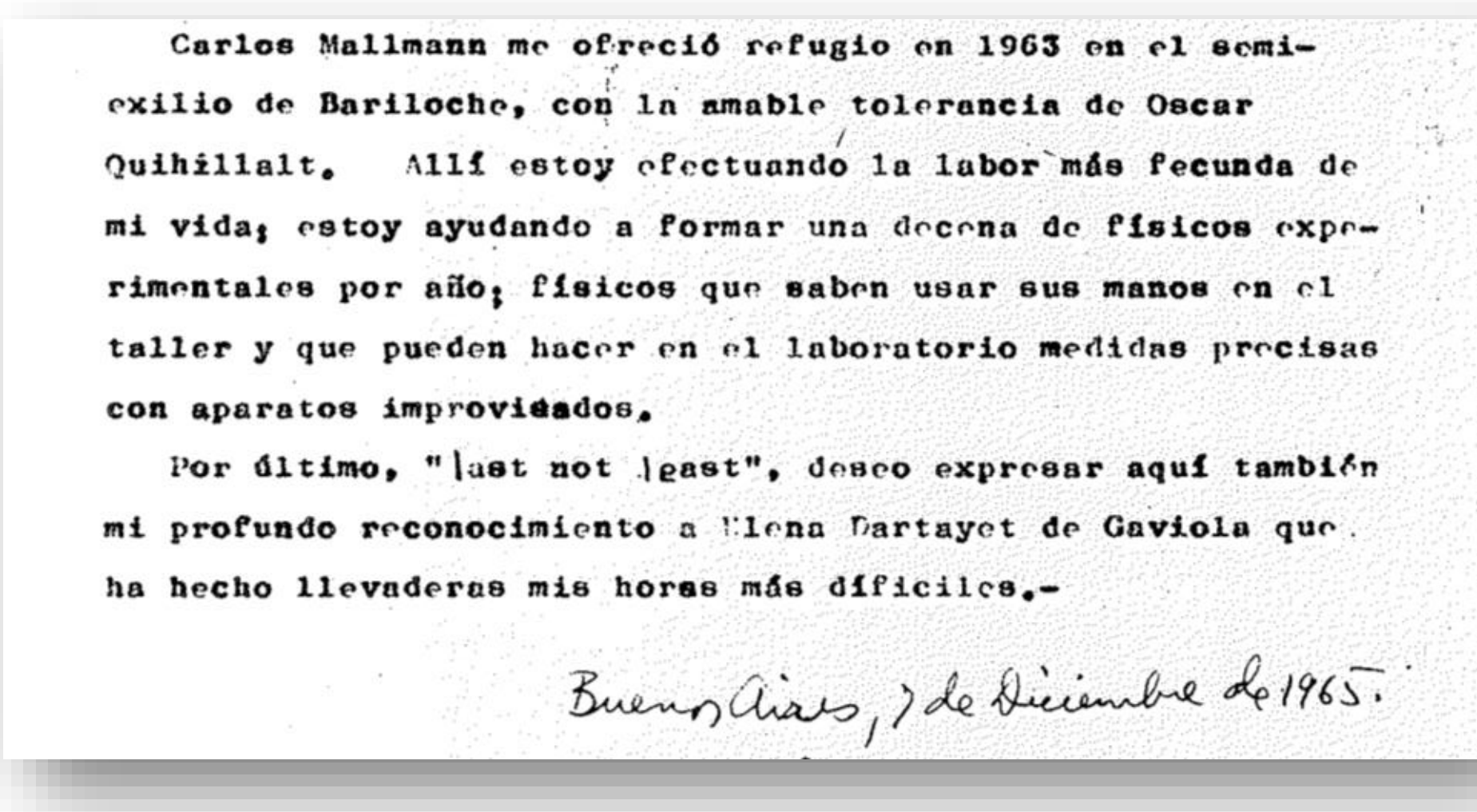

Carlos Mallmann me ofreció refugio en 1963 en el semi-exilio de Bariloche, con la amable tolerancia de Oscar Quihillalt. Allí estoy efectuando la labor más fecunda de mi vida; estoy ayudando a formar una decena de físicos experimentales por año; físicos que saben usar sus manos en el taller y que pueden hacer en el laboratorio medidas precisas con aparatos improvisados.

Por último, "last not least", deseo expresar aquí también mi profundo reconocimiento a Elena Dartayet de Gaviola que ha hecho llevaderas mis horas más difíciles.-

Buenos Aires, 7 de Diciembre de 1965.

Enrique Gaviola no fue el primer físico ni el primer astrónomo de la Argentina. Pero me gusta pensar que fue el padre de la física y la astrofísica modernas. Es una pena que su trabajo y su legado no sean más conocidos, no sólo por el gran público, sino incluso por nuestros colegas. Así que hagamos lo posible por contarlo, como me lo contaron a mí cuando era un estudiante de 21 años, y por difundirlo lo más posible. Muchas gracias.

**Agradecimientos**

Agradezco a la Biblioteca Leo Falicov y al Archivo Histórico Norma Badino, del Instituto Balseiro, y a su personal (en particular a Tamara Cárcamo y Marisa Velazco), por la posibilidad de consultar el Archivo Gaviola y la ayuda que me brindaron. De allí son casi todas las imágenes de documentos históricos. Agradezco también a mis amigos y colegas que conocieron a Gaviola y me transmitieron anécdotas es impresiones: Norma Badino, Abe Kestelman, Pablo Tognetti, Carlos Balseiro, Oscar Bressan, María Elena de la Cruz, Paco de la Cruz y Roberto Iglesias.

La foto de Gaviola y Balseiro es gentileza de M.E. de la Cruz.

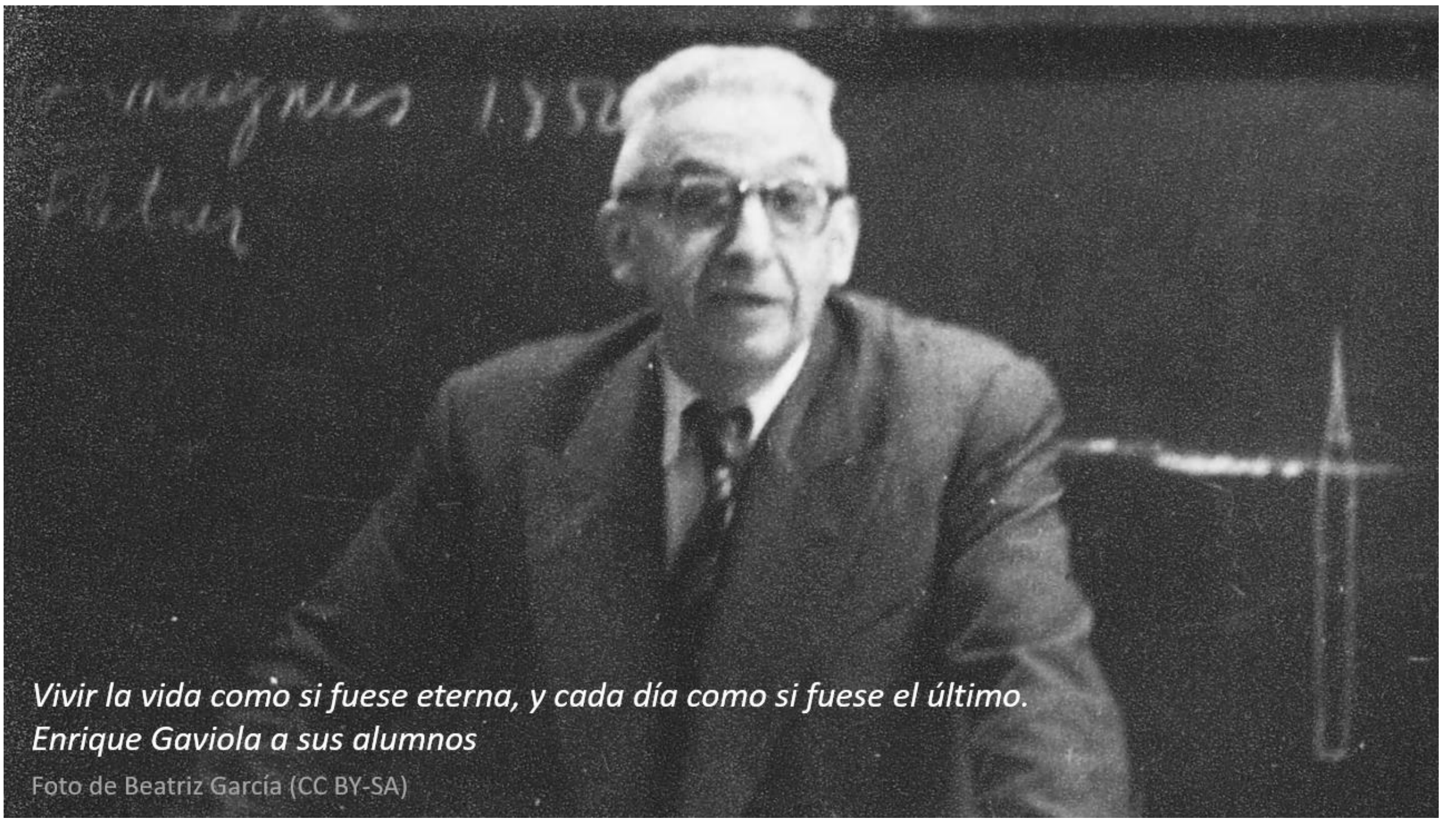

70 AÑOS Instituto Balseiro

BIBLIOTECA LEO FALICOV

AH NB ARCHIVO HISTÓRICO NORMA BADINO INSTITUTO BALSEIRO